\documentclass[8.5pt,twoside,twocolumn]{article}
\usepackage[super,sort&compress,comma]{natbib} 
\usepackage[version=4]{mhchem}
\usepackage{times,mathptmx}
\usepackage[dvipsnames]{xcolor}
\usepackage{sectsty}
\usepackage{titlesec}
\usepackage{balance} 
\usepackage{tabularx,subfigure}
\usepackage{array}      
\usepackage{multirow}   
\usepackage{booktabs} 
\usepackage{dblfloatfix} 
\usepackage{booktabs}  
\usepackage{array}      
\usepackage{xcolor}     
\usepackage{amsmath,graphicx} 
\usepackage{lastpage}
\usepackage[format=plain,justification=raggedright,singlelinecheck=false,font=small,labelfont=bf,labelsep=space]{caption} 
\usepackage{fancyhdr}
\usepackage{dblfloatfix}
\usepackage{xcolor}
\usepackage{graphicx}
\usepackage{etoolbox,hyperref}

\usepackage{parskip}

\begin{document}
\thispagestyle{plain}
\fancypagestyle{plain}{

\renewcommand{\headrulewidth}{1pt}}
\renewcommand{\thefootnote}{\fnsymbol{footnote}}
\renewcommand\footnoterule{\vspace*{1pt}%
\hrule width 3.4in height 0.4pt \vspace*{5pt}} 
\setcounter{secnumdepth}{5}

\captionsetup{justification=justified, singlelinecheck=off} 

\makeatletter 
\def\paragraph{\@startsection{paragraph}{4}{10pt}{-1.25ex plus -1ex minus -.1ex}{0ex plus 0ex}{\normalsize\textit}} 
\renewcommand\@biblabel[1]{#1}            
\renewcommand\@makefntext[1]%
{\noindent\makebox[0pt][r]{\@thefnmark\,}#1}
\makeatother 
\renewcommand{\figurename}{\small{Fig.}~}
\sectionfont{\large}
\subsectionfont{\normalsize} 

\fancyfoot{}
\fancyfoot[RO]{\footnotesize{\sffamily{~\textbar  \hspace{2pt}\thepage}}}
\fancyfoot[LE]{\footnotesize{\sffamily{\thepage~\textbar}}}
\fancyhead{}
\renewcommand{\headrulewidth}{1pt} 
\renewcommand{\footrulewidth}{1pt}
\setlength{\arrayrulewidth}{1pt}
\setlength{\columnsep}{6.5mm}
\setlength\bibsep{1pt}

\clubpenalty = 10000
\widowpenalty = 10000
\displaywidowpenalty = 10000

\twocolumn[
  \begin{@twocolumnfalse}
\noindent\Large{\textbf{Surface Stability Modeling with Universal Machine Learning Interatomic Potentials: A Comprehensive Cleavage Energy Benchmarking Study}}
\vspace{0.6cm}

\noindent\large{\textbf{Ardavan Mehdizadeh\textit{$^{1}$} and Peter Schindler\textit{$^{1,*}$}}}
\vspace{0.5cm}

\noindent\normalsize{Machine learning interatomic potentials (MLIPs) have revolutionized computational materials science by bridging the gap between quantum mechanical accuracy and classical simulation efficiency, enabling unprecedented exploration of materials properties across the periodic table. Despite their remarkable success in predicting bulk properties, no systematic evaluation has assessed how well these universal MLIPs (uMLIPs) can predict cleavage energies, a critical property governing fracture, catalysis, surface stability, and interfacial phenomena. Here, we present a comprehensive benchmark of 19 state-of-the-art uMLIPs for cleavage energy prediction using our previously established density functional theory (DFT) database of 36,718 slab structures spanning elemental, binary, and ternary metallic compounds. We evaluate diverse architectural paradigms, analyzing their performance across chemical compositions, crystal systems, thickness, and surface orientations. Our results reveal that training data composition dominates architectural sophistication: models trained on the Open Materials 2024 (OMat24) dataset, which emphasizes non-equilibrium configurations, achieve mean absolute percentage errors below 6\% and correctly identify the thermodynamically most stable surface terminations in 87\% of cases, without any explicit surface energy training. In contrast, architecturally identical models trained on equilibrium-only datasets show five-fold higher errors, while models trained on surface-adsorbate data fail catastrophically with a 17-fold degradation. Remarkably, simpler architectures trained on appropriate data achieve comparable accuracy to complex transformers while offering 10--100× computational speedup. These findings fundamentally reframe MLIP development priorities: rather than pursuing increasingly complex architectures, the community should focus on strategic training data generation that captures the relevant physical phenomena.}
\vspace{0.5cm}
\end{@twocolumnfalse}
]

\footnotetext{\textit{
$^{1}$~Northeastern University, Department of Mechanical and Industrial Engineering, Boston, MA 02115, USA\\
$^{*}$~Corresponding author: \href{mailto:p.schindler@northeastern.edu}{p.schindler@northeastern.edu}}}

\section{Introduction}

Cleavage and surface energy govern fundamental interfacial processes, including crack tip propagation~\cite{gilman_direct_1960,kawata_experimental_2018}, the brittle-to-ductile behavior of crystalline solids~\cite{rice_ductile_1974,li_cafe_2019}, elemental segregation at grain boundaries~\cite{raabe_grain_2014}, molecular adsorption at surfaces~\cite{schimka_accurate_2010}, heterogeneous catalysis~\cite{zhou_nanomaterials_2011,zhuang_surface_2016,xu_extending_2018}, electron emission devices~\cite{kawano_effective_2008,lin_demonstration_2022,schindler_surface_2019}, photocathodes~\cite{antoniuk_novel_2021}, semiconductor devices~\cite{zhang_surface_2004}, and battery material interfaces~\cite{xiao_understanding_2019,wang_towards_2017}. The cleavage energy is equal to the surface energy for symmetric slabs, which determines the equilibrium shape of nanoparticles and determines the stability of surfaces~\cite{wulff_xxv_1901,mchale_surface_1997}. First principles calculations can accurately predict this critical property; however, its computational expense, combined with the vast structural and chemical space of materials surfaces, limits high-throughput screening efforts.

The grand challenge of computational materials science lies in accurately predicting the properties of materials to enable the discovery and design of new materials that can meet the demanding requirements of modern technological applications~\cite{huang_application_2023}. Over the past decades, computational methods have revolutionized this landscape~\cite{schmidt_recent_2019}, leading to a data-driven discovery approach, the ``fourth paradigm of science'', that uses massive datasets generated by experiments and simulations~\cite{agrawal_perspective_2016}. Machine Learning Interatomic Potentials (MLIPs), also termed ``foundational potentials'', have emerged as a transformative solution, bridging the long-standing gap between accuracy and computational efficiency by emulating density functional theory (DFT) calculations at a fraction of the computational cost~\cite{mishin_machine-learning_2021,deringer_machine_2019,li_deep-learning_2022, anstine_machine_2023,del_rio_deep_2023, fiedler_deep_2022}. This powerful tool has rapidly evolved and enabled remarkable breakthroughs, from million-atom molecular dynamics simulations of crack propagation to the rapid exploration of complex phase diagrams and the accelerated discovery of novel alloys with enhanced mechanical properties~\cite{bartok_machine_2018,shen_development_2021,rosenbrock_machine-learned_2021, handley_next_2014}.\\ 

Early MLIPs typically targeted one or a few chemistries, using hand‐engineered descriptors whose complexity grew combinatorially as new elements were added~\cite{behler_atom-centered_2011}. Architecturally, the field has progressed from fixed descriptor-based functions~\cite{behler_atom-centered_2011,drautz_atomic_2019} to dynamic graph neural networks like SchNet~\cite{schutt_schnet_2017} and DimeNet~\cite{gasteiger_directional_2020} that learn atomic interactions directly from raw coordinates through iterative message-passing~\cite{gilmer_neural_2017}. The advent of equivariant~\cite{batzner_e3-equivariant_2021} and attention-based~\cite{louis_graph_2020} neural networks has further improved data efficiency and long-range interaction modeling, while active-learning workflows~\cite{lookman_active_2019} automatically expand datasets into underrepresented configuration spaces. The recent development of universal MLIPs (uMLIPs) has been enabled by massive computational datasets spanning the periodic table, including Materials Project\cite{jain_commentary_2013}, Open Catalyst\cite{chanussot_open_2021,tran_open_2023}, Alexandria\cite{schmidt_improving_2024}, and Open Materials 2024\cite{barroso-luque_open_2024} databases, which collectively provide millions of DFT calculations across diverse chemical environments. These datasets each contribute unique training regimes, fundamentally shaping uMLIP capabilities (see Supplementary Section~1 for comprehensive dataset descriptions).

Despite training on these comprehensive datasets, the rapid proliferation of uMLIPs has raised important questions about reliability, transferability, and generalizability across different applications. With their varied architectures, descriptors, and training databases, determining which uMLIP framework delivers optimal performance for specific use cases has become a critical challenge for the materials science community. While uMLIPs like MACE~\cite{batatia_foundation_2024}, ORB~\cite{rhodes_orb-v3_2025, neumann_orb_2024}, GRACE~\cite{bochkarev_graph_2024}, and eSEN~\cite{fu_learning_2025} offer the promising vision of a universal force field that could streamline simulations and accelerate materials discovery, their performance must be systematically evaluated across diverse atomic environments beyond their training data. Particularly crucial is assessing how models trained primarily on equilibrium bulk crystal structures perform when predicting properties of materials containing defects, surfaces, or under dynamic conditions such as phonon behavior and thermal expansion that involve near-equilibrium crystalline configurations.\\

Recent benchmarking efforts have systematically evaluated uMLIPs across various materials properties and tasks. The Matbench Discovery framework~\cite{riebesell_framework_2025} established a comprehensive evaluation paradigm addressing four fundamental challenges: prospective benchmarking that mimics real discovery campaigns, thermodynamically relevant targets (convex hull distance rather than formation energies), informative classification metrics beyond regression accuracy, and scalability to large chemical spaces. Complementing this, the MLIP Arena platform~\cite{chiang_mlip_2025} shifts focus from error-based metrics to physics-aware evaluation, revealing that gradient-based force predictions maintain conservative fields essential for stable simulations, while direct force prediction models exhibit center-of-mass drifts exceeding $10^2\;\text{\AA}$, six orders of magnitude larger than gradient-based approaches. Their hydrogen combustion benchmarks exposed critical stability issues, with CHGNet\cite{deng_chgnet_2023}, M3GNet, and EquiformerV2 (OC20) failing to complete 1 ns molecular dynamics trajectories, highlighting that strong regression performance does not guarantee physical reliability.

Property-specific benchmarks have revealed systematic patterns in uMLIP capabilities across diverse applications. For phonon anharmonicity, graph neural networks and artificial neural networks trained on ab initio molecular dynamics achieve agreement up to fifth-order irreducible derivatives~\cite{bandi_benchmarking_2024}. Reactive hydrogen dynamics at metal surfaces show that REANN and MACE provide an optimal balance between accuracy and computational efficiency, achieving sub-millisecond force evaluations while maintaining chemical accuracy~\cite{stark_benchmarking_2024}. Harmonic phonon properties across 10,000 ab initio calculations demonstrate that models with non-conservative force predictions excel at equilibrium geometry optimization yet catastrophically fail at phonon properties, while gradient-based forces show better phonon accuracy~\cite{loew_universal_2025}. Comprehensive assessments across high-entropy alloys and disordered oxides revealed that equivariant uMLIPs offer 1.5–-2× improvements over non-equivariant architectures, though this advantage requires an order of magnitude more training data to achieve comparable accuracy for simpler systems~\cite{maxson_ms25_2025}. Analyses of over 2,300 MLIP models have identified fundamental trade-offs, with 35 challenging properties, particularly defect formation energies, elastic constants, and energy rankings, where fewer than 15\% of models achieve errors below DFT-derived or practically-motivated thresholds (e.g., 0.1 eV for defect formation energies).

\cite{liu_learning_2024}. Intriguingly, these benchmarking efforts hint at a pattern we explore systematically: model performance on out-of-distribution tasks appears more strongly influenced by training data composition than architectural sophistication.

The computational cost of generating comprehensive surface data, which requires numerous orientations, terminations, and convergence tests per material, has historically prevented its systematic inclusion in existing training sets. Hence, the training data landscape was heavily skewed toward bulk configurations, with surface structures appearing only in specialized catalysis datasets (OC20/OC22) that focus on adsorbate-surface interactions rather than surface creation itself~\cite{chanussot_open_2021,tran_open_2023}. While Focassio et al. recently assessed earlier uMLIPs (MACE-MP-0, CHGNet-0.3.0, and M3GNet) for surface energy predictions, their evaluation was limited to elemental systems (73 elements, 1,497 surfaces) and revealed systematic underestimation requiring extensive fine-tuning, with root mean squared error improvements from 25\% to 6\%~\cite{focassio_performance_2024,zguns_benchmarking_2025}. The performance of modern uMLIPs on chemically complex surface energies remains unexplored.

Here, we present a comprehensive benchmark of 19 state-of-the-art uMLIPs for cleavage energy prediction using our previously established high-throughput DFT database of 36,718 surface structures spanning elemental, binary, and ternary metallic compounds. Our benchmarking study, comprising over 36,000 surface evaluations for each of the 19 models (totaling over 1.3 million individual energy predictions), represents the largest systematic evaluation of uMLIP transferability to surface properties to date. Our evaluation encompasses diverse architectural paradigms, including equivariant transformers (EquiformerV2, UMA), graph neural networks (GRACE, eSEN), orbital-based representations (ORB), and multi-scale approaches (MatterSim), all tested on an out-of-distribution task for which they were not explicitly trained. We systematically analyze prediction accuracy across chemical compositions, crystal systems, thickness, and surface orientations to identify the structural and electronic factors that govern uMLIP transferability. Beyond accuracy metrics, we evaluate each model's ability to correctly identify the thermodynamically most stable surface terminations, a critical requirement for materials design applications. This work establishes cleavage energy as a rigorous benchmark for assessing uMLIP generalization to surface phenomena and provides actionable insights for developing next-generation foundational potentials capable of accurately modeling interfacial properties across the periodic table.

This paper is organized as follows: Section~\ref{sec:methods} details our high-throughput DFT benchmark dataset of 36,718 surface structures and describes the 19 evaluated uMLIPs spanning diverse architectural paradigms, as illustrated in Figure~\ref{fig:graphical_abstract}. Section~\ref{sec:results} presents comprehensive performance metrics, analyzing prediction accuracy across chemical compositions, crystal systems, slab thicknesses, and surface orientations, and discusses the implications of our findings for uMLIP development and deployment. Section~\ref{sec:conclusions} concludes with recommendations for developing next-generation universal potentials capable of accurately modeling surface and interfacial properties.

\section{Methods}
\label{sec:methods}

\begin{figure}[ht]
\centering
\includegraphics[width=0.7\columnwidth]{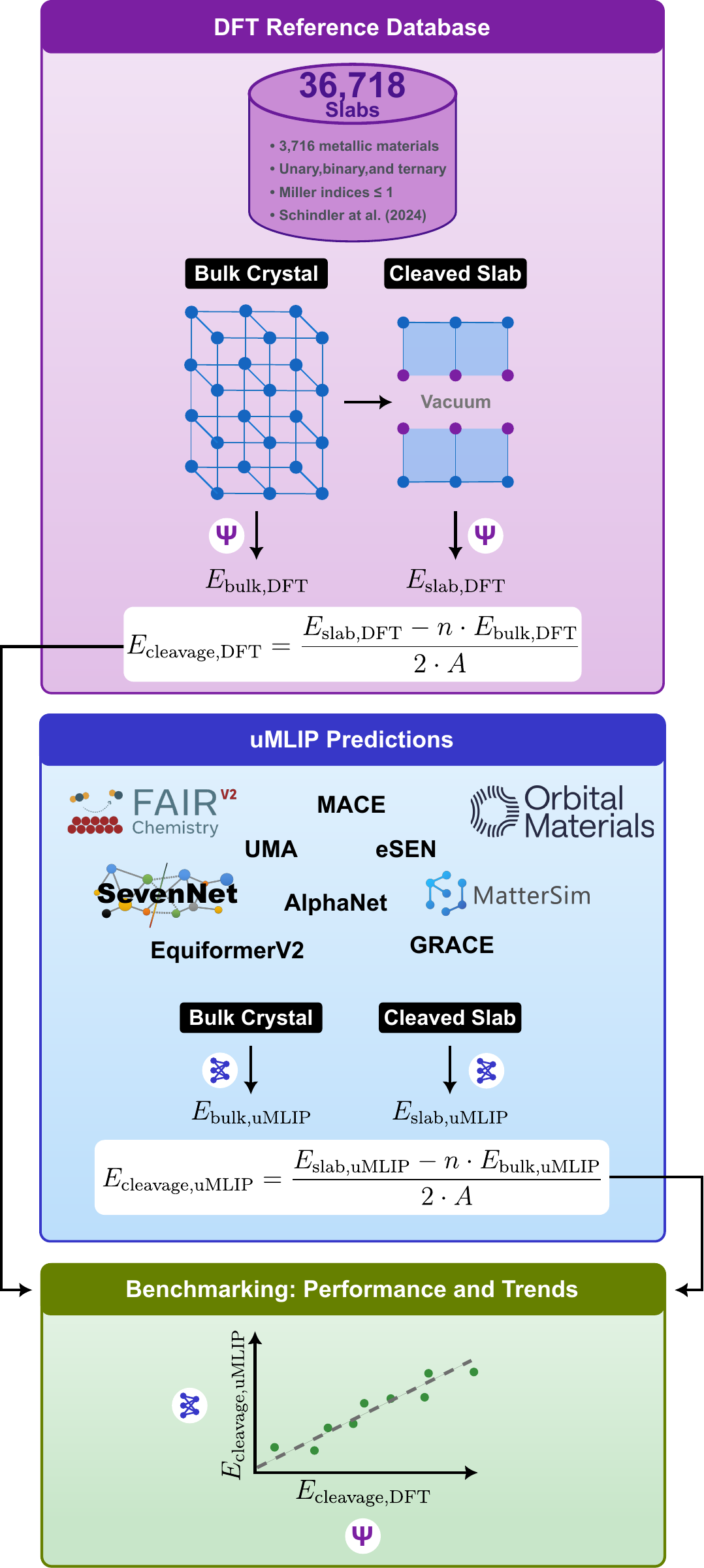}
\caption{Overview illustration of our cleavage energy benchmarking study highlighting the DFT-based cleavage energy database, uMLIP models considered, and performance benchmarking undertaken.}
\label{fig:graphical_abstract}
\end{figure}

We present a comprehensive benchmarking framework to evaluate how well uMLIPs, trained primarily on bulk structures, can predict cleavage energies. Figure~\ref{fig:graphical_abstract} illustrates our benchmarking workflow: we consider DFT-computed cleavage energies of surface structures from 3,716 bulk materials and use 19 uMLIPs representing different architectural families and training strategies for cleavage energy predictions, then compare the predictions to the DFT ground truth, finally, we conduct extensive analyses and quantitative assessment. The following subsections detail each aspect of this methodology, from the computational challenges that have historically limited surface property datasets (Section~\ref{sec:computational_challenges}) through our statistical analysis framework (Section~\ref{sec:Statistical_Analysis}).

\subsection{Computational Challenges in Cleavage Energy Calculations}
\label{sec:computational_challenges}

The computational demands of comprehensive cleavage energy calculations have historically limited their application in high-throughput materials screening. A rigorous assessment requires evaluating all symmetrically distinct crystallographic planes, which scales drastically with decreasing crystal symmetry; even restricting to low Miller indices generates substantial computational demands~\cite{tran_surface_2016, xiao_high-index-facet-_2020}. Furthermore, surfaces with Miller indices up to 3 frequently contribute to Wulff constructions, directly influencing nanoparticle shapes and catalytic properties~\cite{barmparis_nanoparticle_2015}, making their evaluation essential for materials design applications.
 
Recent high-throughput studies have demonstrated that each orientation may exhibit multiple terminations with different stoichiometries and atomic arrangements, particularly for oxide materials, where polar compensation mechanisms can generate dozens of distinct surface configurations per facet~\cite{barroso-luque_open_2024}. The computational workflow for each surface involves iterative optimization of slab thickness (typically $15$–$25\,\text{\AA}$ to achieve bulk-like interiors), vacuum spacing ($>15\,\text{\AA}$
 to prevent inter-slab interactions), and surface reconstruction patterns that may involve complex (2×2), $(\sqrt{3}\times\sqrt{3})$, or even (7×7) superstructures~\cite{sun_efficient_2013}. The \emph{Open Catalyst 2022} dataset exemplifies the computational challenge of surface-adsorbate systems specifically, requiring over $2\times10^8\,\mathrm{CPU\,hours}$ to generate 62,331 calculations, though the majority (69\%) were computationally intensive adsorbate+slab relaxations rather than bare surfaces~\cite{tran_open_2023}. Machine-learning approximations can accelerate such calculations by factors of $10^4$–-$10^6$, yet training these models requires extensive DFT data~\cite{broberg_high-throughput_2023}. For comprehensive characterization, including multiple orientations, terminations, and adsorbate configurations, computational costs can be enormous even for a single material, creating a critical bottleneck in developing universal MLIPs for surface-property prediction.

\subsection{DFT-Derived Cleavage Energy Dataset}

Several databases of surface and cleavage energies have been developed to date, each with distinct scopes and limitations. Tran et al.~\cite{tran_surface_2016} provided one of the earliest systematic collections, computing surface energies for over 100 polymorphs of approximately 70 elemental crystals with Miller indices up to 2 for non-cubic and 3 for cubic systems, totaling approximately 1,500 surfaces. While comprehensive for pure elements, it lacks the chemical complexity of multicomponent systems needed to evaluate uMLIP performance on diverse materials. Palizhati et al.~\cite{palizhati_toward_2019} extended surface property calculations to intermetallic compounds, generating 3,033 cleavage energies across 36 elements and 47 space groups, with 72\% bimetallic, 25\% trimetallic, and 3\% monometallic surfaces. While this introduced compositional complexity beyond pure elements, the dataset focused exclusively on intermetallic systems. Shi et al.~\cite{shi_surface_2024} recently developed a comprehensive dataset of 3,526 surface energies and work functions for binary magnesium intermetallics, focusing on low-index surfaces (Miller indices up to 1) across 188 binary Mg compounds. This work specifically addressed the importance of surface properties in understanding corrosion behavior and mechanical performance of Mg alloys.

The Open Catalyst datasets represent significant contributions with OC20~\cite{chanussot_open_2021} containing $\sim 1.28$ million DFT relaxations (265 million single-point calculations) focused on adsorbate-metal interactions with Miller indices $\leq 2$, while OC22~\cite{tran_open_2023} specifically addresses oxide electrocatalysts with 62,331 systems (19,142 slabs and 43,189 adsorbate+slabs) and Miller indices up to 3. Though these datasets are extensive, they primarily focus on adsorbate-surface interactions rather than systematic clean surface energetics across diverse material classes.

For our benchmark, we selected the high-throughput DFT database previously developed by our group~\cite{schindler_discovery_2024}. This database contains 36,718 surface structures spanning elemental, binary, and ternary metallic compounds, stored as \texttt{pymatgen} structure dictionaries~\cite{ong_python_2013}, providing substantial chemical diversity and scale for comprehensive uMLIP evaluation. The database focuses on Miller index 1 surfaces, including both symmetric and asymmetric surface terminations, and focuses on metallic systems (band gap $< 0.1$ eV), providing a well-defined material class for benchmarking.

The cleavage energy, defined as the energy required to create new surfaces by separating a crystal along specific crystallographic planes, was calculated using:

\begin{equation}\label{eq:ce}
E_{\text{cleavage}} = \frac{E_{\text{slab}} - n_{\text{bulk}} \cdot E_{\text{bulk}}}{2 \cdot A_{\text{slab}}}
\end{equation}

\noindent where $E_{\text{slab}}$ and $E_{\text{bulk}}$ represent the total energies of the slab and bulk configurations, respectively, $n_{\text{bulk}}$ is the number of bulk unit cells contained in the slab, and $A_{\text{slab}}$ is the surface area. All DFT calculations employed the Perdew-Burke-Ernzerhof (PBE) exchange-correlation functional~\cite{perdew_generalized_1996} with plane-wave basis sets and a kinetic energy cutoff of 550 eV, as detailed in the original work~\cite{schindler_discovery_2024}.

\subsection{Dataset Preparation and Filtering}

To ensure computational tractability and maintain consistency in the DFT reference calculations, we applied a systematic filtering protocol to the original dataset. We excluded materials containing both oxide or fluoride anions (O, F) and specific transition metals (Co, Cr, Fe, Mn, Mo, Ni, V, W), as these combinations typically require Hubbard U corrections (DFT+U) to accurately describe their strongly correlated d-electrons and localized electronic states. Since the original DFT database employed standard PBE functionals without U corrections, these materials would have unreliable reference energies, potentially introducing systematic errors in our benchmarking analysis. This filtering removed only two bulk materials out of 3,718 total (0.05\%): mp-19094 (V$_2$O$_4$) and mp-555934 (V$_2$F$_4$), both vanadium compounds that would require DFT+U corrections. These materials corresponded to 17 surface structures (6 surfaces for V$_2$O$_4$ and 11 for V$_2$F$_4$), resulting in the retention of 36,718 surface structures for benchmarking.

The resulting database provides cleavage energies ranging from 1.0 to 397.3 meV/\AA$^2$, with an average of 100.0 meV/\AA$^2$ for symmetric and asymmetric slabs. The filtered dataset was organized with comprehensive metadata including material identifiers (Materials Project IDs), Miller indices, surface terminations, slab thickness ratios ($n_{\text{slab}}/n_{\text{bulk}}$), surface areas, and both bulk and slab atomic structures.

While uMLIPs are computationally efficient, where completing each evaluation takes milliseconds rather than minutes/hours required by DFT, systematic organization was essential for managing the large amount of predictions across 19 models.

\subsection{uMLIP Benchmarking Protocol}

We evaluated 19 state-of-the-art universal MLIPs spanning diverse architectural paradigms and training strategies, none of which were exposed to our cleavage energy dataset during training. Table~\ref{tab:mlips} summarizes the evaluated models, the number of their parameters, and the training datasets they utilized. These models represent the current landscape of foundation potentials in computational materials science, thus providing a comprehensive generalization test. The evaluated models were categorized into nine families based on their underlying architectures. The uMLIP training datasets each contribute unique aspects: equilibrium properties (MP), surface structures (OC20/OC22), dynamic trajectories (MPtrj\cite{deng_materials_2023, deng_chgnet_2023}), dimensional diversity (Alexandria), and non-equilibrium configurations (OMat24). See Supplementary Section~1 for comprehensive descriptions of all training datasets.

\begin{table*}[htbp]
\centering
\footnotesize
\caption{\textbf{The 19 state-of-the-art uMLIPs considered in this study, representing a diverse set of architectures.} The number of model parameters and the training sets are summarized. Training datasets vary from equilibrium-focused collections (Materials Project) to non-equilibrium configurations (OMat24) and surface-adsorbate systems (OC20), enabling assessment of how training data composition affects out-of-distribution generalization. Parameter counts range from 4.55M to 1.4B, highlighting the diversity of model complexity.}
\label{tab:mlips}
\begin{tabular}{p{2.3cm}p{3.8cm}p{1.5cm}p{6.7cm}}
\toprule
\textbf{Model Family} & \textbf{Model Name} & \textbf{\#Params} & \textbf{Training Data Description} \\
\midrule
\multirow{3}{*}{\textbf{eSEN}} 
& eSEN-30M-MP & 30.1M & MPtrj: 1.58M equilibrium structures \\
& eSEN-30M-OMAT24 & 30.0M & OMat24: 101M non-equilibrium structures \\
& eSEN-30M-OAM & 30.2M & Combined: OMat24 + MPtrj + sAlex (10.4M) \\
\midrule
\multirow{3}{*}{\textbf{EquiformerV2}} 
& Eqf.V2-153M-OMAT24 & 153M & OMat24: Non-equilibrium configurations \\
& Eqf.V2-153M-OAM & 153M & Combined: OMat24 + MPtrj + sAlex \\
& Eqf.V2-153M-OC20 & 153M & OC20: Surface-adsorbate interactions \\
\midrule
\textbf{UMA} & UMA-m-1p1-OMAT24 & 1.4B & OMat24: Full non-equilibrium dataset \\
\midrule
\multirow{3}{*}{\textbf{GRACE}} 
& GRACE-2L-OMAT & 12.6M & OMat24: Non-equilibrium structures \\
& GRACE-2L-OAM & 12.6M & Combined: OMat24 + MPtrj + sAlex \\
& GRACE-2L-MP & 12.6M & MPtrj: Equilibrium MD trajectories \\
\midrule
\multirow{3}{*}{\textbf{MACE}} 
& MACE-MP-0b3 & 4.69M & Materials Project: Equilibrium only \\
& MACE-MPA & 9.06M & MP + Alexandria (subset) \\
& MACE-OMAT & 9.06M & OMat24: Non-equilibrium focus \\
\midrule
\textbf{SevenNet} & SevenNet-MF-OMPA & 25.7M & Combined: MPtrj + sAlex + OMat24 \\
\midrule
\multirow{3}{*}{\textbf{ORB}} 
& ORB-v3-OMAT & 25.5M & OMat24: Non-equilibrium \\
& ORB-v3-MPA & 25.5M & MP + Alexandria (subset) \\
& ORB-v2 & 25.2M & MPtrj + full Alexandria \\
\midrule
\textbf{MatterSim} & MatterSim-v1 & 4.55M & Proprietary workflow-generated data \\
\midrule
\textbf{AlphaNet} & AlphaNet-OMA & 4.65M & Combined: OMat24 + MPtrj + sAlex \\
\bottomrule
\end{tabular}
\end{table*}

For each uMLIP, we performed single-point energy calculations on both the slab and bulk structures from our database, maintaining identical atomic configurations (positions and cell parameters) as used in the original DFT calculations. The total energies of both slab and bulk configurations predicted by uMLIPs are used to calculate the cleavage energy with Equation~\ref{eq:ce}, comparing against the ground truth DFT value contained in the dataset. This approach enabled direct comparison between uMLIP-predicted and DFT-reference cleavage energies. Additionally, we recorded energy prediction times for each slab structure using uMLIPs to assess the computational efficiency of different model architectures.

To handle the computational demands of evaluating 19 models on $\sim$36,000 structures, we utilized a high-performance computing node equipped with four NVIDIA L40S GPUs, each featuring 45 GB of GDDR6 memory (135 GB total) and 18,176 CUDA cores running on the Ada Lovelace architecture. The system, powered by an Intel Xeon Gold 6448Y processor with 502 GB of system memory and running Rocky Linux 9.3 with CUDA 12.3, provided the necessary computational resources for our large-scale evaluation. We implemented a parallel processing framework with checkpoint-based recovery, distributing the 19 models across the available GPUs. The GPUs' high memory bandwidth (864 GB/s per GPU) and compute capability enabled efficient processing, while intermediate results were saved at regular intervals (128 structures initially, then every 4,000 structures) to enable fault-tolerant execution.

\subsection{Computational Implementation}
\label{sec:software_implementation}

All uMLIP evaluations were conducted using the official implementations of each model architecture with rigorously controlled software environments. The FAIRChem family of models employed two distinct versions: \texttt{fairchem-core} v1.10.0 for EquiformerV2 and eSEN architectures, and \texttt{fairchem-core} v2.3.0 for the UMA model, reflecting the major architectural changes between FAIRChem generations~\cite{chanussot_open_2021,tran_open_2023}. The ORB models utilized \texttt{orb-models} v0.5.4~\cite{neumann_orb_2024,rhodes_orb-v3_2025}, while MACE implementations employed \texttt{mace-torch} v0.3.13~\cite{batatia_foundation_2024}. SevenNet calculations were performed using \texttt{sevenn} v0.11.2.post1~\cite{kim_data-efficient_2025}, and GRACE models were evaluated through \texttt{grace-tensorpotential} v0.5.1 with TensorFlow 2.16.2 backend~\cite{bochkarev_graph_2024}. MatterSim predictions utilized the \texttt{mattersim} package v1.2.0~\cite{yang_mattersim_2024}, while AlphaNet employed a custom implementation (v0.0.1) built on PyTorch 2.1.2 with PyTorch Geometric dependencies~\cite{yin_alphanet_2025}. Each uMLIP was installed in an isolated conda environment to prevent dependency conflicts and ensure computational reproducibility. All models were executed on NVIDIA L40S GPUs with CUDA 12.3, utilizing model-specific optimizations where available. The software versions and environment specifications used for each uMLIP are detailed above to ensure computational reproducibility. The filtered dataset and all uMLIP predictions are publicly available (see Data Availability Statement section).

We developed a unified Python framework to orchestrate the computational workflow, abstracting model-specific initialization and prediction interfaces across all 19 uMLIPs. The framework managed GPU memory allocation and cleanup between predictions, tracked detailed timing information for performance analysis, and saved results in hierarchical directory structures organized by model family.

\subsection{Statistical Analysis}
\label{sec:Statistical_Analysis}

We evaluated model performance using multiple error metrics to capture different aspects of prediction accuracy. Mean Absolute Error (MAE) provided the average magnitude of errors in meV/Å², while the Absolute Percentage Error (APE) and Mean APE (MAPE) provide normalized errors relative to the true cleavage energy values, enabling comparison across different energy scales. Root Mean Square Error (RMSE) emphasized larger deviations by squaring errors before averaging, helping identify models with significant outliers. We also calculated the coefficient of determination ($R^2$) to assess the linear correlation between uMLIP predictions and DFT reference values. For computational efficiency analysis, we measured the wall-clock time required for each model to complete single-point energy calculations on slab structures, providing a practical comparison of model speeds across different architectures.
The error distributions were examined to identify systematic biases or limitations in specific chemical environments, with particular attention to outliers that might indicate fundamental limitations in model transferability.

\section{Results and Discussion}
\label{sec:results}

First, we analyze the overall performance of uMLIPs for cleavage energy prediction across our benchmark dataset of 36,718 surface structures. We then examine how training data composition influences model accuracy, followed by a detailed analysis of the best-performing models (see Supplementary Information for all other models). Finally, we investigate chemical and structural dependencies in prediction errors and discuss the computational efficiency trade-offs between different architectures.

\subsection{Overall Performance of uMLIPs}

\begin{figure*}[ht]
\centering
\includegraphics[width=\textwidth]{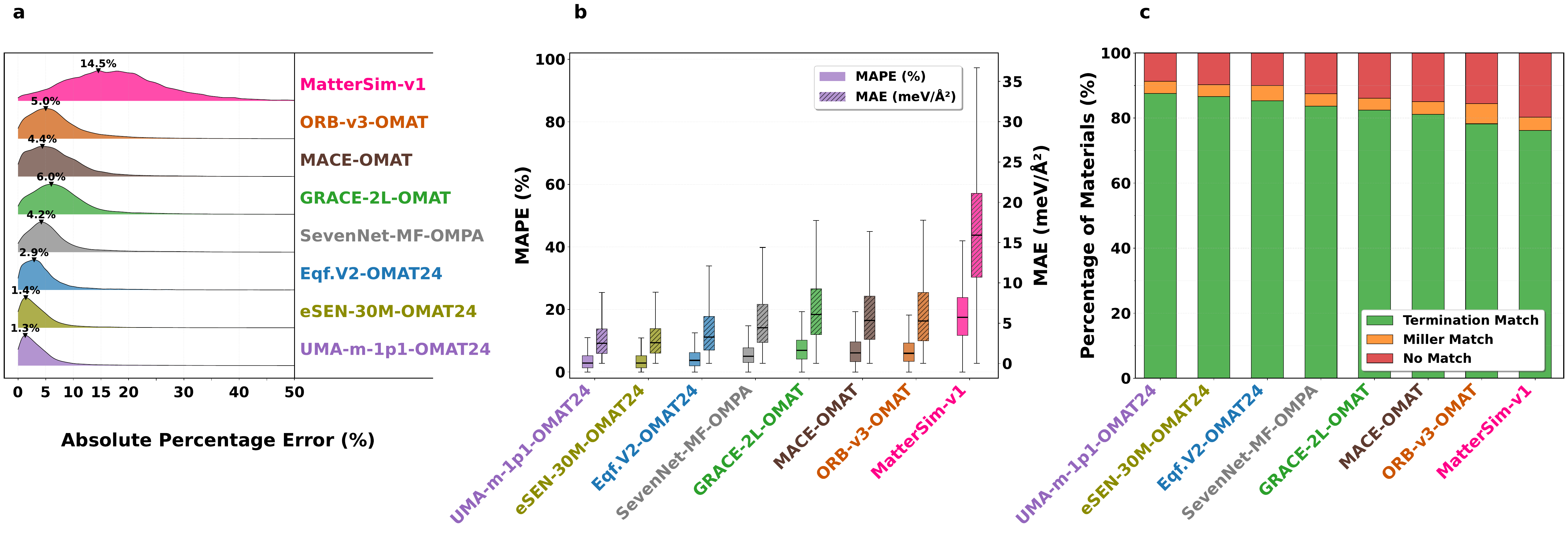}
\caption{\textbf{Comparative performance analysis of eight selected uMLIPs for cleavage energy prediction.} 
\textbf{(a)} Ridge plot of APE distributions for 36,718 surface structures, ordered by median APE. Numbers indicate the mode of each distribution. The $x$-axis is truncated at 50\%. 
\textbf{(b)} Dual-axis box plots showing MAPE (solid boxes, left axis) and MAE (hatched boxes, right axis in meV/\AA$^2$). 
\textbf{(c)} Stacked bar chart showing agreement with DFT for identifying thermodynamically most stable surface terminations across 3,699 materials. Green: exact termination match; yellow: correct Miller plane, wrong termination; red: complete disagreement.}
\label{fig:model_comparison}
\end{figure*}

We evaluated 19 state-of-the-art uMLIPs on cleavage energy prediction across 36,718 surface structures (see Supplementary Figure~1 for a complete comparison of all 19 models). Figure~\ref{fig:model_comparison} presents a comprehensive overview of select model performance, revealing substantial variation in prediction accuracy across different uMLIP architectures.

The ridge plots (Figure~\ref{fig:model_comparison}a) illustrate the distribution of APEs for each model, with numerical annotations indicating the mode (most frequent error value) of each distribution. The $x$-axis truncation at 50\% emphasizes the primary error distribution while excluding extreme outliers. Lower peak values indicate that models typically produce smaller errors, with UMA-m-1p1-OMAT24 achieving the lowest mode at 1.3\% compared to MatterSim-v1 at 14.5\%. The median APE values range from 2.85\% (UMA-m-1p1-OMAT24) to 17.50\% (MatterSim-v1), representing a six-fold variation in performance.

The dual-axis representation (Figure~\ref{fig:model_comparison}b) reveals that transformer-based (UMA) and equivariant (eSEN, EquiformerV2) architectures trained on OMat24 achieve median MAEs below 3.5 meV/\AA$^2$, while the multi-scale MatterSim model exhibits approximately five-fold higher median errors. This stark difference, despite MatterSim's sophisticated multi-scale architecture, underscores the dominant influence of training data composition on model performance for out-of-distribution tasks (see Supplementary Tables~2--5 for complete performance metrics, outlier statistics, and summary statistics across all models).

Beyond energy prediction accuracy, we evaluated each model's ability to identify the thermodynamically most stable surface terminations (Figure~\ref{fig:model_comparison}c), a critical requirement for practical materials design applications. For each of the 3,699 bulk materials in our dataset, we compared whether the uMLIPs correctly identified the lowest-energy surface among all possible crystallographic orientations (up to a Miller index of 1) and atomic terminations, as predicted by DFT. We classified the following three outcomes: (1) exact termination match, where the model correctly identified both the Miller index (crystallographic plane) and the specific atomic termination (i.e., which atoms form the outermost layer) that yields the minimum cleavage energy; (2) Miller match only, where the model identified the correct crystallographic plane but selected the wrong atomic termination among the possible terminations for that plane; and (3) no match, where the model predicted an entirely different crystallographic orientation as most stable one. The analysis across 3,699 materials reveals a clear performance hierarchy: UMA (87.5\%) $>$ eSEN (86.6\%) $>$ EquiformerV2 (85.3\%) for exact termination matches. Remarkably, all OMat24-trained models achieve greater than 76\% exact termination match accuracy without explicit surface energy training, validating that exposure to non-equilibrium configurations during training enables models to learn bond-breaking energetics relevant to surface stability. 

To provide context for model performance, we establish a trivial baseline using the mean cleavage energy across the entire database (100.0 meV/Å²) as a constant prediction for all structures. This mean baseline has a MAE of 33.8 meV/Å² and MAPE of 48.7\%, with 2,816 structures (7.7\%) showing errors exceeding 100\%. The top-performing models like UMA-m-1p1-OMAT24 achieve up to nine-fold reduction in error compared to this mean baseline (see Supplementary Table~10 for complete baseline performance metrics and improvement factors).

Quantitative analysis reveals an observable correlation between prediction accuracy and termination identification capability (as shown in Supplementary Figure~5). Models with lower MAE consistently achieve higher termination match rates ($R^2= 0.542$), with the relationship strengthening for low-energy surfaces (0--50 meV/Å²) where small energy differences make correct ranking most challenging. This correlation confirms that accurate energy predictions translate directly to reliable identification of stable surface configurations.

Note that surface properties obtained from DFT, such as the cleavage energy, can depend on slab thickness and may further exhibit oscillations depending on whether the slab contains an even or odd number of atomic layers. A well-established strategy to mitigate this issue is to perform the bulk total energy calculation using a conventional unit cell that is reoriented along the relevant Miller index, thereby ensuring consistency in the $k$-point sampling between bulk and slab calculations.\cite{sun_efficient_2013} This protocol was also adopted in constructing the DFT database employed for benchmarking in this work~\cite{schindler_discovery_2024}. By contrast, it is less clear whether uMLIP predictions capture such effects, since uMLIPs do not involve a $k$-point mesh, and the corresponding training datasets typically rely on primitive bulk unit cells rather than conventional, reoriented cells aligned with the surface orientation.

\begin{table*}[htbp]
\centering
\caption{\textbf{Comprehensive performance evaluation of 19 uMLIPs for cleavage energy prediction.} Models are grouped by architectural family and sorted within each family by mean APE from lowest to highest error. Error metrics quantify prediction accuracy across 36,718 surface structures, with MAPE $>$100\% indicating the count of structures with Mean Absolute Percentage Error exceeding 100\% (outliers). Stability agreement indicates the percentage of 3,699 bulk materials where models correctly identify the thermodynamically most stable surface termination (Term.), Miller plane only (Miller), or neither (None). Computational time represents single-point energy evaluation on slab structures. Bold values indicate the best performance across all models for each metric (lowest values for error metrics, including MAPE $>$100\%, highest for stability agreement).}
\label{tab:performance_grouped}
\footnotesize
\begin{tabular}{lccccccccccc}
\toprule
 & \multicolumn{6}{c}{Error Metrics} & \multicolumn{3}{c}{Stability Agreement (\%)} & \multicolumn{2}{c}{} \\
\cmidrule(lr){2-7} \cmidrule(lr){8-10}
Model & APE Mean & APE Med. & MAE & RMSE & R$^2$ & MAPE & Term. & Miller & None & Params & Time \\
 & (\%) & (\%) & (meV/\AA$^2$) & (meV/\AA$^2$) & & $>$100\% & & & & & (ms) \\
\midrule
\multicolumn{12}{l}{\textbf{Mean Baseline (Reference)}} \\
\textcolor{gray}{Mean Baseline} & 48.7 & 28.6 & 33.8 & 46 & 0.000 & 2816 & --- & --- & --- & 1 & --- \\
\midrule
\multicolumn{12}{l}{\textbf{UMA}} \\
\textcolor[HTML]{9467bd}{UMA-m-1p1-OMAT24} & 5.8 & 2.8 & 5.2 & 16 & 0.882 & 145 & \textbf{87.5} & 3.8 & \textbf{8.7} & 1.4B & 215 \\
\midrule
\multicolumn{12}{l}{\textbf{EquiformerV2}} \\
\textcolor[HTML]{5aa3d1}{Eqf.V2-OMAT24-MP-sAlex} & \textbf{5.2} & \textbf{2.5} & \textbf{4.8} & \textbf{9} & \textbf{0.961} & 24 & 84.0 & 4.8 & 11.2 & 153M & 267 \\
\textcolor[HTML]{1f77b4}{Eqf.V2-OMAT24} & 6.3 & 3.7 & 6.4 & 15 & 0.895 & 113 & 85.3 & 4.6 & 10.0 & 153M & 153 \\
\textcolor[HTML]{87ceeb}{Eqf.V2-OC20} & 71.1 & 62.7 & 68.1 & 87 & -2.562 & 6641 & 16.2 & \textbf{8.8} & 75.1 & 153M & 130 \\
\midrule
\multicolumn{12}{l}{\textbf{eSEN}} \\
\textcolor[HTML]{8b8c00}{eSEN-30M-OMAT24} & 5.7 & 2.9 & 5.0 & 15 & 0.898 & 140 & 86.6 & 3.6 & 9.8 & 30M & 125 \\
\textcolor[HTML]{bcbd22}{eSEN-30M-OAM} & 6.3 & 4.1 & 5.6 & \textbf{9} & 0.959 & 18 & 85.7 & 4.2 & 10.1 & 30.2M & 121 \\
\textcolor[HTML]{d4d555}{eSEN-30M-MP} & 15.6 & 14.7 & 14.7 & 18 & 0.839 & 27 & 77.7 & 4.3 & 18.0 & 30.1M & 121 \\
\midrule
\multicolumn{12}{l}{\textbf{GRACE}} \\
\textcolor[HTML]{5cbf5c}{GRACE-2L-OAM} & 8.3 & 6.4 & 7.6 & 11 & 0.944 & 17 & 82.4 & 4.1 & 13.5 & 12.6M & 8 \\
\textcolor[HTML]{2ca02c}{GRACE-2L-OMAT} & 8.4 & 6.9 & 7.7 & 10 & 0.948 & \textbf{15} & 82.5 & 3.6 & 13.9 & 12.6M & 8 \\
\textcolor[HTML]{90ee90}{GRACE-2L-MP} & 15.3 & 13.6 & 14.5 & 19 & 0.835 & 29 & 70.9 & 4.2 & 24.9 & 12.6M & 9 \\
\midrule
\multicolumn{12}{l}{\textbf{ORB}} \\
\textcolor[HTML]{ff7f0e}{ORB-v3-MPA} & 7.4 & 5.1 & 6.8 & 11 & 0.947 & 17 & 77.8 & 8.9 & 13.2 & 25.5M & 11 \\
\textcolor[HTML]{cc5500}{ORB-v3-OMAT} & 7.8 & 6.0 & 7.3 & 11 & 0.946 & 23 & 78.2 & 6.2 & 15.5 & 25.5M & 17 \\
\textcolor[HTML]{ffb366}{ORB-v2} & 9.2 & 6.3 & 8.8 & 14 & 0.906 & 39 & 69.3 & 9.2 & 21.5 & 25.2M & \textbf{6} \\
\midrule
\multicolumn{12}{l}{\textbf{MACE}} \\
\textcolor[HTML]{5d3a2f}{MACE-OMAT} & 7.8 & 6.1 & 6.9 & 10 & 0.955 & \textbf{15} & 81.2 & 3.9 & 14.9 & 9.06M & 20 \\
\textcolor[HTML]{b8906b}{MACE-MPA} & 9.7 & 8.0 & 9.1 & 13 & 0.922 & 22 & 79.0 & 3.8 & 17.2 & 9.06M & 20 \\
\textcolor[HTML]{d2b48c}{MACE-MP-0b3} & 25.2 & 24.6 & 24.7 & 30 & 0.584 & 17 & 69.9 & 4.5 & 25.6 & 4.69M & 21 \\
\midrule
\multicolumn{12}{l}{\textbf{SevenNet}} \\
\textcolor[HTML]{7f7f7f}{SevenNet-MF-OMPA} & 7.1 & 5.0 & 6.6 & 10 & 0.953 & 18 & 83.6 & 3.8 & 12.5 & 25.7M & 84 \\
\midrule
\multicolumn{12}{l}{\textbf{MatterSim}} \\
\textcolor[HTML]{ff0088}{MatterSim-v1} & 19.3 & 17.5 & 17.4 & 21 & 0.782 & 126 & 76.2 & 4.1 & 19.7 & 4.55M & 53 \\
\midrule
\multicolumn{12}{l}{\textbf{AlphaNet}} \\
\textcolor[HTML]{d62728}{AlphaNet-OMA} & 11.1 & 9.3 & 10.3 & 14 & 0.908 & 21 & 77.6 & 4.5 & 18.0 & 4.65M & 21 \\
\bottomrule
\end{tabular}
\end{table*}

\subsection{Outlier Analysis}

While median errors provide the primary performance metric, examining outliers reveals important failure modes across architectures (as shown in Supplementary Table~5). We define outliers as predictions with MAPE exceeding 100\%, representing cases where models predict cleavage energies at least twice the DFT reference value. Most models exhibit remarkably low outlier rates ($<0.5\%$ of structures), with EquiformerV2-OC20 as a notable exception (18.1\% outlier rate), confirming its unsuitability for general surface predictions. Among well-performing models, an interesting pattern emerges: models trained on OMat24 alone (eSEN-30M-OMAT24, EquiformerV2-153M-OMAT24, UMA-m-1p1-OMAT24) show slightly higher outlier counts (140, 113, and 145 structures, respectively) compared to their counterparts trained on combined datasets, yet maintain lower mean outlier APE (181--252\% vs 695--878\%). This suggests that pure OMat24 training produces more frequent but less severe failures, while combined dataset training yields fewer but more catastrophic errors when models fail. The median error of outliers provides additional insight—negative values around $-125$ meV/Å² for OMat24-trained models indicate systematic underestimation for these challenging cases, likely corresponding to high-energy surfaces or unusual coordination environments underrepresented in training data. In contrast, models trained on combined datasets show positive median outlier errors ($\sim 30$ meV/Å²), suggesting overestimation when encountering unfamiliar configurations. These outlier characteristics, while affecting less than 0.5\% of predictions for most models, could be critical for applications requiring reliable uncertainty quantification or when screening for extreme surface properties.

\subsection{Impact of Training Data versus Architecture}

Interestingly, we observe that the choice of training dataset exerts a more profound influence on cleavage energy prediction accuracy than architectural design. This observation highlights the critical importance of training data curation for out-of-distribution tasks.

Consider the eSEN architecture family: eSEN-30M-OMAT24 (trained on OMat24) achieves a median APE of 2.9\%, while the architecturally identical eSEN-30M-MP (trained on MPtrj) shows 14.7\% median APE, a five-fold degradation despite using the same model architecture. Similarly, within the EquiformerV2 family, the model trained on OMat24 (3.7\% median APE) dramatically outperforms its OC20-trained counterpart (62.7\% median APE), representing a 17-fold improvement solely through training data selection.

This pattern persists across all architectural families examined. The GRACE models show median APE ranging from 6.9\% (OMat24-trained) to 13.6\% (MPtrj-trained), while maintaining identical 2-layer GNN architectures. Even the sophisticated MACE architecture cannot overcome poor training data: MACE-OMAT achieves 6.1\% median APE compared to MACE-MP's 24.6\%, despite MACE's advanced equivariant message-passing design.

The superiority of OMat24-trained models likely stems from their emphasis on non-equilibrium configurations. By sampling structures with Gaussian perturbations ($\sigma = 0.5\,\text{\AA}$) and at elevated temperatures (up to 3000K), OMat24 exposes models to atomic environments with forces and stresses spanning orders of magnitude beyond equilibrium values, precisely the regime relevant for bond-breaking during cleavage. In contrast, models trained exclusively on equilibrium structures (MP) or surface-adsorbate interactions (OC20) lack exposure to the intermediate states critical for cleavage energy prediction.

This finding has profound implications for uMLIP development strategies. While architectural innovations, from simple graph networks to sophisticated equivariant transformers, contribute to performance improvements, our results demonstrate that a well-designed training dataset can provide an order-of-magnitude greater impact on out-of-distribution generalization. Hence, computational resources may yield the greatest returns if directed toward systematic dataset design that captures relevant physical phenomena.

The $R^2$ values across models trained on OMat24 consistently exceed 0.88, as shown in Table~\ref{tab:performance_grouped} (see also Supplementary Figures~2b and 4 for a complete $R^2$ comparison and pairwise metric correlations, respectively), confirming strong linear correlation with DFT predictions. In stark contrast, the negative $R^2$ for EquiformerV2-OC20 ($-2.562$) indicates predictions worse than using the mean value, mathematically confirming complete failure for out-of-distribution generalization from catalyst-specific to general surface data. We hypothesize that a fundamental issue is that OC20-trained models cannot accurately predict bulk energies, as the OC20 dataset contains only surface-adsorbate configurations without any bulk structures, rendering cleavage energy calculations, which require both bulk and surface energy predictions, inaccurate. This statistical validation underscores that training data relevance outweighs architectural sophistication for successful transfer to new property domains.

\subsection{Impact of Architecture on Computational Efficiency}

However, the choice of model architecture critically affects the computational cost during inference, with a moderate speed-accuracy trade-off. As shown in Table~\ref{tab:performance_grouped} (and Supplementary Figure~3), prediction times vary up to 45-fold across architectures: ORB-v2 and GRACE models require only 6--8 ms per structure, while transformer-based models like UMA-m-1p1-OMAT24 (215 ms) and EquiformerV2-MP-sAlex (267 ms) are 36× and 45× slower, respectively. This speedup translates to computational durations of days instead of months for high-throughput screening of the order of tens of millions of surfaces. Remarkably, simpler architectures like GRACE-2L-OAM achieve reasonable accuracy (8.3\% MAPE) at 8 ms per structure, providing a 33-fold speed advantage over the most accurate EquiformerV2-MP-sAlex while maintaining sub-10\% errors. SevenNet-MF-OMPA offers a middle ground with competitive accuracy (7.1\% MAPE) at 84 ms per structure. The fastest models (ORB at 6--17 ms, GRACE at 8--9 ms, MACE at 20--21 ms) still maintain sub-10\% MAPE when trained on OMat24, suggesting that for large-scale screening, these simpler architectures provide an optimal balance. This 10--45× speed advantage, combined with comparable accuracy when trained on appropriate data, questions whether architectural complexity is justified for routine materials screening applications.

\subsection{Detailed Analysis of Optimal Model Performance}

\begin{figure*}[t]
  \centering
  \includegraphics[width=\linewidth]{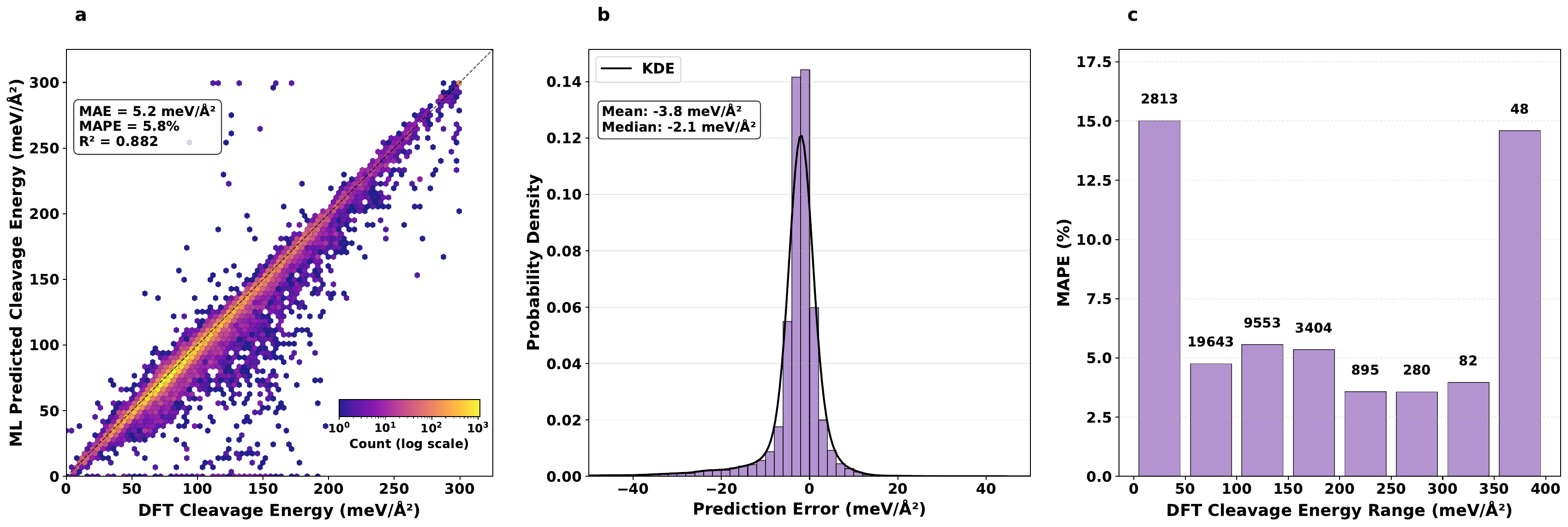}
  \caption{\textbf{Detailed performance analysis of UMA--m--1p1--OMat24 for cleavage energy prediction.}
  \textbf{(a)} Hexagonal density parity plot comparing uMLIP-predicted versus DFT-calculated cleavage energies. Note that the color bar is logarithmic.
  \textbf{(b)} Probability density distribution of prediction errors (uMLIP $-$ DFT), showing a kernel density estimate (KDE) curve calculated using a Gaussian kernel with Scott's rule bandwidth selection. The mean and median error values are displayed in the text box.
  \textbf{(c)} MAPE as a function of DFT cleavage energy bins. Numbers above bars indicate the sample count in each bin.}
  \label{fig:best_model_analysis}
\end{figure*}

Among all evaluated models, UMA-m-1p1-OMAT24 demonstrated superior performance with a mean absolute error (MAE) of 5.2 meV/Å² and MAPE of 5.8\%, and most importantly, achieved the highest termination match rate of 87.5\%, correctly identifying the thermodynamically most stable surface configuration for nearly 9 out of 10 materials. Figure~\ref{fig:best_model_analysis} provides a detailed analysis of this top-performing model across 36,718 surface structures. 

The parity plot (Figure~\ref{fig:best_model_analysis}a) reveals excellent agreement between uMLIP-predicted and DFT-calculated cleavage energies, with $R^2=0.882$. The hexagonal binning visualization highlights the high density of accurate predictions along the ideal prediction line, with most structures concentrated in the 0--150 meV/Å² range. Notably, the model maintains good correlation even for high-energy surfaces above 200 meV/Å², though with increased scatter, suggesting robust extrapolation to extreme cases.

The error distribution analysis (Figure~\ref{fig:best_model_analysis}b) reveals a slight systematic underestimation bias, with the distribution centered at a mean error of $-3.8$ meV/Å² and a median of $-2.1$ meV/Å². The leptokurtic profile with high kurtosis indicates a sharp central peak with heavier tails than a Gaussian distribution, suggesting the model makes highly accurate predictions for most structures while producing larger errors for a small subset of challenging cases. Despite these outliers, the narrow central peak with most errors falling within $\pm 25$ meV/Å² demonstrates consistent model behavior across the majority of surface chemistries and crystallographic orientations.

As shown in Figure~\ref{fig:best_model_analysis}c (see Supplementary Table~1 for detailed statistics across all models and energy ranges), the elevated MAPE ($\sim$15\%) for low-energy surfaces (0--50 meV/\AA$^2$) is primarily a mathematical artifact of relative error calculation---small absolute errors become large percentage errors when divided by small cleavage energy values. This affects only 2,813 structures (7.7\% of the dataset). Conversely, high-energy surfaces (350--400 meV/\AA$^2$) maintain MAPE below 15\%, demonstrating robust performance across the full energy spectrum, though this bin contains only 48 structures (0.1\% of the dataset), limiting statistical significance.

While we present detailed analysis for the top-performing UMA-m-1p1-OMAT24 model here, similar comprehensive performance analyses, including parity plots, error distributions, and energy-dependent MAPE for all other 18 evaluated models, are provided in Supplementary Figures~6--23.

\subsection{Chemical and Structural Dependencies}

To understand the sources of prediction errors, we analyzed model performance of UMA-m-1p1-OMAT24 across different chemical and structural categories (Figure~\ref{fig:error_decomposition}; similar decomposition analyses for all other models are provided in Supplementary Figures~24--41).

Analysis of element-specific performance (Figure~\ref{fig:error_decomposition}a) reveals striking patterns in the model's predictive accuracy. Alkaline earth metals demonstrate exceptional accuracy ($\mathrm{Be}$: 1.1\%, $\mathrm{Mg}$: 3.8\%, $\mathrm{Ca}$: 3.6\%, $\mathrm{Sr}$: 3.4\%), along with select transition metals ($\mathrm{Hf}$: 1.4\%, $\mathrm{Re}$: 1.2\%, $\mathrm{Ta}$: 1.9\%, $\mathrm{Ni}$: 3.2\%). In stark contrast, halogens exhibit the poorest performance ($\mathrm{Cl}$: 39.3\%, $\mathrm{Br}$: 80.2\%, $\mathrm{I}$: 74.6\%), while alkali metals show variable accuracy ($\mathrm{Na}$: 6.2\%, $\mathrm{K}$: 9.7\%, $\mathrm{Cs}$: 34.0\%). Lanthanides display intermediate but elevated errors, particularly $\mathrm{Eu}$ at 26.7\%. 

These patterns provide insights into the limitations of current training datasets. The excellent performance for transition metals likely reflects their prevalence in the OMat24 dataset, while the systematically higher errors for halogens and heavy alkali metals suggest these chemistries are underrepresented and/or exhibit complex bonding characteristics difficult to capture by the uMLIPs. The anomalously high error for Eu among lanthanides may stem from its variable oxidation states ($+2/+3$), indicating that elements with complex f-orbital occupancy remain challenging for uMLIPs (see Supplementary Table~8 for mean errors across the ten most common elements in the dataset).

Crystal system analysis (Figure~\ref{fig:error_decomposition}b, detailed in Supplementary Table~7) demonstrates that prediction accuracy correlates with structural symmetry. Hexagonal and orthorhombic structures are predicted most accurately ($\mathrm{MAPE}\approx 5\%$), while triclinic and trigonal systems present greater challenges ($\mathrm{MAPE}$ between $10$–-$15\%$). This pattern persists across all models, suggesting that low-symmetry structures represent a fundamental challenge rather than an architecture-specific limitation.

Slab thickness analysis reveals robust model performance across typical computational thickness ranges (Figure~\ref{fig:error_decomposition}c). UMA-m-1p1-OMAT24 maintains consistent accuracy for slabs up to 25 Å (MAPE: $\sim 6\%$), demonstrating reliable predictions across typical DFT slab thicknesses. Performance degrades for very thick slabs ($\geq 25$ Å), showing elevated errors (10.3\% for UMA), likely reflecting receptive field limitations of the graph neural network architectures, where messages from the top side of the slab no longer reach the bottom side. Notably, all OMat24-trained models show similar trends (Supplementary Table~9), with MAPE increasing by approximately two-fold for slabs $\geq 25$ Å, suggesting this represents a systematic prediction challenge.

Additional structural analysis reveals that compositional complexity and surface symmetry have a limited impact on prediction accuracy for UMA-m-1p1-OMAT24. Unary compounds show slightly higher MAPE (9.8\%) compared to binary (6.6\%) and ternary (5.4\%) systems (Supplementary Table~6), potentially reflecting the smaller sample size of elemental structures (236 vs.\ 10,185 and 26,297, respectively), making errors more sensitive to outliers. Surface symmetry shows a negligible effect, with symmetric and asymmetric slabs exhibiting comparable error rates, indicating models handle both slab types equally well. This pattern persists across all evaluated models except EquiformerV2-OC20, which exhibits catastrophic failure for unary compounds (273\% MAPE) while maintaining its already poor performance for binary and ternary systems (see Supplementary Figures~43--60 for composition and symmetry analyses of all 19 models).

\begin{figure}[ht]
\centering
\includegraphics[width=\columnwidth]{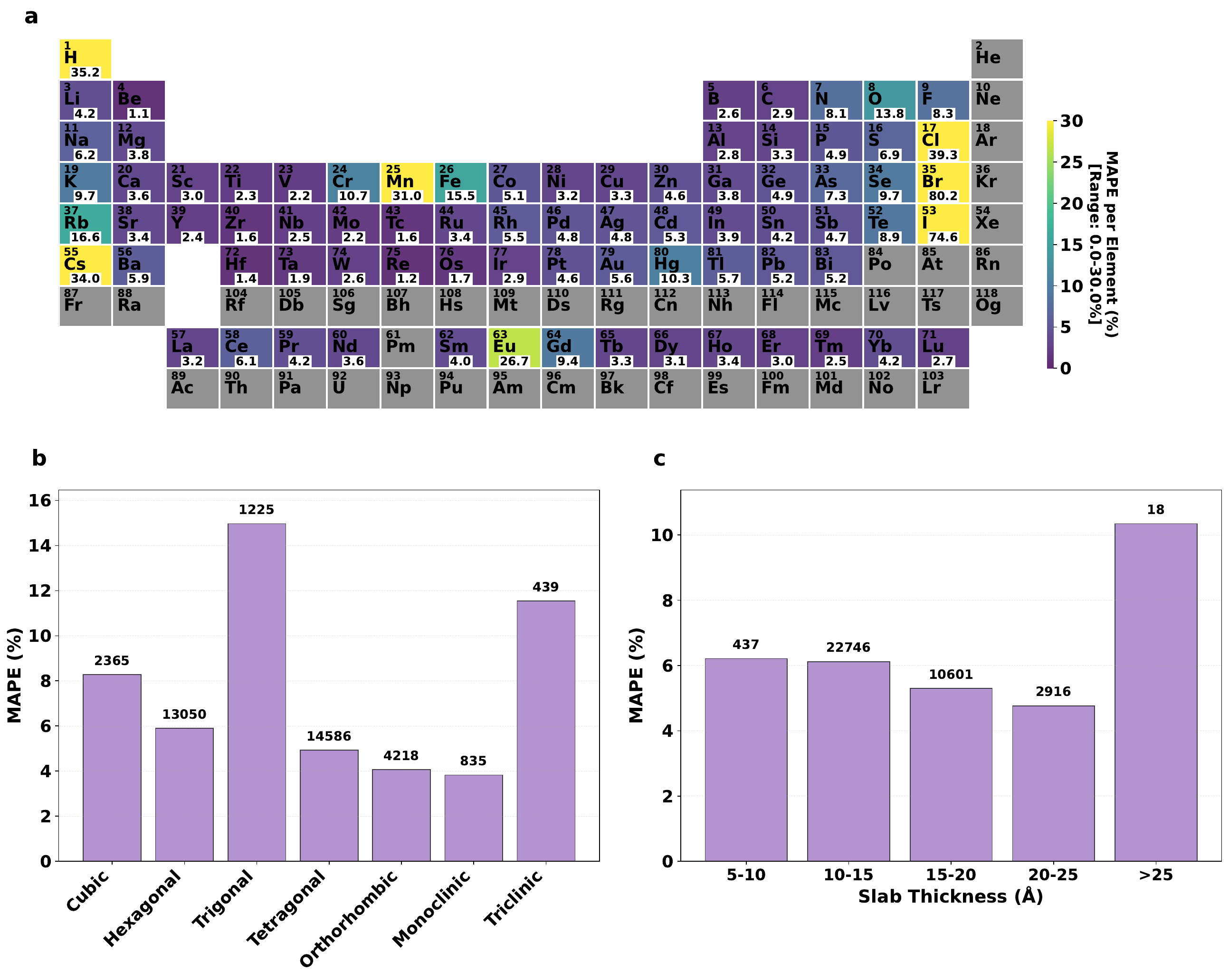}
\caption{\textbf{Decomposition of UMA-m-1p1-OMAT24 prediction errors by element, crystal system, and slab thickness.} 
\textbf{(a)} Periodic table heat map showing MAPE for surfaces containing each element, with viridis colormap normalized to 0--30\% range. 
\textbf{(b)} MAPE distribution across the seven crystal systems with sample counts indicated above bars. 
\textbf{(c)} MAPE dependency on slab thickness, binned into 5~Å intervals, showing consistent accuracy across different slab thicknesses with sample counts per bin.}
\label{fig:error_decomposition}
\end{figure}

\section{Conclusions}
\label{sec:conclusions}

Our comprehensive benchmark of 19 uMLIPs on 36,718 surface structures reveals fundamental insights that reframe the challenge of uMLIP development. Rather than pursuing increasingly complex architectures, our results demonstrate that strategic training data generation, capturing the full spectrum of atomic environments, is paramount for accurate property prediction.

We show that training data diversity trumps architectural sophistication. The five-fold to seventeen-fold improvement from dataset selection (OMat24 vs.\ MPtrj) overshadows differences between architectural families. Moreover, exposure to non-equilibrium configurations proves essential for learning bond-breaking physics, with OMat24-trained models achieving sub-6\% MAPE without explicit surface training, suggesting that sufficiently diverse bulk training data incorporating non-equilibrium states can enable accurate surface property prediction. Hence, models can identify stable configurations without expensive surface-specific training data. Finally, simpler architectures (ORB, MACE) offer 10--100× speed advantages over transformer-based models while maintaining comparable accuracy when trained on appropriate data, questioning whether architectural complexity is justified for routine materials screening.

While our study provides the most comprehensive surface property benchmark to date, the following limitations are worth noting. Our evaluation uses fixed DFT geometries, thus not testing the surface relaxation capabilities of uMLIPs. Further, the dataset focuses on materials with a near-zero bandgap and surface with a Miller index up to 1; extending this to higher indices and semiconducting or insulating materials would provide additional generalization insights. Additionally, all DFT references use the PBE functional, and exploring functional-dependent performance remains for future work (e.g., $r^2$SCAN). We have recently also developed a new equivariant model with symmetry breaking and uMLIP-informed features to directly train on the same database (predicting both the cleavage energy and work function). The best model in that work achieves a test-MAPE of 3.2\%, which, surprisingly, is not even two times better than UMA-m-1p1-OMAT24, which was not directly trained on our dataset~\cite{hsu_fire-gnn_2025}.

The impact of this work extends beyond surface energy prediction. By establishing cleavage energy as a rigorous out-of-distribution benchmark, we provide a framework for evaluating uMLIP transferability to other challenging properties. Our finding that training data quality dominates over architectural complexity suggests reallocating computational resources from architecture development to strategic dataset generation. Recent advances in simulation infrastructure, such as the TorchSim package~\cite{cohen_torchsim_2025} enabling batched GPU-accelerated simulations, could further accelerate uMLIP evaluation, potentially reducing benchmarking time by orders of magnitude. Additional future approaches may rely on a fine-tuning approach using MatterTune~\cite{kong_mattertune_2025}, or the recently established fully automated framework to capture the relevant potential energy surface~\cite{liu_automated_2025}.

Future directions should prioritize expanding training datasets to include underrepresented chemistries (particularly halogens and f-block elements), low-symmetry structures, and slab structures. In addition, deliberate sampling of transition states, metastable configurations, and other high-energy structures relevant to the intended applications will be essential for improving model transferability and robustness. The success of OMat24-trained models across diverse architectures demonstrates that current MLIP architectures are sufficient; the bottleneck lies in training data that captures relevant physics. As the field moves toward foundational potentials for materials design, our work establishes that a systematic evaluation on out-of-distribution tasks is essential for developing truly generalizable uMLIPs capable of predicting the full spectrum of materials properties.

\section*{Data Availability Statement}
The filtered dataset and predictions for all slab cleavage energies for every uMLIP model will be available, open-source, on Zenodo at \href{https://doi.org/10.5281/zenodo.16970767}{doi.org/10.5281/zenodo.16970767} after peer-review.

\section*{Conflict of Interest}
The authors declare that they have no conflicts of interest relevant to the content of this manuscript.

\section*{Acknowledgments}

P.S. acknowledges the start-up funding from the Department of Mechanical and Industrial Engineering at Northeastern University and funding through the TIER1 Seed Grant. This work was completed in part using the \textit{Explorer} cluster, supported by Northeastern University’s Research Computing team.

\section*{Author Contributions}

A.\ M.\ contributed to Writing – Original Draft (Lead), Writing – Review \& Editing (Lead), Conceptualization (Equal), Software (Lead), Investigation (Lead), Data Curation (Lead), and Visualization (Lead). P.\ S.\ contributed to Writing – Original Draft (Supporting), Writing – Review \& Editing (Supporting), Conceptualization (Equal), Visualization (Supporting), Resources (Lead), Supervision (Lead), and Funding acquisition (Lead). Both authors have read and approved this manuscript.

\footnotesize{
\bibliography{references} 

\providecommand*{\mcitethebibliography}{\thebibliography}
\csname @ifundefined\endcsname{endmcitethebibliography}
{\let\endmcitethebibliography\endthebibliography}{}
\begin{mcitethebibliography}{77}
\providecommand*{\natexlab}[1]{#1}
\providecommand*{\mciteSetBstSublistMode}[1]{}
\providecommand*{\mciteSetBstMaxWidthForm}[2]{}
\providecommand*{\mciteBstWouldAddEndPuncttrue}
  {\def\EndOfBibitem{\unskip.}}
\providecommand*{\mciteBstWouldAddEndPunctfalse}
  {\let\EndOfBibitem\relax}
\providecommand*{\mciteSetBstMidEndSepPunct}[3]{}
\providecommand*{\mciteSetBstSublistLabelBeginEnd}[3]{}
\providecommand*{\EndOfBibitem}{}
\mciteSetBstSublistMode{f}
\mciteSetBstMaxWidthForm{subitem}
{(\emph{\alph{mcitesubitemcount}})}
\mciteSetBstSublistLabelBeginEnd{\mcitemaxwidthsubitemform\space}
{\relax}{\relax}

\bibitem[Gilman(1960)]{gilman_direct_1960}
J.~J. Gilman, \emph{Journal of Applied Physics}, 1960, \textbf{31}, 2208--2218\relax
\mciteBstWouldAddEndPuncttrue
\mciteSetBstMidEndSepPunct{\mcitedefaultmidpunct}
{\mcitedefaultendpunct}{\mcitedefaultseppunct}\relax
\EndOfBibitem
\bibitem[Kawata \emph{et~al.}(2018)Kawata, Nakai, and Aihara]{kawata_experimental_2018}
I.~Kawata, H.~Nakai and S.~Aihara, \emph{Acta Materialia}, 2018, \textbf{150}, 40--52\relax
\mciteBstWouldAddEndPuncttrue
\mciteSetBstMidEndSepPunct{\mcitedefaultmidpunct}
{\mcitedefaultendpunct}{\mcitedefaultseppunct}\relax
\EndOfBibitem
\bibitem[Rice and Thomson(1974)]{rice_ductile_1974}
J.~R. Rice and R.~Thomson, \emph{The Philosophical Magazine: A Journal of Theoretical Experimental and Applied Physics}, 1974, \textbf{29}, 73--97\relax
\mciteBstWouldAddEndPuncttrue
\mciteSetBstMidEndSepPunct{\mcitedefaultmidpunct}
{\mcitedefaultendpunct}{\mcitedefaultseppunct}\relax
\EndOfBibitem
\bibitem[Li \emph{et~al.}(2019)Li, Shterenlikht, Ren, He, and Zhang]{li_cafe_2019}
Y.~Li, A.~Shterenlikht, X.~Ren, J.~He and Z.~Zhang, \emph{Materials Science and Engineering: A}, 2019, \textbf{755}, 220--230\relax
\mciteBstWouldAddEndPuncttrue
\mciteSetBstMidEndSepPunct{\mcitedefaultmidpunct}
{\mcitedefaultendpunct}{\mcitedefaultseppunct}\relax
\EndOfBibitem
\bibitem[Raabe \emph{et~al.}(2014)Raabe, Herbig, Sandlöbes, Li, Tytko, Kuzmina, Ponge, and Choi]{raabe_grain_2014}
D.~Raabe, M.~Herbig, S.~Sandlöbes, Y.~Li, D.~Tytko, M.~Kuzmina, D.~Ponge and P.-P. Choi, \emph{Current Opinion in Solid State and Materials Science}, 2014, \textbf{18}, 253--261\relax
\mciteBstWouldAddEndPuncttrue
\mciteSetBstMidEndSepPunct{\mcitedefaultmidpunct}
{\mcitedefaultendpunct}{\mcitedefaultseppunct}\relax
\EndOfBibitem
\bibitem[Schimka \emph{et~al.}(2010)Schimka, Harl, Stroppa, Grüneis, Marsman, Mittendorfer, and Kresse]{schimka_accurate_2010}
L.~Schimka, J.~Harl, A.~Stroppa, A.~Grüneis, M.~Marsman, F.~Mittendorfer and G.~Kresse, \emph{Nature Materials}, 2010, \textbf{9}, 741--744\relax
\mciteBstWouldAddEndPuncttrue
\mciteSetBstMidEndSepPunct{\mcitedefaultmidpunct}
{\mcitedefaultendpunct}{\mcitedefaultseppunct}\relax
\EndOfBibitem
\bibitem[Zhou \emph{et~al.}(2011)Zhou, Tian, Li, Broadwell, and Sun]{zhou_nanomaterials_2011}
Z.-Y. Zhou, N.~Tian, J.-T. Li, I.~Broadwell and S.-G. Sun, \emph{Chemical Society Reviews}, 2011, \textbf{40}, 4167\relax
\mciteBstWouldAddEndPuncttrue
\mciteSetBstMidEndSepPunct{\mcitedefaultmidpunct}
{\mcitedefaultendpunct}{\mcitedefaultseppunct}\relax
\EndOfBibitem
\bibitem[Zhuang \emph{et~al.}(2016)Zhuang, Tkalych, and Carter]{zhuang_surface_2016}
H.~Zhuang, A.~J. Tkalych and E.~A. Carter, \emph{The Journal of Physical Chemistry C}, 2016, \textbf{120}, 23698--23706\relax
\mciteBstWouldAddEndPuncttrue
\mciteSetBstMidEndSepPunct{\mcitedefaultmidpunct}
{\mcitedefaultendpunct}{\mcitedefaultseppunct}\relax
\EndOfBibitem
\bibitem[Xu \emph{et~al.}(2018)Xu, Kim, Park, Higgins, Shen, Schindler, Thian, Provine, Torgersen, Graf, Schladt, Orazov, Liu, Jaramillo, and Prinz]{xu_extending_2018}
S.~Xu, Y.~Kim, J.~Park, D.~Higgins, S.-J. Shen, P.~Schindler, D.~Thian, J.~Provine, J.~Torgersen, T.~Graf, T.~D. Schladt, M.~Orazov, B.~H. Liu, T.~F. Jaramillo and F.~B. Prinz, \emph{Nature Catalysis}, 2018, \textbf{1}, 624--630\relax
\mciteBstWouldAddEndPuncttrue
\mciteSetBstMidEndSepPunct{\mcitedefaultmidpunct}
{\mcitedefaultendpunct}{\mcitedefaultseppunct}\relax
\EndOfBibitem
\bibitem[Kawano(2008)]{kawano_effective_2008}
H.~Kawano, \emph{Progress in Surface Science}, 2008, \textbf{83}, 1--165\relax
\mciteBstWouldAddEndPuncttrue
\mciteSetBstMidEndSepPunct{\mcitedefaultmidpunct}
{\mcitedefaultendpunct}{\mcitedefaultseppunct}\relax
\EndOfBibitem
\bibitem[Lin \emph{et~al.}(2022)Lin, Jacobs, Chen, Vlahos, Lu‐Steffes, Alonso, Morgan, and Booske]{lin_demonstration_2022}
L.~Lin, R.~Jacobs, D.~Chen, V.~Vlahos, O.~Lu‐Steffes, J.~A. Alonso, D.~Morgan and J.~Booske, \emph{Advanced Functional Materials}, 2022, \textbf{32}, 2203703\relax
\mciteBstWouldAddEndPuncttrue
\mciteSetBstMidEndSepPunct{\mcitedefaultmidpunct}
{\mcitedefaultendpunct}{\mcitedefaultseppunct}\relax
\EndOfBibitem
\bibitem[Schindler \emph{et~al.}(2019)Schindler, Riley, Bargatin, Sahasrabuddhe, Schwede, Sun, Pianetta, Shen, Howe, and Melosh]{schindler_surface_2019}
P.~Schindler, D.~C. Riley, I.~Bargatin, K.~Sahasrabuddhe, J.~W. Schwede, S.~Sun, P.~Pianetta, Z.-X. Shen, R.~T. Howe and N.~A. Melosh, \emph{ACS Energy Letters}, 2019, \textbf{4}, 2436--2443\relax
\mciteBstWouldAddEndPuncttrue
\mciteSetBstMidEndSepPunct{\mcitedefaultmidpunct}
{\mcitedefaultendpunct}{\mcitedefaultseppunct}\relax
\EndOfBibitem
\bibitem[Antoniuk \emph{et~al.}(2021)Antoniuk, Schindler, Schroeder, Dunham, Pianetta, Vecchione, and Reed]{antoniuk_novel_2021}
E.~R. Antoniuk, P.~Schindler, W.~A. Schroeder, B.~Dunham, P.~Pianetta, T.~Vecchione and E.~J. Reed, \emph{Advanced Materials}, 2021, \textbf{33}, 2104081\relax
\mciteBstWouldAddEndPuncttrue
\mciteSetBstMidEndSepPunct{\mcitedefaultmidpunct}
{\mcitedefaultendpunct}{\mcitedefaultseppunct}\relax
\EndOfBibitem
\bibitem[Zhang and Wei(2004)]{zhang_surface_2004}
S.~B. Zhang and S.-H. Wei, \emph{Physical Review Letters}, 2004, \textbf{92}, 086102\relax
\mciteBstWouldAddEndPuncttrue
\mciteSetBstMidEndSepPunct{\mcitedefaultmidpunct}
{\mcitedefaultendpunct}{\mcitedefaultseppunct}\relax
\EndOfBibitem
\bibitem[Xiao \emph{et~al.}(2019)Xiao, Wang, Bo, Kim, Miara, and Ceder]{xiao_understanding_2019}
Y.~Xiao, Y.~Wang, S.-H. Bo, J.~C. Kim, L.~J. Miara and G.~Ceder, \emph{Nature Reviews Materials}, 2019, \textbf{5}, 105--126\relax
\mciteBstWouldAddEndPuncttrue
\mciteSetBstMidEndSepPunct{\mcitedefaultmidpunct}
{\mcitedefaultendpunct}{\mcitedefaultseppunct}\relax
\EndOfBibitem
\bibitem[Wang \emph{et~al.}(2017)Wang, Zhang, Zheng, Cui, Rojo, and Zhang]{wang_towards_2017}
D.~Wang, W.~Zhang, W.~Zheng, X.~Cui, T.~Rojo and Q.~Zhang, \emph{Advanced Science}, 2017, \textbf{4}, 1600168\relax
\mciteBstWouldAddEndPuncttrue
\mciteSetBstMidEndSepPunct{\mcitedefaultmidpunct}
{\mcitedefaultendpunct}{\mcitedefaultseppunct}\relax
\EndOfBibitem
\bibitem[Wulff(1901)]{wulff_xxv_1901}
G.~Wulff, \emph{Zeitschrift für Kristallographie - Crystalline Materials}, 1901, \textbf{34}, 449--530\relax
\mciteBstWouldAddEndPuncttrue
\mciteSetBstMidEndSepPunct{\mcitedefaultmidpunct}
{\mcitedefaultendpunct}{\mcitedefaultseppunct}\relax
\EndOfBibitem
\bibitem[McHale \emph{et~al.}(1997)McHale, Auroux, Perrotta, and Navrotsky]{mchale_surface_1997}
J.~M. McHale, A.~Auroux, A.~J. Perrotta and A.~Navrotsky, \emph{Science}, 1997, \textbf{277}, 788--791\relax
\mciteBstWouldAddEndPuncttrue
\mciteSetBstMidEndSepPunct{\mcitedefaultmidpunct}
{\mcitedefaultendpunct}{\mcitedefaultseppunct}\relax
\EndOfBibitem
\bibitem[Huang \emph{et~al.}(2023)Huang, Guo, Chen, and Nie]{huang_application_2023}
G.~Huang, Y.~Guo, Y.~Chen and Z.~Nie, \emph{Materials}, 2023, \textbf{16}, 5977\relax
\mciteBstWouldAddEndPuncttrue
\mciteSetBstMidEndSepPunct{\mcitedefaultmidpunct}
{\mcitedefaultendpunct}{\mcitedefaultseppunct}\relax
\EndOfBibitem
\bibitem[Schmidt \emph{et~al.}(2019)Schmidt, Marques, Botti, and Marques]{schmidt_recent_2019}
J.~Schmidt, M.~R.~G. Marques, S.~Botti and M.~A.~L. Marques, \emph{npj Computational Materials}, 2019, \textbf{5}, 83\relax
\mciteBstWouldAddEndPuncttrue
\mciteSetBstMidEndSepPunct{\mcitedefaultmidpunct}
{\mcitedefaultendpunct}{\mcitedefaultseppunct}\relax
\EndOfBibitem
\bibitem[Agrawal and Choudhary(2016)]{agrawal_perspective_2016}
A.~Agrawal and A.~Choudhary, \emph{APL Materials}, 2016, \textbf{4}, 053208\relax
\mciteBstWouldAddEndPuncttrue
\mciteSetBstMidEndSepPunct{\mcitedefaultmidpunct}
{\mcitedefaultendpunct}{\mcitedefaultseppunct}\relax
\EndOfBibitem
\bibitem[Mishin(2021)]{mishin_machine-learning_2021}
Y.~Mishin, \emph{Acta Materialia}, 2021, \textbf{214}, 116980\relax
\mciteBstWouldAddEndPuncttrue
\mciteSetBstMidEndSepPunct{\mcitedefaultmidpunct}
{\mcitedefaultendpunct}{\mcitedefaultseppunct}\relax
\EndOfBibitem
\bibitem[Deringer \emph{et~al.}(2019)Deringer, Caro, and Csányi]{deringer_machine_2019}
V.~L. Deringer, M.~A. Caro and G.~Csányi, \emph{Advanced Materials}, 2019, \textbf{31}, 1902765\relax
\mciteBstWouldAddEndPuncttrue
\mciteSetBstMidEndSepPunct{\mcitedefaultmidpunct}
{\mcitedefaultendpunct}{\mcitedefaultseppunct}\relax
\EndOfBibitem
\bibitem[Li \emph{et~al.}(2022)Li, Wang, Zou, Ye, Xu, Gong, Duan, and Xu]{li_deep-learning_2022}
H.~Li, Z.~Wang, N.~Zou, M.~Ye, R.~Xu, X.~Gong, W.~Duan and Y.~Xu, \emph{Nature Computational Science}, 2022, \textbf{2}, 367--377\relax
\mciteBstWouldAddEndPuncttrue
\mciteSetBstMidEndSepPunct{\mcitedefaultmidpunct}
{\mcitedefaultendpunct}{\mcitedefaultseppunct}\relax
\EndOfBibitem
\bibitem[Anstine and Isayev(2023)]{anstine_machine_2023}
D.~M. Anstine and O.~Isayev, \emph{The Journal of Physical Chemistry A}, 2023, \textbf{127}, 2417--2431\relax
\mciteBstWouldAddEndPuncttrue
\mciteSetBstMidEndSepPunct{\mcitedefaultmidpunct}
{\mcitedefaultendpunct}{\mcitedefaultseppunct}\relax
\EndOfBibitem
\bibitem[Del~Rio \emph{et~al.}(2023)Del~Rio, Phan, and Ramprasad]{del_rio_deep_2023}
B.~G. Del~Rio, B.~Phan and R.~Ramprasad, \emph{npj Computational Materials}, 2023, \textbf{9}, 158\relax
\mciteBstWouldAddEndPuncttrue
\mciteSetBstMidEndSepPunct{\mcitedefaultmidpunct}
{\mcitedefaultendpunct}{\mcitedefaultseppunct}\relax
\EndOfBibitem
\bibitem[Fiedler \emph{et~al.}(2022)Fiedler, Shah, Bussmann, and Cangi]{fiedler_deep_2022}
L.~Fiedler, K.~Shah, M.~Bussmann and A.~Cangi, \emph{Physical Review Materials}, 2022, \textbf{6}, 040301\relax
\mciteBstWouldAddEndPuncttrue
\mciteSetBstMidEndSepPunct{\mcitedefaultmidpunct}
{\mcitedefaultendpunct}{\mcitedefaultseppunct}\relax
\EndOfBibitem
\bibitem[Bartók \emph{et~al.}(2018)Bartók, Kermode, Bernstein, and Csányi]{bartok_machine_2018}
A.~P. Bartók, J.~Kermode, N.~Bernstein and G.~Csányi, \emph{Physical Review X}, 2018, \textbf{8}, 041048\relax
\mciteBstWouldAddEndPuncttrue
\mciteSetBstMidEndSepPunct{\mcitedefaultmidpunct}
{\mcitedefaultendpunct}{\mcitedefaultseppunct}\relax
\EndOfBibitem
\bibitem[Shen \emph{et~al.}(2021)Shen, Wang, and Lai]{shen_development_2021}
L.~Shen, Y.~Wang and W.~Lai, \emph{International Journal of Pressure Vessels and Piping}, 2021, \textbf{194}, 104514\relax
\mciteBstWouldAddEndPuncttrue
\mciteSetBstMidEndSepPunct{\mcitedefaultmidpunct}
{\mcitedefaultendpunct}{\mcitedefaultseppunct}\relax
\EndOfBibitem
\bibitem[Rosenbrock \emph{et~al.}(2021)Rosenbrock, Gubaev, Shapeev, Pártay, Bernstein, Csányi, and Hart]{rosenbrock_machine-learned_2021}
C.~W. Rosenbrock, K.~Gubaev, A.~V. Shapeev, L.~B. Pártay, N.~Bernstein, G.~Csányi and G.~L.~W. Hart, \emph{npj Computational Materials}, 2021, \textbf{7}, 24\relax
\mciteBstWouldAddEndPuncttrue
\mciteSetBstMidEndSepPunct{\mcitedefaultmidpunct}
{\mcitedefaultendpunct}{\mcitedefaultseppunct}\relax
\EndOfBibitem
\bibitem[Handley and Behler(2014)]{handley_next_2014}
C.~M. Handley and J.~Behler, \emph{The European Physical Journal B}, 2014, \textbf{87}, 152\relax
\mciteBstWouldAddEndPuncttrue
\mciteSetBstMidEndSepPunct{\mcitedefaultmidpunct}
{\mcitedefaultendpunct}{\mcitedefaultseppunct}\relax
\EndOfBibitem
\bibitem[Behler(2011)]{behler_atom-centered_2011}
J.~Behler, \emph{The Journal of Chemical Physics}, 2011, \textbf{134}, 074106\relax
\mciteBstWouldAddEndPuncttrue
\mciteSetBstMidEndSepPunct{\mcitedefaultmidpunct}
{\mcitedefaultendpunct}{\mcitedefaultseppunct}\relax
\EndOfBibitem
\bibitem[Drautz(2019)]{drautz_atomic_2019}
R.~Drautz, \emph{Physical Review B}, 2019, \textbf{99}, 014104\relax
\mciteBstWouldAddEndPuncttrue
\mciteSetBstMidEndSepPunct{\mcitedefaultmidpunct}
{\mcitedefaultendpunct}{\mcitedefaultseppunct}\relax
\EndOfBibitem
\bibitem[Schütt \emph{et~al.}(2017)Schütt, Kindermans, Sauceda~Felix, Chmiela, Tkatchenko, and Müller]{schutt_schnet_2017}
K.~Schütt, P.-J. Kindermans, H.~E. Sauceda~Felix, S.~Chmiela, A.~Tkatchenko and K.-R. Müller, Advances in {Neural} {Information} {Processing} {Systems}, 2017\relax
\mciteBstWouldAddEndPuncttrue
\mciteSetBstMidEndSepPunct{\mcitedefaultmidpunct}
{\mcitedefaultendpunct}{\mcitedefaultseppunct}\relax
\EndOfBibitem
\bibitem[Gasteiger \emph{et~al.}(2020)Gasteiger, Groß, and Günnemann]{gasteiger_directional_2020}
J.~Gasteiger, J.~Groß and S.~Günnemann, \emph{Directional {Message} {Passing} for {Molecular} {Graphs}}, 2020, \url{https://arxiv.org/abs/2003.03123}\relax
\mciteBstWouldAddEndPuncttrue
\mciteSetBstMidEndSepPunct{\mcitedefaultmidpunct}
{\mcitedefaultendpunct}{\mcitedefaultseppunct}\relax
\EndOfBibitem
\bibitem[Gilmer \emph{et~al.}(2017)Gilmer, Schoenholz, Riley, Vinyals, and Dahl]{gilmer_neural_2017}
J.~Gilmer, S.~S. Schoenholz, P.~F. Riley, O.~Vinyals and G.~E. Dahl, \emph{Neural {Message} {Passing} for {Quantum} {Chemistry}}, 2017, \url{https://arxiv.org/abs/1704.01212}\relax
\mciteBstWouldAddEndPuncttrue
\mciteSetBstMidEndSepPunct{\mcitedefaultmidpunct}
{\mcitedefaultendpunct}{\mcitedefaultseppunct}\relax
\EndOfBibitem
\bibitem[Batzner \emph{et~al.}(2021)Batzner, Musaelian, Sun, Geiger, Mailoa, Kornbluth, Molinari, Smidt, and Kozinsky]{batzner_e3-equivariant_2021}
S.~Batzner, A.~Musaelian, L.~Sun, M.~Geiger, J.~P. Mailoa, M.~Kornbluth, N.~Molinari, T.~E. Smidt and B.~Kozinsky, \emph{E(3)-{Equivariant} {Graph} {Neural} {Networks} for {Data}-{Efficient} and {Accurate} {Interatomic} {Potentials}}, 2021, \url{http://arxiv.org/abs/2101.03164}, arXiv:2101.03164\relax
\mciteBstWouldAddEndPuncttrue
\mciteSetBstMidEndSepPunct{\mcitedefaultmidpunct}
{\mcitedefaultendpunct}{\mcitedefaultseppunct}\relax
\EndOfBibitem
\bibitem[Louis \emph{et~al.}(2020)Louis, Zhao, Nasiri, Wang, Song, Liu, and Hu]{louis_graph_2020}
S.-Y. Louis, Y.~Zhao, A.~Nasiri, X.~Wang, Y.~Song, F.~Liu and J.~Hu, \emph{Physical Chemistry Chemical Physics}, 2020, \textbf{22}, 18141--18148\relax
\mciteBstWouldAddEndPuncttrue
\mciteSetBstMidEndSepPunct{\mcitedefaultmidpunct}
{\mcitedefaultendpunct}{\mcitedefaultseppunct}\relax
\EndOfBibitem
\bibitem[Lookman \emph{et~al.}(2019)Lookman, Balachandran, Xue, and Yuan]{lookman_active_2019}
T.~Lookman, P.~V. Balachandran, D.~Xue and R.~Yuan, \emph{npj Computational Materials}, 2019, \textbf{5}, 21\relax
\mciteBstWouldAddEndPuncttrue
\mciteSetBstMidEndSepPunct{\mcitedefaultmidpunct}
{\mcitedefaultendpunct}{\mcitedefaultseppunct}\relax
\EndOfBibitem
\bibitem[Jain \emph{et~al.}(2013)Jain, Ong, Hautier, Chen, Richards, Dacek, Cholia, Gunter, Skinner, Ceder, and Persson]{jain_commentary_2013}
A.~Jain, S.~P. Ong, G.~Hautier, W.~Chen, W.~D. Richards, S.~Dacek, S.~Cholia, D.~Gunter, D.~Skinner, G.~Ceder and K.~A. Persson, \emph{APL Materials}, 2013, \textbf{1}, 011002\relax
\mciteBstWouldAddEndPuncttrue
\mciteSetBstMidEndSepPunct{\mcitedefaultmidpunct}
{\mcitedefaultendpunct}{\mcitedefaultseppunct}\relax
\EndOfBibitem
\bibitem[Chanussot \emph{et~al.}(2021)Chanussot, Das, Goyal, Lavril, Shuaibi, Riviere, Tran, Heras-Domingo, Ho, Hu, Palizhati, Sriram, Wood, Yoon, Parikh, Zitnick, and Ulissi]{chanussot_open_2021}
L.~Chanussot, A.~Das, S.~Goyal, T.~Lavril, M.~Shuaibi, M.~Riviere, K.~Tran, J.~Heras-Domingo, C.~Ho, W.~Hu, A.~Palizhati, A.~Sriram, B.~Wood, J.~Yoon, D.~Parikh, C.~L. Zitnick and Z.~Ulissi, \emph{ACS Catalysis}, 2021, \textbf{11}, 6059--6072\relax
\mciteBstWouldAddEndPuncttrue
\mciteSetBstMidEndSepPunct{\mcitedefaultmidpunct}
{\mcitedefaultendpunct}{\mcitedefaultseppunct}\relax
\EndOfBibitem
\bibitem[Tran \emph{et~al.}(2023)Tran, Lan, Shuaibi, Wood, Goyal, Das, Heras-Domingo, Kolluru, Rizvi, Shoghi, Sriram, Therrien, Abed, Voznyy, Sargent, Ulissi, and Zitnick]{tran_open_2023}
R.~Tran, J.~Lan, M.~Shuaibi, B.~M. Wood, S.~Goyal, A.~Das, J.~Heras-Domingo, A.~Kolluru, A.~Rizvi, N.~Shoghi, A.~Sriram, F.~Therrien, J.~Abed, O.~Voznyy, E.~H. Sargent, Z.~Ulissi and C.~L. Zitnick, \emph{ACS Catalysis}, 2023, \textbf{13}, 3066--3084\relax
\mciteBstWouldAddEndPuncttrue
\mciteSetBstMidEndSepPunct{\mcitedefaultmidpunct}
{\mcitedefaultendpunct}{\mcitedefaultseppunct}\relax
\EndOfBibitem
\bibitem[Schmidt \emph{et~al.}(2024)Schmidt, Cerqueira, Romero, Loew, Jäger, Wang, Botti, and Marques]{schmidt_improving_2024}
J.~Schmidt, T.~F. Cerqueira, A.~H. Romero, A.~Loew, F.~Jäger, H.-C. Wang, S.~Botti and M.~A. Marques, \emph{Materials Today Physics}, 2024, \textbf{48}, 101560\relax
\mciteBstWouldAddEndPuncttrue
\mciteSetBstMidEndSepPunct{\mcitedefaultmidpunct}
{\mcitedefaultendpunct}{\mcitedefaultseppunct}\relax
\EndOfBibitem
\bibitem[Barroso-Luque \emph{et~al.}(2024)Barroso-Luque, Shuaibi, Fu, Wood, Dzamba, Gao, Rizvi, Zitnick, and Ulissi]{barroso-luque_open_2024}
L.~Barroso-Luque, M.~Shuaibi, X.~Fu, B.~M. Wood, M.~Dzamba, M.~Gao, A.~Rizvi, C.~L. Zitnick and Z.~W. Ulissi, \emph{Open {Materials} 2024 ({OMat24}) {Inorganic} {Materials} {Dataset} and {Models}}, 2024, \url{http://arxiv.org/abs/2410.12771}, arXiv:2410.12771\relax
\mciteBstWouldAddEndPuncttrue
\mciteSetBstMidEndSepPunct{\mcitedefaultmidpunct}
{\mcitedefaultendpunct}{\mcitedefaultseppunct}\relax
\EndOfBibitem
\bibitem[Batatia \emph{et~al.}(2024)Batatia, Benner, Chiang, Elena, Kovács, Riebesell, Advincula, Asta, Avaylon, Baldwin, Berger, Bernstein, Bhowmik, Blau, Cărare, Darby, De, Della~Pia, Deringer, Elijošius, El-Machachi, Falcioni, Fako, Ferrari, Genreith-Schriever, George, Goodall, Grey, Grigorev, Han, Handley, Heenen, Hermansson, Holm, Jaafar, Hofmann, Jakob, Jung, Kapil, Kaplan, Karimitari, Kermode, Kroupa, Kullgren, Kuner, Kuryla, Liepuoniute, Margraf, Magdău, Michaelides, Moore, Naik, Niblett, Norwood, O'Neill, Ortner, Persson, Reuter, Rosen, Schaaf, Schran, Shi, Sivonxay, Stenczel, Svahn, Sutton, Swinburne, Tilly, van~der Oord, Varga-Umbrich, Vegge, Vondrák, Wang, Witt, Zills, and Csányi]{batatia_foundation_2024}
I.~Batatia, P.~Benner, Y.~Chiang, A.~M. Elena, D.~P. Kovács, J.~Riebesell, X.~R. Advincula, M.~Asta, M.~Avaylon, W.~J. Baldwin, F.~Berger, N.~Bernstein, A.~Bhowmik, S.~M. Blau, V.~Cărare, J.~P. Darby, S.~De, F.~Della~Pia, V.~L. Deringer, R.~Elijošius, Z.~El-Machachi, F.~Falcioni, E.~Fako, A.~C. Ferrari, A.~Genreith-Schriever, J.~George, R.~E.~A. Goodall, C.~P. Grey, P.~Grigorev, S.~Han, W.~Handley, H.~H. Heenen, K.~Hermansson, C.~Holm, J.~Jaafar, S.~Hofmann, K.~S. Jakob, H.~Jung, V.~Kapil, A.~D. Kaplan, N.~Karimitari, J.~R. Kermode, N.~Kroupa, J.~Kullgren, M.~C. Kuner, D.~Kuryla, G.~Liepuoniute, J.~T. Margraf, I.-B. Magdău, A.~Michaelides, J.~H. Moore, A.~A. Naik, S.~P. Niblett, S.~W. Norwood, N.~O'Neill, C.~Ortner, K.~A. Persson, K.~Reuter, A.~S. Rosen, L.~L. Schaaf, C.~Schran, B.~X. Shi, E.~Sivonxay, T.~K. Stenczel, V.~Svahn, C.~Sutton, T.~D. Swinburne, J.~Tilly, C.~van~der Oord, E.~Varga-Umbrich, T.~Vegge, M.~Vondrák, Y.~Wang, W.~C. Witt, F.~Zills and G.~Csányi, \emph{A foundation model for atomistic
  materials chemistry}, 2024, \url{http://arxiv.org/abs/2401.00096}, arXiv:2401.00096 [cond-mat, physics:physics]\relax
\mciteBstWouldAddEndPuncttrue
\mciteSetBstMidEndSepPunct{\mcitedefaultmidpunct}
{\mcitedefaultendpunct}{\mcitedefaultseppunct}\relax
\EndOfBibitem
\bibitem[Rhodes \emph{et~al.}(2025)Rhodes, Vandenhaute, Šimkus, Gin, Godwin, Duignan, and Neumann]{rhodes_orb-v3_2025}
B.~Rhodes, S.~Vandenhaute, V.~Šimkus, J.~Gin, J.~Godwin, T.~Duignan and M.~Neumann, \emph{Orb-v3: atomistic simulation at scale}, 2025, \url{https://arxiv.org/abs/2504.06231}\relax
\mciteBstWouldAddEndPuncttrue
\mciteSetBstMidEndSepPunct{\mcitedefaultmidpunct}
{\mcitedefaultendpunct}{\mcitedefaultseppunct}\relax
\EndOfBibitem
\bibitem[Neumann \emph{et~al.}(2024)Neumann, Gin, Rhodes, Bennett, Li, Choubisa, Hussey, and Godwin]{neumann_orb_2024}
M.~Neumann, J.~Gin, B.~Rhodes, S.~Bennett, Z.~Li, H.~Choubisa, A.~Hussey and J.~Godwin, \emph{Orb: {A} {Fast}, {Scalable} {Neural} {Network} {Potential}}, 2024, \url{https://arxiv.org/abs/2410.22570}\relax
\mciteBstWouldAddEndPuncttrue
\mciteSetBstMidEndSepPunct{\mcitedefaultmidpunct}
{\mcitedefaultendpunct}{\mcitedefaultseppunct}\relax
\EndOfBibitem
\bibitem[Bochkarev \emph{et~al.}(2024)Bochkarev, Lysogorskiy, and Drautz]{bochkarev_graph_2024}
A.~Bochkarev, Y.~Lysogorskiy and R.~Drautz, \emph{Physical Review X}, 2024, \textbf{14}, 021036\relax
\mciteBstWouldAddEndPuncttrue
\mciteSetBstMidEndSepPunct{\mcitedefaultmidpunct}
{\mcitedefaultendpunct}{\mcitedefaultseppunct}\relax
\EndOfBibitem
\bibitem[Fu \emph{et~al.}(2025)Fu, Wood, Barroso-Luque, Levine, Gao, Dzamba, and Zitnick]{fu_learning_2025}
X.~Fu, B.~M. Wood, L.~Barroso-Luque, D.~S. Levine, M.~Gao, M.~Dzamba and C.~L. Zitnick, \emph{Learning {Smooth} and {Expressive} {Interatomic} {Potentials} for {Physical} {Property} {Prediction}}, 2025, \url{https://arxiv.org/abs/2502.12147}\relax
\mciteBstWouldAddEndPuncttrue
\mciteSetBstMidEndSepPunct{\mcitedefaultmidpunct}
{\mcitedefaultendpunct}{\mcitedefaultseppunct}\relax
\EndOfBibitem
\bibitem[Riebesell \emph{et~al.}(2025)Riebesell, Goodall, Benner, Chiang, Deng, Ceder, Asta, Lee, Jain, and Persson]{riebesell_framework_2025}
J.~Riebesell, R.~E.~A. Goodall, P.~Benner, Y.~Chiang, B.~Deng, G.~Ceder, M.~Asta, A.~A. Lee, A.~Jain and K.~A. Persson, \emph{Nature Machine Intelligence}, 2025, \textbf{7}, 836--847\relax
\mciteBstWouldAddEndPuncttrue
\mciteSetBstMidEndSepPunct{\mcitedefaultmidpunct}
{\mcitedefaultendpunct}{\mcitedefaultseppunct}\relax
\EndOfBibitem
\bibitem[Chiang \emph{et~al.}(2025)Chiang, Kreiman, Weaver, Amin, Kuner, Zhang, Kaplan, Chrzan, Blau, Krishnapriyan, and Asta]{chiang_mlip_2025}
Y.~Chiang, T.~Kreiman, E.~Weaver, I.~Amin, M.~Kuner, C.~Zhang, A.~Kaplan, D.~Chrzan, S.~M. Blau, A.~S. Krishnapriyan and M.~Asta, {AI4MAT}-{ICLR} 2025 {Spotlight}, 2025\relax
\mciteBstWouldAddEndPuncttrue
\mciteSetBstMidEndSepPunct{\mcitedefaultmidpunct}
{\mcitedefaultendpunct}{\mcitedefaultseppunct}\relax
\EndOfBibitem
\bibitem[Deng \emph{et~al.}(2023)Deng, Zhong, Jun, Riebesell, Han, Bartel, and Ceder]{deng_chgnet_2023}
B.~Deng, P.~Zhong, K.~Jun, J.~Riebesell, K.~Han, C.~J. Bartel and G.~Ceder, \emph{Nature Machine Intelligence}, 2023, \textbf{5}, 1031--1041\relax
\mciteBstWouldAddEndPuncttrue
\mciteSetBstMidEndSepPunct{\mcitedefaultmidpunct}
{\mcitedefaultendpunct}{\mcitedefaultseppunct}\relax
\EndOfBibitem
\bibitem[Bandi \emph{et~al.}(2024)Bandi, Jiang, and Marianetti]{bandi_benchmarking_2024}
S.~Bandi, C.~Jiang and C.~A. Marianetti, \emph{Machine Learning: Science and Technology}, 2024, \textbf{5}, 030502\relax
\mciteBstWouldAddEndPuncttrue
\mciteSetBstMidEndSepPunct{\mcitedefaultmidpunct}
{\mcitedefaultendpunct}{\mcitedefaultseppunct}\relax
\EndOfBibitem
\bibitem[Stark \emph{et~al.}(2024)Stark, Van Der~Oord, Batatia, Zhang, Jiang, Csányi, and Maurer]{stark_benchmarking_2024}
W.~G. Stark, C.~Van Der~Oord, I.~Batatia, Y.~Zhang, B.~Jiang, G.~Csányi and R.~J. Maurer, \emph{Machine Learning: Science and Technology}, 2024, \textbf{5}, 030501\relax
\mciteBstWouldAddEndPuncttrue
\mciteSetBstMidEndSepPunct{\mcitedefaultmidpunct}
{\mcitedefaultendpunct}{\mcitedefaultseppunct}\relax
\EndOfBibitem
\bibitem[Loew \emph{et~al.}(2025)Loew, Sun, Wang, Botti, and Marques]{loew_universal_2025}
A.~Loew, D.~Sun, H.-C. Wang, S.~Botti and M.~A.~L. Marques, \emph{npj Computational Materials}, 2025, \textbf{11}, 178\relax
\mciteBstWouldAddEndPuncttrue
\mciteSetBstMidEndSepPunct{\mcitedefaultmidpunct}
{\mcitedefaultendpunct}{\mcitedefaultseppunct}\relax
\EndOfBibitem
\bibitem[Maxson \emph{et~al.}(2025)Maxson, Soyemi, Zhang, Chen, and Szilvási]{maxson_ms25_2025}
T.~Maxson, A.~Soyemi, X.~Zhang, B.~W.~J. Chen and T.~Szilvási, \emph{Journal of Chemical Information and Modeling}, 2025, \textbf{65}, 8097--8112\relax
\mciteBstWouldAddEndPuncttrue
\mciteSetBstMidEndSepPunct{\mcitedefaultmidpunct}
{\mcitedefaultendpunct}{\mcitedefaultseppunct}\relax
\EndOfBibitem
\bibitem[Liu and Mo(2024)]{liu_learning_2024}
Y.~Liu and Y.~Mo, \emph{npj Computational Materials}, 2024, \textbf{10}, 159\relax
\mciteBstWouldAddEndPuncttrue
\mciteSetBstMidEndSepPunct{\mcitedefaultmidpunct}
{\mcitedefaultendpunct}{\mcitedefaultseppunct}\relax
\EndOfBibitem
\bibitem[Focassio \emph{et~al.}(2024)Focassio, Freitas, and Schleder]{focassio_performance_2024}
B.~Focassio, L.~P.~M. Freitas and G.~R. Schleder, \emph{Performance {Assessment} of {Universal} {Machine} {Learning} {Interatomic} {Potentials}: {Challenges} and {Directions} for {Materials}' {Surfaces}}, 2024, \url{http://arxiv.org/abs/2403.04217}, arXiv:2403.04217 [cond-mat, physics:physics]\relax
\mciteBstWouldAddEndPuncttrue
\mciteSetBstMidEndSepPunct{\mcitedefaultmidpunct}
{\mcitedefaultendpunct}{\mcitedefaultseppunct}\relax
\EndOfBibitem
\bibitem[Žguns \emph{et~al.}(2025)Žguns, Pudza, and Kuzmin]{zguns_benchmarking_2025}
P.~Žguns, I.~Pudza and A.~Kuzmin, \emph{Journal of Chemical Theory and Computation}, 2025,  acs.jctc.5c00955\relax
\mciteBstWouldAddEndPuncttrue
\mciteSetBstMidEndSepPunct{\mcitedefaultmidpunct}
{\mcitedefaultendpunct}{\mcitedefaultseppunct}\relax
\EndOfBibitem
\bibitem[Tran \emph{et~al.}(2016)Tran, Xu, Radhakrishnan, Winston, Sun, Persson, and Ong]{tran_surface_2016}
R.~Tran, Z.~Xu, B.~Radhakrishnan, D.~Winston, W.~Sun, K.~A. Persson and S.~P. Ong, \emph{Scientific Data}, 2016, \textbf{3}, 160080\relax
\mciteBstWouldAddEndPuncttrue
\mciteSetBstMidEndSepPunct{\mcitedefaultmidpunct}
{\mcitedefaultendpunct}{\mcitedefaultseppunct}\relax
\EndOfBibitem
\bibitem[Xiao \emph{et~al.}(2020)Xiao, Lu, Xue, Tian, Zhou, Lin, Lin, and Sun]{xiao_high-index-facet-_2020}
C.~Xiao, B.-A. Lu, P.~Xue, N.~Tian, Z.-Y. Zhou, X.~Lin, W.-F. Lin and S.-G. Sun, \emph{Joule}, 2020, \textbf{4}, 2562--2598\relax
\mciteBstWouldAddEndPuncttrue
\mciteSetBstMidEndSepPunct{\mcitedefaultmidpunct}
{\mcitedefaultendpunct}{\mcitedefaultseppunct}\relax
\EndOfBibitem
\bibitem[Barmparis \emph{et~al.}(2015)Barmparis, Lodziana, Lopez, and Remediakis]{barmparis_nanoparticle_2015}
G.~D. Barmparis, Z.~Lodziana, N.~Lopez and I.~N. Remediakis, \emph{Beilstein Journal of Nanotechnology}, 2015, \textbf{6}, 361--368\relax
\mciteBstWouldAddEndPuncttrue
\mciteSetBstMidEndSepPunct{\mcitedefaultmidpunct}
{\mcitedefaultendpunct}{\mcitedefaultseppunct}\relax
\EndOfBibitem
\bibitem[Sun and Ceder(2013)]{sun_efficient_2013}
W.~Sun and G.~Ceder, \emph{Surface Science}, 2013, \textbf{617}, 53--59\relax
\mciteBstWouldAddEndPuncttrue
\mciteSetBstMidEndSepPunct{\mcitedefaultmidpunct}
{\mcitedefaultendpunct}{\mcitedefaultseppunct}\relax
\EndOfBibitem
\bibitem[Broberg \emph{et~al.}(2023)Broberg, Bystrom, Srivastava, Dahliah, Williamson, Weston, Scanlon, Rignanese, Dwaraknath, Varley, Persson, Asta, and Hautier]{broberg_high-throughput_2023}
D.~Broberg, K.~Bystrom, S.~Srivastava, D.~Dahliah, B.~A.~D. Williamson, L.~Weston, D.~O. Scanlon, G.-M. Rignanese, S.~Dwaraknath, J.~Varley, K.~A. Persson, M.~Asta and G.~Hautier, \emph{npj Computational Materials}, 2023, \textbf{9}, 1--12\relax
\mciteBstWouldAddEndPuncttrue
\mciteSetBstMidEndSepPunct{\mcitedefaultmidpunct}
{\mcitedefaultendpunct}{\mcitedefaultseppunct}\relax
\EndOfBibitem
\bibitem[Palizhati \emph{et~al.}(2019)Palizhati, Zhong, Tran, Back, and Ulissi]{palizhati_toward_2019}
A.~Palizhati, W.~Zhong, K.~Tran, S.~Back and Z.~W. Ulissi, \emph{Journal of Chemical Information and Modeling}, 2019, \textbf{59}, 4742--4749\relax
\mciteBstWouldAddEndPuncttrue
\mciteSetBstMidEndSepPunct{\mcitedefaultmidpunct}
{\mcitedefaultendpunct}{\mcitedefaultseppunct}\relax
\EndOfBibitem
\bibitem[Shi \emph{et~al.}(2024)Shi, Wang, Yang, Qiu, Zhu, and Zeng]{shi_surface_2024}
G.~Shi, Y.~Wang, K.~Yang, Y.~Qiu, H.~Zhu and X.~Zeng, \emph{Journal of Magnesium and Alloys}, 2024,  S221395672400389X\relax
\mciteBstWouldAddEndPuncttrue
\mciteSetBstMidEndSepPunct{\mcitedefaultmidpunct}
{\mcitedefaultendpunct}{\mcitedefaultseppunct}\relax
\EndOfBibitem
\bibitem[Schindler \emph{et~al.}(2024)Schindler, Antoniuk, Cheon, Zhu, and Reed]{schindler_discovery_2024}
P.~Schindler, E.~R. Antoniuk, G.~Cheon, Y.~Zhu and E.~J. Reed, \emph{Advanced Functional Materials}, 2024, \textbf{34}, 2401764\relax
\mciteBstWouldAddEndPuncttrue
\mciteSetBstMidEndSepPunct{\mcitedefaultmidpunct}
{\mcitedefaultendpunct}{\mcitedefaultseppunct}\relax
\EndOfBibitem
\bibitem[Ong \emph{et~al.}(2013)Ong, Richards, Jain, Hautier, Kocher, Cholia, Gunter, Chevrier, Persson, and Ceder]{ong_python_2013}
S.~P. Ong, W.~D. Richards, A.~Jain, G.~Hautier, M.~Kocher, S.~Cholia, D.~Gunter, V.~L. Chevrier, K.~A. Persson and G.~Ceder, \emph{Computational Materials Science}, 2013, \textbf{68}, 314--319\relax
\mciteBstWouldAddEndPuncttrue
\mciteSetBstMidEndSepPunct{\mcitedefaultmidpunct}
{\mcitedefaultendpunct}{\mcitedefaultseppunct}\relax
\EndOfBibitem
\bibitem[Perdew \emph{et~al.}(1996)Perdew, Burke, and Ernzerhof]{perdew_generalized_1996}
J.~P. Perdew, K.~Burke and M.~Ernzerhof, \emph{Physical Review Letters}, 1996, \textbf{77}, 3865--3868\relax
\mciteBstWouldAddEndPuncttrue
\mciteSetBstMidEndSepPunct{\mcitedefaultmidpunct}
{\mcitedefaultendpunct}{\mcitedefaultseppunct}\relax
\EndOfBibitem
\bibitem[Deng(2023)]{deng_materials_2023}
B.~Deng, \emph{Materials {Project} {Trajectory} ({MPtrj}) {Dataset}}, 2023, \url{https://figshare.com/articles/dataset/Materials_Project_Trjectory_MPtrj_Dataset/23713842/2}\relax
\mciteBstWouldAddEndPuncttrue
\mciteSetBstMidEndSepPunct{\mcitedefaultmidpunct}
{\mcitedefaultendpunct}{\mcitedefaultseppunct}\relax
\EndOfBibitem
\bibitem[Kim \emph{et~al.}(2025)Kim, Kim, Kim, Lee, Park, Kang, and Han]{kim_data-efficient_2025}
J.~Kim, J.~Kim, J.~Kim, J.~Lee, Y.~Park, Y.~Kang and S.~Han, \emph{Journal of the American Chemical Society}, 2025, \textbf{147}, 1042--1054\relax
\mciteBstWouldAddEndPuncttrue
\mciteSetBstMidEndSepPunct{\mcitedefaultmidpunct}
{\mcitedefaultendpunct}{\mcitedefaultseppunct}\relax
\EndOfBibitem
\bibitem[Yang \emph{et~al.}(2024)Yang, Hu, Zhou, Liu, Shi, Li, Li, Chen, Chen, Zeni, Horton, Pinsler, Fowler, Zügner, Xie, Smith, Sun, Wang, Kong, Liu, Hao, and Lu]{yang_mattersim_2024}
H.~Yang, C.~Hu, Y.~Zhou, X.~Liu, Y.~Shi, J.~Li, G.~Li, Z.~Chen, S.~Chen, C.~Zeni, M.~Horton, R.~Pinsler, A.~Fowler, D.~Zügner, T.~Xie, J.~Smith, L.~Sun, Q.~Wang, L.~Kong, C.~Liu, H.~Hao and Z.~Lu, \emph{{MatterSim}: {A} {Deep} {Learning} {Atomistic} {Model} {Across} {Elements}, {Temperatures} and {Pressures}}, 2024, \url{http://arxiv.org/abs/2405.04967}, arXiv:2405.04967\relax
\mciteBstWouldAddEndPuncttrue
\mciteSetBstMidEndSepPunct{\mcitedefaultmidpunct}
{\mcitedefaultendpunct}{\mcitedefaultseppunct}\relax
\EndOfBibitem
\bibitem[Yin \emph{et~al.}(2025)Yin, Wang, Du, Wang, Ying, Jia, Zhang, Du, Gomes, Duan, Henkelman, and Xiao]{yin_alphanet_2025}
B.~Yin, J.~Wang, W.~Du, P.~Wang, P.~Ying, H.~Jia, Z.~Zhang, Y.~Du, C.~P. Gomes, C.~Duan, G.~Henkelman and H.~Xiao, \emph{{AlphaNet}: {Scaling} {Up} {Local}-frame-based {Atomistic} {Interatomic} {Potential}}, 2025, \url{https://arxiv.org/abs/2501.07155}\relax
\mciteBstWouldAddEndPuncttrue
\mciteSetBstMidEndSepPunct{\mcitedefaultmidpunct}
{\mcitedefaultendpunct}{\mcitedefaultseppunct}\relax
\EndOfBibitem
\bibitem[Hsu \emph{et~al.}(2025)Hsu, Schlesinger, Mudaliar, Leung, Walters, and Schindler]{hsu_fire-gnn_2025}
C.~Hsu, C.~Schlesinger, K.~Mudaliar, J.~Leung, R.~Walters and P.~Schindler, \emph{{FIRE}-{GNN}: {Force}-informed, {Relaxed} {Equivariance} {Graph} {Neural} {Network} for {Rapid} and {Accurate} {Prediction} of {Surface} {Properties}}, 2025, \url{https://arxiv.org/abs/2508.16012}\relax
\mciteBstWouldAddEndPuncttrue
\mciteSetBstMidEndSepPunct{\mcitedefaultmidpunct}
{\mcitedefaultendpunct}{\mcitedefaultseppunct}\relax
\EndOfBibitem
\bibitem[Cohen \emph{et~al.}(2025)Cohen, Riebesell, Goodall, Kolluru, Falletta, Krause, Colindres, Ceder, and Gangan]{cohen_torchsim_2025}
O.~Cohen, J.~Riebesell, R.~Goodall, A.~Kolluru, S.~Falletta, J.~Krause, J.~Colindres, G.~Ceder and A.~S. Gangan, \emph{{TorchSim}: {An} efficient atomistic simulation engine in {PyTorch}}, 2025, \url{http://arxiv.org/abs/2508.06628}, arXiv:2508.06628 [physics]\relax
\mciteBstWouldAddEndPuncttrue
\mciteSetBstMidEndSepPunct{\mcitedefaultmidpunct}
{\mcitedefaultendpunct}{\mcitedefaultseppunct}\relax
\EndOfBibitem
\bibitem[Kong \emph{et~al.}(2025)Kong, Shoghi, Hu, Li, and Fung]{kong_mattertune_2025}
L.~Kong, N.~Shoghi, G.~Hu, P.~Li and V.~Fung, \emph{Digital Discovery}, 2025, \textbf{4}, 2253--2262\relax
\mciteBstWouldAddEndPuncttrue
\mciteSetBstMidEndSepPunct{\mcitedefaultmidpunct}
{\mcitedefaultendpunct}{\mcitedefaultseppunct}\relax
\EndOfBibitem
\bibitem[Liu \emph{et~al.}(2025)Liu, Morrow, Ertural, Fragapane, Gardner, Naik, Zhou, George, and Deringer]{liu_automated_2025}
Y.~Liu, J.~D. Morrow, C.~Ertural, N.~L. Fragapane, J.~L.~A. Gardner, A.~A. Naik, Y.~Zhou, J.~George and V.~L. Deringer, \emph{Nature Communications}, 2025, \textbf{16}, 7666\relax
\mciteBstWouldAddEndPuncttrue
\mciteSetBstMidEndSepPunct{\mcitedefaultmidpunct}
{\mcitedefaultendpunct}{\mcitedefaultseppunct}\relax
\EndOfBibitem
\end{mcitethebibliography}
\bibliographystyle{rsc} 
}

\end{document}



\sloppy

\maketitle


\section{MLIP Training Datasets}

The evolution of computational materials datasets has been fundamental to the transformation of MLIPs from specialized tools to universal predictive frameworks. Here we detail the key datasets used to train the MLIPs evaluated in this study.

\textbf{Materials Project (MP):} Launched in 2011 as one of the pioneering open-access databases, the Materials Project \cite{jain_commentary_2013} democratized computational materials data by providing DFT calculations for over 146,000 inorganic compounds spanning 89 elements. The database employs standardized protocols using the generalized gradient approximation (GGA) within the Perdew-Burke-Ernzerhof (PBE) functional to ensure consistency across diverse material chemistries. While foundational for data-driven materials discovery, its focus on equilibrium structures at 0 K limited applications to static property predictions.

\textbf{Open Catalyst 2020 (OC20):} The OC20 dataset \cite{chanussot_open_2021} revolutionized surface modeling by introducing 1,281,040 DFT relaxations (~264,890,000 single-point evaluations) focused on adsorbate-catalyst interactions across nitrogen, carbon, and oxygen chemistries. Crucially, OC20 implemented a systematic surface enumeration strategy, cleaving ~15,000 bulk materials along all symmetrically distinct crystallographic planes with Miller indices up to 2, generating up to 125 different surfaces per bulk structure. Combined with ~30 unique adsorption sites per surface identified through Delaunay triangulation, this produced over 2.5 million initial adsorbate configurations. The dataset captures the full complexity of surface-adsorbate interactions from heuristically determined initial states to relaxed local energy minima, providing crucial training data for predicting surface energies and cleavage energies.

\textbf{Open Catalyst 2022 (OC22):} The OC22 dataset \cite{tran_open_2023} addressed fundamental limitations in oxide surface modeling by contributing 62,331 DFT relaxations (~9,854,504 single-point calculations) with Miller indices extended up to 3. Each crystallographic plane can exhibit multiple thermodynamically stable terminations—metal-terminated, oxygen-terminated, or mixed stoichiometry surfaces. By explicitly modeling all possible surface terminations for each Miller index and incorporating Hubbard U corrections for accurate electronic structure description, OC22 captured the complexity of polar oxide surfaces requiring charge compensation. The inclusion of surface energy calculations and Pourbaix stability diagrams directly addresses the thermodynamic aspects of surface creation, making it particularly relevant for training MLIPs to predict cleavage energies.

\textbf{Materials Project Trajectory Dataset (MPtrj):} MPtrj \cite{deng_materials_2023, deng_chgnet_2023} leveraged over a decade of accumulated DFT molecular dynamics simulations from the Materials Project, encompassing 1,580,395 structures with associated energies, forces, stresses, and crucially, magnetic moments, totaling 7,944,833 magnetic moment labels. By extracting trajectories from ~1.37 million structure relaxation and static calculation tasks performed with both GGA and GGA+U exchange-correlation functionals, MPtrj captured the dynamic evolution of materials far from equilibrium. The inclusion of magnetic moments as explicit labels enabled the development of charge-informed MLIPs like CHGNet, which learns not only the potential-energy surface but also the orbital occupancy of electrons through its explicit modeling of magnetic moments, bridging the gap between classical force fields and ab initio molecular dynamics and enabling large-scale simulations of complex phenomena.

\textbf{Alexandria Database:} Alexandria \cite{schmidt_improving_2024} represents a paradigmatic advancement through its systematic organization by material dimensionality, containing over 5 million DFT calculations strategically partitioned into 3D, 2D, and 1D compounds. The 3D compounds were calculated with multiple exchange-correlation functionals (PBE, PBEsol, and SCAN), providing a comprehensive view of bulk material properties across different levels of theory. The 2D and 1D compounds utilized PBE calculations to capture the unique physics of reduced dimensionality. This dimensional stratification enables MLIPs to learn fundamentally different bonding characteristics—from twelve-fold coordination typical of bulk materials to exposed dangling bonds in 2D sheets and highly directional bonding in 1D chains. Alexandria's development employed crystal graph attention networks to predict compounds near the convex hull of stability and universal MLIPs for preliminary geometry optimization, accelerating the identification of synthesizable materials.

\textbf{Open Materials 2024 (OMat24):} OMat24 \cite{barroso-luque_open_2024} represents an unprecedented expansion, delivering over 110 million DFT calculations spanning single-point evaluations, structural relaxations, and molecular dynamics snapshots for inorganic bulk materials across 1-100 atoms. The dataset implements three complementary sampling strategies: (1) Boltzmann sampling of rattled structures at multiple temperatures (300K, 500K, 1000K), (2) ab initio molecular dynamics trajectories at extreme conditions (1000K and 3000K in both NVT and NPT ensembles), and (3) systematic relaxations of perturbed structures with Gaussian perturbations of atomic positions ($\sigma = 0.5\,\mathrm{\AA}$) and cell deformations ($\sigma = 5\%$). This emphasis on non-equilibrium configurations generates structures with forces and stresses spanning orders of magnitude beyond typical equilibrium values. With most structures containing fewer than 20 atoms (though ranging from 1-100 atoms) while maintaining broad elemental coverage, OMat24 captures a dramatically wider distribution of energies, forces, and stresses than previous datasets, providing computationally tractable yet comprehensive training data enabling MLIPs to accurately predict energetics in the far-from-equilibrium regime where cleavage occurs.

These datasets collectively provide the foundation for training universal MLIPs, with each contributing unique aspects: equilibrium properties (MP), surface structures (OC20/OC22), dynamic trajectories (MPtrj), dimensional diversity (Alexandria), and non-equilibrium configurations (OMat24). The models evaluated in our study leverage various combinations of these datasets, as detailed in Table 1.

\begin{figure*}[p]  
\centering
\begin{minipage}{\textwidth}
    \centering \includegraphics[height=0.85\textheight,width=\textwidth,keepaspectratio]{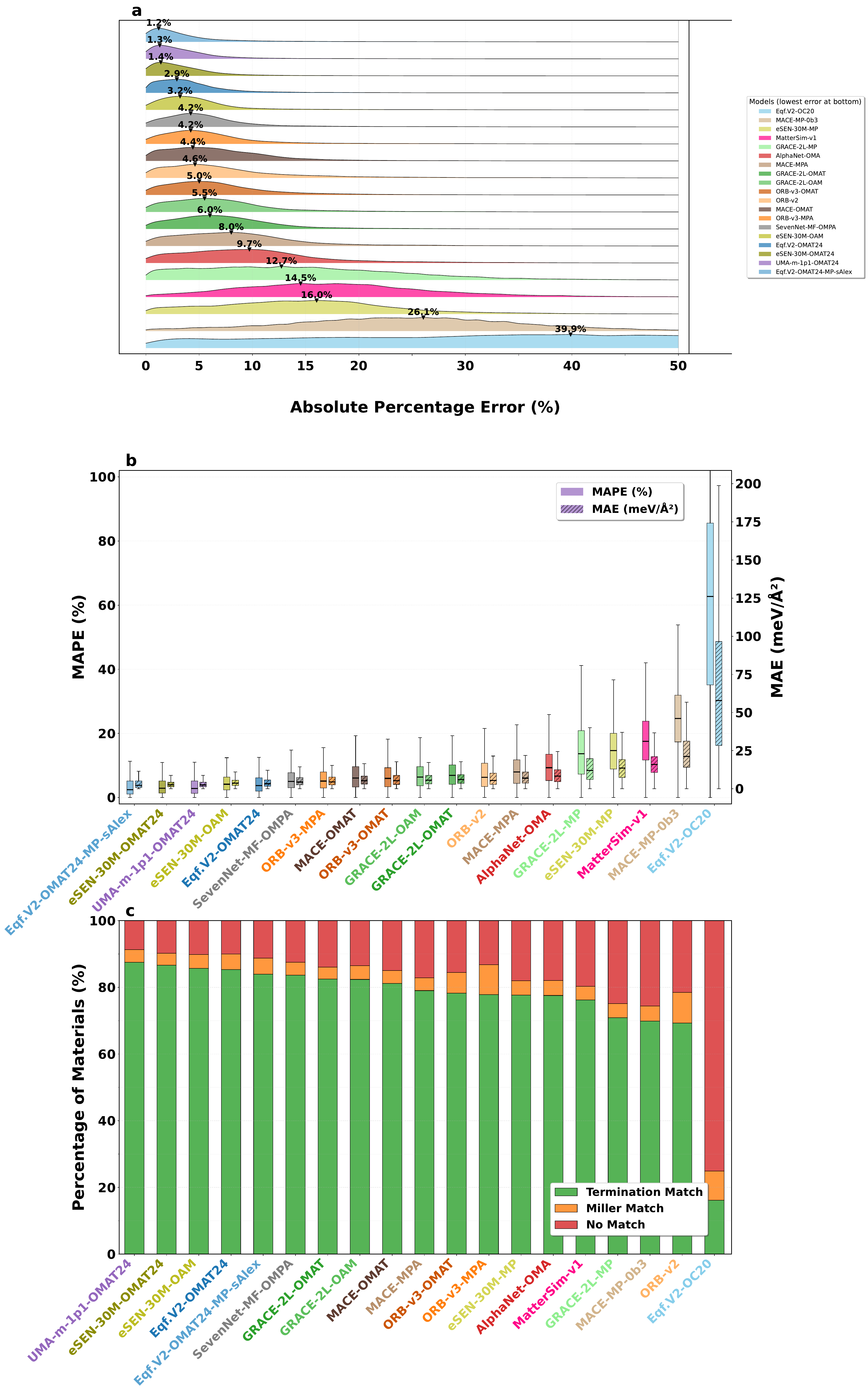}
    \caption{\textbf{Performance comparison of 19 universal MLIPs for cleavage energy prediction across 36,718 surface structures.}
    \textbf{(a)} Ridge plots showing absolute percentage error distributions ordered by median performance (best to worst). Median values displayed on each distribution. 
    \textbf{(b)} Box plots of MAPE (left axis) and MAE (right axis) distributions. 
    \textbf{(c)} Agreement between MLIP and DFT for identifying the most stable surface across $3{,}699$ materials: green = exact termination match, yellow = Miller-index match only, red = no match.}
    \label{fig:model_comparison_complete}
\end{minipage}
\end{figure*}

\begin{figure*}[ht]
\centering
\includegraphics[width=\textwidth]{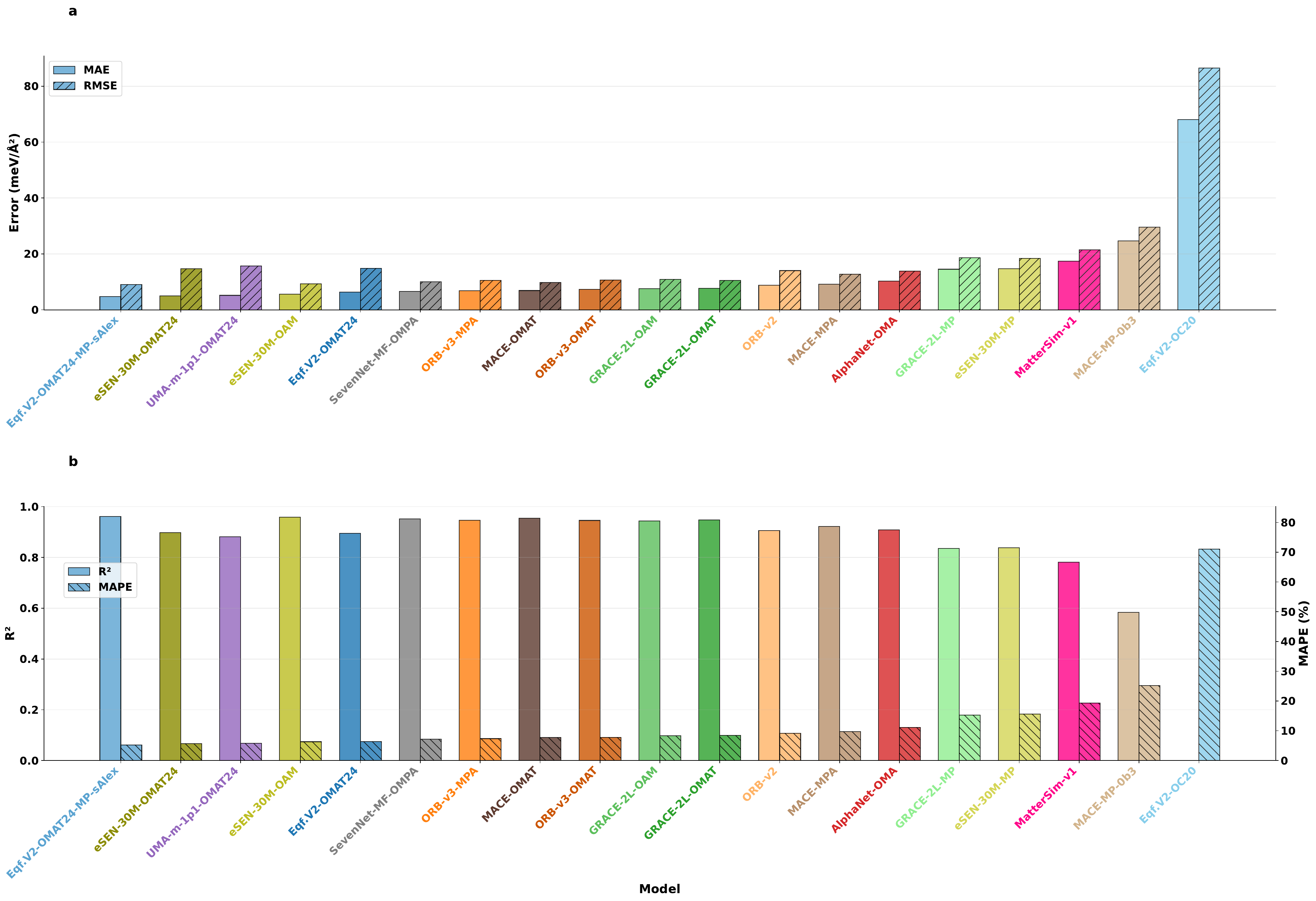}
\caption{\textbf{Comprehensive error metrics comparison across 19 universal MLIPs for cleavage energy prediction.} 
(\textbf{a}) MAE and RMSE in eV/\AA$^2$ for each model, ordered by increasing MAE.
(\textbf{b}) Coefficient of determination (R$^2$) and MAPE for each model. The R$^2$ values (left axis) indicate the quality of linear correlation between MLIP and DFT predictions, while MAPE values (right axis, hatched bars) quantify relative prediction errors. Models are arranged in the same order as panel (a). Note the logarithmic scale break in the MAPE axis to accommodate the wide range of values.}
\label{fig:model_metrics_comparison}
\end{figure*}

\begin{figure*}[ht]
\centering
\includegraphics[width=\textwidth]{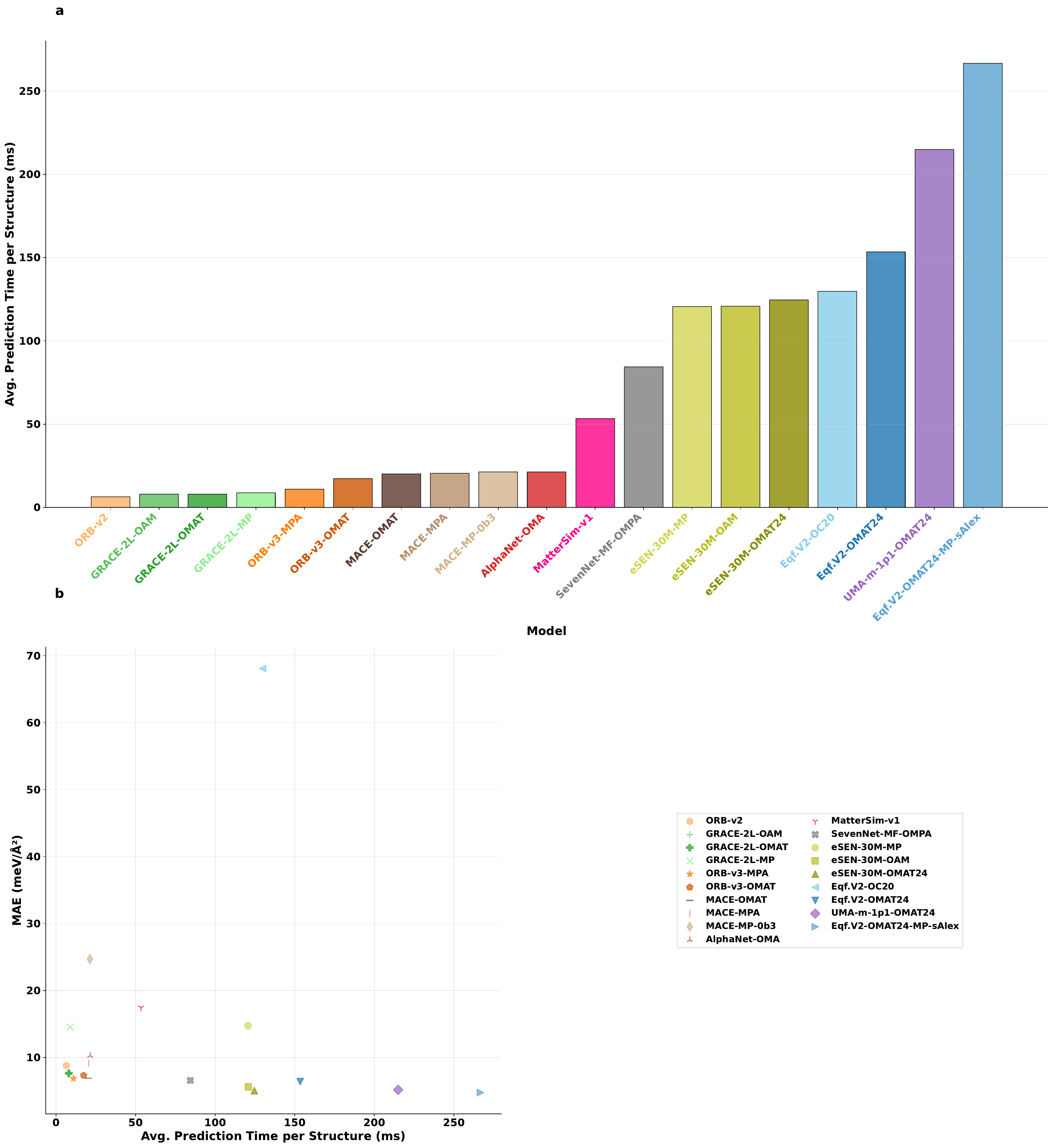}
\caption{\textbf{Computational efficiency analysis of universal MLIPs for cleavage energy prediction.}
(\textbf{a}) Average wall-time per slab structure prediction on NVIDIA L40S GPUs, ordered by computational speed. The fastest models (ORB-v2, GRACE family, and MACE variants) achieve sub-20 ms predictions, while transformer-based architectures (UMA-m-1p1, EquiformerV2 variants) require 200+ ms per structure.
(\textbf{b}) Speed-accuracy trade-off analysis showing the Pareto frontier of model performance. Models are colored by architecture. The ideal region (fast and accurate) is in the bottom-left corner.}
\label{fig:computational_efficiency}
\end{figure*}

\begin{figure*}[ht]
\centering
\includegraphics[width=\textwidth]{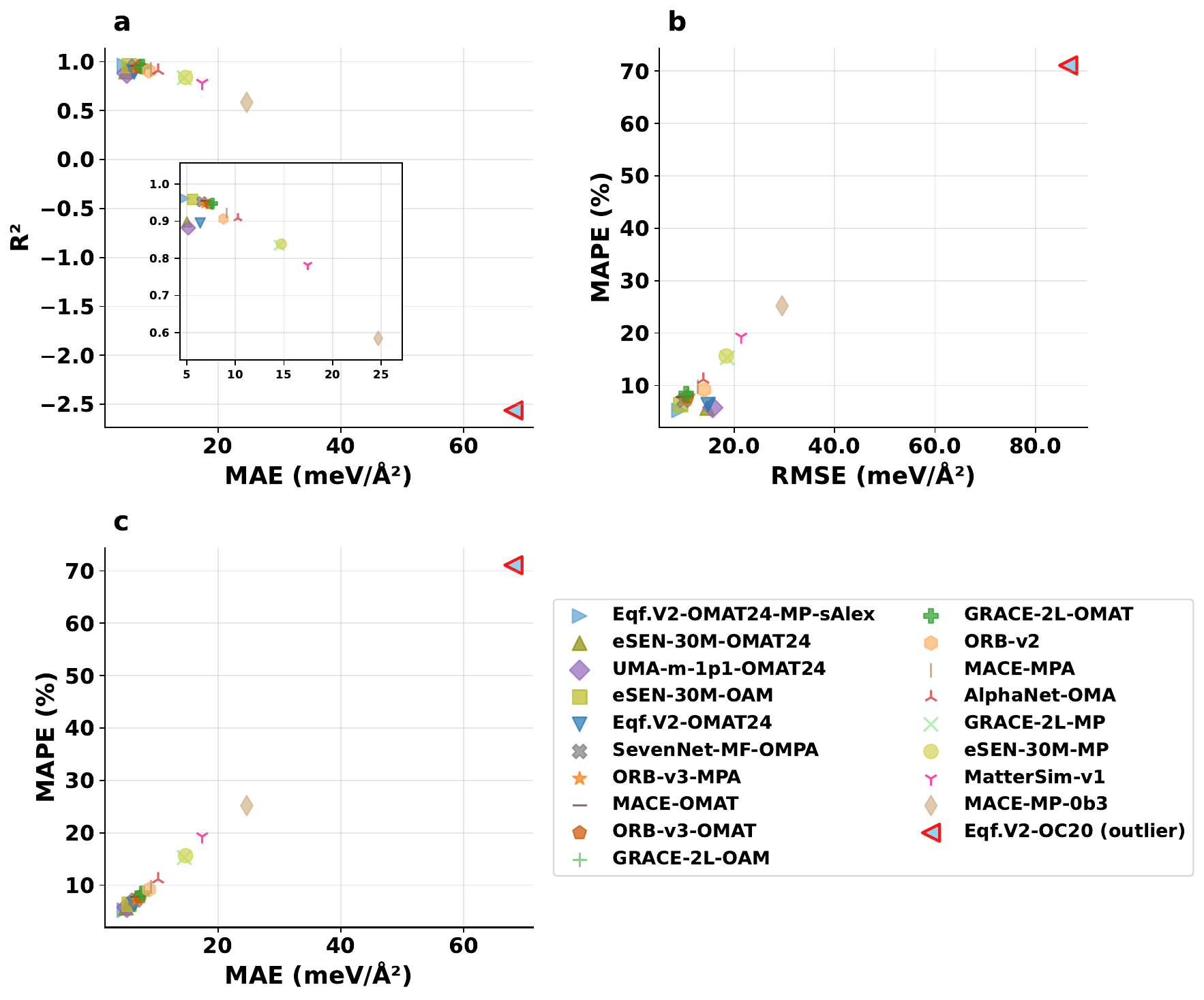}
\caption{\textbf{Correlation analysis of performance metrics across universal MLIPs.}
Scatter plots showing pairwise relationships between model performance metrics: mean absolute error (MAE), root mean square error (RMSE), coefficient of determination (R$^2$), and mean absolute percentage error (MAPE). Each point represents one of the 19 evaluated models, with marker shapes and colors indicating different model families and architectures. The inset in the MAE vs R$^2$ panel provides a magnified view of the main cluster of models, excluding the outlier model (EquiformerV2-OC20, marked with red triangle). Notable observations include: (i) near-perfect correlation between MAE and RMSE, confirming consistent error distributions; (ii) strong negative correlation between R$^2$ and error metrics, with most models achieving $(\mathrm{R}^2 > 90\%)$ except for eSEN-30M-MP, MatterSim-v1, MACE-MP-0b3 and EquiformerV2-OC20 outlier; (iii) EquiformerV2-OC20 exhibits anomalously poor performance (R$^2$ = -2.5, $\mathrm{MAPE} > 70\%$) due to training on catalysis-specific data rather than materials datasets featuring comprehensive elemental coverage spanning the periodic table, diverse structural representations including relaxed geometries and dynamic trajectories, and a full spectrum of configurational states from equilibrium ground-state structures through near-equilibrium thermal fluctuations to far-from-equilibrium high-energy configurations across three-dimensional bulk crystals, two-dimensional layered materials, and one-dimensional chain structures. The clustering of models by architecture family suggests that training data quality and relevance outweigh architectural differences for cleavage energy prediction.}
\label{fig:metrics_scatter}
\end{figure*}

\begin{figure}[htbp]
\centering
\includegraphics[width=\textwidth]{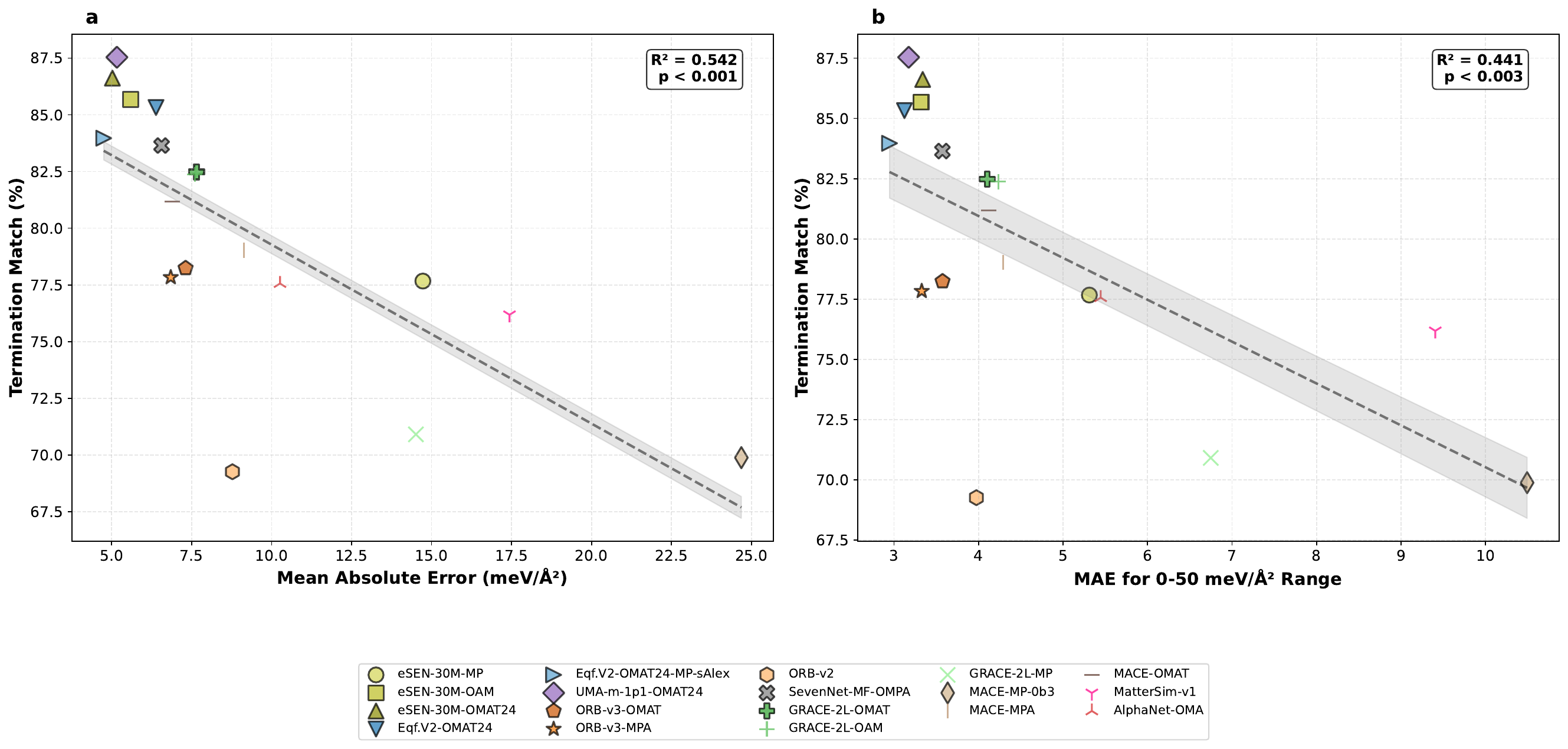}
\caption{\textbf{Correlation between prediction accuracy metrics and surface termination identification accuracy across MLIP models, excluding EquiformerV2-153M-S2EF-OC20-All+MD.} 
\textbf{(a)} Relationship between mean absolute error (MAE) across all cleavage energies and the percentage of materials where the MLIP correctly identifies the same minimum-energy surface termination as DFT. A significant negative correlation (R² = 0.542, p < 0.001, Pearson's correlation) demonstrates that models with lower overall prediction errors more accurately identify stable surface terminations. The regression line (dashed) with 95\% confidence interval (shaded region) indicates a decrease of 0.79\% in termination match accuracy per meV/Å² increase in MAE. 
\textbf{(b)} Correlation between MAE for low-energy surfaces (0--50 meV/Å² range) and termination match percentage. The relationship remains significant (R² = 0.441, p = 0.003) with a steeper slope (-1.74\% per meV/Å²), suggesting that accuracy in the low-energy regime is particularly important for correct surface stability ranking. The exclusion of EquiformerV2-OC20, which was trained on a distinct dataset (Open Catalyst 2020) compared to the materials-focused training sets of other models, reveals a more robust correlation pattern among MLIPs trained on similar data distributions. Each point represents one MLIP model (n = 18), with markers and colors consistent with the main text figures.}
\label{fig:correlation_excluding_oc20}
\end{figure}


\begin{landscape}
\begin{table}[htbp]
\centering
\tiny
\caption{MAPE statistics for each cleavage energy range}
\label{tab:binned_performance}
\begin{tabular}{lcccccccccccccccccccccccc}
\toprule
Model & \multicolumn{3}{c}{0.00-0.05 eV/\AA$^2$} & \multicolumn{3}{c}{0.05-0.10 eV/\AA$^2$} & \multicolumn{3}{c}{0.10-0.15 eV/\AA$^2$} & \multicolumn{3}{c}{0.15-0.20 eV/\AA$^2$} & \multicolumn{3}{c}{0.20-0.25 eV/\AA$^2$} & \multicolumn{3}{c}{0.25-0.30 eV/\AA$^2$} & \multicolumn{3}{c}{0.30-0.35 eV/\AA$^2$} & \multicolumn{3}{c}{0.35-0.40 eV/\AA$^2$} \\
 & N & Mean & Std & N & Mean & Std & N & Mean & Std & N & Mean & Std & N & Mean & Std & N & Mean & Std & N & Mean & Std & N & Mean & Std \\
\midrule
eSEN-30M-MP & 2813 & 19.3 & 119.2 & 19643 & 16.2 & 9.3 & 9553 & 14.9 & 8.9 & 3404 & 13.7 & 10.7 & 895 & 9.1 & 6.3 & 280 & 8.4 & 6.8 & 82 & 9.5 & 6.9 & 48 & 9.6 & 7.9 \\
eSEN-30M-OAM & 2813 & 14.5 & 142.7 & 19643 & 6.0 & 6.6 & 9553 & 5.3 & 6.3 & 3404 & 5.3 & 6.8 & 895 & 4.5 & 5.1 & 280 & 4.8 & 6.6 & 82 & 5.6 & 6.7 & 48 & 6.4 & 7.7 \\
eSEN-30M-OMAT24 & 2813 & 14.2 & 121.8 & 19643 & 5.0 & 9.5 & 9553 & 4.9 & 13.9 & 3404 & 5.3 & 15.4 & 895 & 3.5 & 4.7 & 280 & 3.7 & 6.3 & 82 & 4.6 & 6.5 & 48 & 5.3 & 7.3 \\
EquiformerV2-153M-OMAT24 & 2813 & 11.3 & 52.9 & 19643 & 4.9 & 8.4 & 9553 & 7.1 & 14.0 & 3404 & 7.9 & 12.9 & 895 & 7.2 & 7.3 & 280 & 6.9 & 7.5 & 82 & 5.8 & 6.8 & 48 & 6.7 & 8.2 \\
EquiformerV2-MP-sAlex & 2813 & 13.1 & 145.1 & 19643 & 4.4 & 7.2 & 9553 & 4.7 & 6.3 & 3404 & 5.1 & 6.5 & 895 & 5.1 & 5.5 & 280 & 4.9 & 6.8 & 82 & 4.9 & 6.9 & 48 & 5.9 & 8.3 \\
EquiformerV2-OC20 & 2813 & 109.3 & 223.8 & 19643 & 70.7 & 53.6 & 9553 & 62.5 & 35.9 & 3404 & 65.8 & 31.8 & 895 & 69.5 & 30.4 & 280 & 71.8 & 22.3 & 82 & 85.1 & 23.3 & 48 & 85.5 & 30.8 \\
UMA-m-1p1-OMAT24 & 2813 & 15.0 & 117.0 & 19643 & 4.8 & 8.3 & 9553 & 5.6 & 15.3 & 3404 & 5.4 & 12.7 & 895 & 3.6 & 5.1 & 280 & 3.6 & 6.4 & 82 & 4.0 & 6.4 & 48 & 14.6 & 49.3 \\
ORB-v3-OMAT & 2813 & 15.2 & 131.5 & 19643 & 7.2 & 6.5 & 9553 & 6.9 & 6.0 & 3404 & 7.8 & 6.4 & 895 & 8.0 & 5.7 & 280 & 8.4 & 5.9 & 82 & 8.2 & 5.9 & 48 & 8.4 & 6.9 \\
ORB-v3-MPA & 2813 & 14.8 & 152.5 & 19643 & 6.7 & 6.8 & 9553 & 6.5 & 6.6 & 3404 & 7.3 & 6.7 & 895 & 7.5 & 5.7 & 280 & 7.0 & 6.6 & 82 & 7.4 & 6.5 & 48 & 8.4 & 7.3 \\
ORB-v2 & 2813 & 15.9 & 121.7 & 19643 & 8.3 & 9.1 & 9553 & 8.9 & 9.3 & 3404 & 9.7 & 10.0 & 895 & 8.7 & 7.9 & 280 & 6.8 & 7.3 & 82 & 7.0 & 6.7 & 48 & 8.4 & 8.0 \\
SevenNet-MF-OMPA & 2813 & 15.1 & 138.8 & 19643 & 6.5 & 6.5 & 9553 & 6.3 & 6.2 & 3404 & 6.7 & 6.2 & 895 & 6.6 & 5.6 & 280 & 5.9 & 7.1 & 82 & 6.7 & 6.7 & 48 & 7.4 & 7.9 \\
GRACE-2L-OMAT & 2813 & 16.4 & 110.8 & 19643 & 8.3 & 6.1 & 9553 & 7.0 & 5.5 & 3404 & 7.2 & 6.0 & 895 & 7.2 & 5.4 & 280 & 6.8 & 7.1 & 82 & 7.1 & 6.9 & 48 & 8.4 & 8.1 \\
GRACE-2L-OAM & 2813 & 16.8 & 127.9 & 19643 & 8.1 & 6.9 & 9553 & 6.9 & 6.3 & 3404 & 7.2 & 6.3 & 895 & 7.8 & 5.7 & 280 & 7.9 & 6.3 & 82 & 7.7 & 6.7 & 48 & 9.3 & 7.8 \\
GRACE-2L-MP & 2813 & 22.6 & 83.5 & 19643 & 15.3 & 10.3 & 9553 & 13.7 & 8.7 & 3404 & 14.2 & 9.2 & 895 & 13.3 & 8.2 & 280 & 13.8 & 7.8 & 82 & 12.9 & 8.1 & 48 & 17.1 & 7.8 \\
MACE-MP-0b3 & 2813 & 32.0 & 72.9 & 19643 & 24.8 & 11.3 & 9553 & 23.8 & 10.4 & 3404 & 26.0 & 11.1 & 895 & 25.3 & 9.8 & 280 & 24.2 & 9.4 & 82 & 21.8 & 9.2 & 48 & 23.0 & 10.0 \\
MACE-MPA & 2813 & 18.0 & 156.4 & 19643 & 9.1 & 7.0 & 9553 & 8.6 & 6.9 & 3404 & 8.7 & 6.9 & 895 & 10.1 & 6.2 & 280 & 11.9 & 7.2 & 82 & 12.4 & 7.9 & 48 & 14.3 & 8.1 \\
MACE-OMAT & 2813 & 16.7 & 110.1 & 19643 & 7.6 & 5.9 & 9553 & 6.2 & 5.4 & 3404 & 6.0 & 5.9 & 895 & 6.3 & 5.4 & 280 & 7.0 & 6.8 & 82 & 8.2 & 6.3 & 48 & 8.7 & 7.3 \\
MatterSim-v1 & 2813 & 31.1 & 103.5 & 19643 & 20.4 & 12.1 & 9553 & 16.5 & 12.5 & 3404 & 13.9 & 8.1 & 895 & 11.9 & 7.0 & 280 & 10.9 & 7.3 & 82 & 13.1 & 6.6 & 48 & 13.8 & 6.6 \\
AlphaNet-OMA & 2813 & 21.5 & 156.6 & 19643 & 10.7 & 7.5 & 9553 & 9.7 & 7.7 & 3404 & 9.8 & 7.6 & 895 & 10.1 & 6.8 & 280 & 8.9 & 7.2 & 82 & 10.5 & 8.3 & 48 & 11.9 & 7.8 \\
\bottomrule
\end{tabular}
\end{table}
\end{landscape}

\begin{table}[htbp]
\centering
\small
\caption{Complete performance metrics for MLIP models on cleavage energy prediction}
\label{tab:mlip_complete_metrics}
\begin{tabular}{lcccccccc}
\toprule
Model and Family & MAE & RMSE & MAPE & R$^2$ & Med. Err & Q1 & Q3 \\
 & (meV/\AA$^2$) & (meV/\AA$^2$) & (\%) & & (meV/\AA$^2$) & (meV/\AA$^2$) & (meV/\AA$^2$) \\
\midrule
\multicolumn{8}{l}{\textbf{FAIRCHEM V1}} \\
  EquiformerV2-MP-sAlex & 4.8 & 9.0 & 5.2 & 0.961 & -1.5 & -4.2 & 0.1 \\
  eSEN-30M-OMAT24 & 5.0 & 14.6 & 5.7 & 0.898 & -2.2 & -3.9 & -0.5 \\
  eSEN-30M-OAM & 5.6 & 9.3 & 6.3 & 0.959 & -3.4 & -5.4 & -1.7 \\
  EquiformerV2-153M-OMAT24 & 6.4 & 14.8 & 6.3 & 0.895 & -3.0 & -5.5 & -1.2 \\
  eSEN-30M-MP & 14.7 & 18.4 & 15.6 & 0.839 & -13.4 & -19.1 & -6.9 \\
  EquiformerV2-OC20 & 68.1 & 86.5 & 71.1 & -2.562 & -52.1 & -93.0 & -16.3 \\
\addlinespace
\multicolumn{8}{l}{\textbf{FAIRCHEM V2}} \\
  UMA-m-1p1-OMAT24 & 5.2 & 15.7 & 5.8 & 0.882 & -2.1 & -3.9 & -0.5 \\
\addlinespace
\multicolumn{8}{l}{\textbf{SEVENNET}} \\
  SevenNet-MF-OMPA & 6.6 & 10.0 & 7.1 & 0.953 & -4.1 & -7.0 & -2.0 \\
\addlinespace
\multicolumn{8}{l}{\textbf{ORB}} \\
  ORB-v3-MPA & 6.8 & 10.6 & 7.4 & 0.947 & -4.2 & -7.3 & -1.9 \\
  ORB-v3-OMAT & 7.3 & 10.7 & 7.8 & 0.946 & -5.0 & -8.6 & -2.4 \\
  ORB-v2 & 8.8 & 14.0 & 9.2 & 0.906 & -5.1 & -9.8 & -2.0 \\
\addlinespace
\multicolumn{8}{l}{\textbf{MACE}} \\
  MACE-OMAT & 6.9 & 9.7 & 7.8 & 0.955 & -4.9 & -8.1 & -2.1 \\
  MACE-MPA & 9.1 & 12.8 & 9.7 & 0.922 & -6.6 & -10.7 & -2.9 \\
  MACE-MP-0b3 & 24.7 & 29.6 & 25.2 & 0.584 & -21.2 & -31.2 & -14.1 \\
\addlinespace
\multicolumn{8}{l}{\textbf{GRACE}} \\
  GRACE-2L-OAM & 7.6 & 10.9 & 8.3 & 0.944 & -5.3 & -8.6 & -2.5 \\
  GRACE-2L-OMAT & 7.7 & 10.5 & 8.4 & 0.948 & -5.8 & -8.9 & -3.0 \\
  GRACE-2L-MP & 14.5 & 18.6 & 15.3 & 0.835 & -11.4 & -19.3 & -4.7 \\
\addlinespace
\multicolumn{8}{l}{\textbf{ALPHANET}} \\
  AlphaNet-OMA & 10.3 & 13.9 & 11.1 & 0.908 & -8.2 & -12.3 & -3.9 \\
\addlinespace
\multicolumn{8}{l}{\textbf{MATTERSIM}} \\
  MatterSim-v1 & 17.4 & 21.4 & 19.3 & 0.782 & -15.8 & -21.0 & -10.6 \\
\addlinespace
\bottomrule
\end{tabular}
\end{table}

\begin{table}[htbp]
\centering
\small
\caption{Outlier statistics and error ranges for MLIP models}
\label{tab:mlip_outlier_stats}
\begin{tabular}{lcccc}
\toprule
Model & Min Error & Max Error & Error Range & Std Error \\
 & (meV/\AA$^2$) & (meV/\AA$^2$) & (meV/\AA$^2$) & (meV/\AA$^2$) \\
\midrule
Eqf.V2-OMAT24-MP-sAlex & -151.6 & 52.8 & 204.3 & 8.4 \\
eSEN-30M-OMAT24 & -516.0 & 182.8 & 698.8 & 14.2 \\
UMA-m-1p1-OMAT24 & -385.7 & 892.9 & 1278.6 & 15.3 \\
eSEN-30M-OAM & -175.2 & 50.1 & 225.3 & 8.0 \\
Eqf.V2-OMAT24 & -242.1 & 71.1 & 313.2 & 13.7 \\
SevenNet-MF-OMPA & -153.8 & 43.8 & 197.5 & 8.2 \\
ORB-v3-MPA & -150.1 & 48.7 & 198.8 & 8.8 \\
MACE-OMAT & -148.2 & 44.9 & 193.2 & 7.9 \\
ORB-v3-OMAT & -144.1 & 41.1 & 185.2 & 8.3 \\
GRACE-2L-OAM & -154.5 & 41.6 & 196.2 & 8.7 \\
GRACE-2L-OMAT & -160.9 & 37.2 & 198.1 & 8.1 \\
ORB-v2 & -146.9 & 401.6 & 548.5 & 11.9 \\
MACE-MPA & -157.3 & 87.7 & 245.0 & 9.9 \\
AlphaNet-OMA & -146.3 & 80.4 & 226.7 & 10.3 \\
GRACE-2L-MP & -151.5 & 106.7 & 258.1 & 13.7 \\
eSEN-30M-MP & -274.0 & 68.0 & 342.0 & 12.1 \\
MatterSim-v1 & -214.7 & 197.2 & 411.9 & 13.2 \\
MACE-MP-0b3 & -181.0 & 35.1 & 216.1 & 16.6 \\
Eqf.V2-OC20 & -847.4 & 482.2 & 1329.6 & 67.5 \\
\bottomrule
\end{tabular}
\end{table}


\begin{table}[htbp]
\centering
\caption{Summary statistics across all MLIP models}
\label{tab:summary_stats}
\begin{tabular}{lcccc}
\toprule
Metric & Mean & Std & Min & Max \\
\midrule
MAE (meV/\AA$^2$) & 12.5 & 14.4 & 4.8 & 68.1 \\
RMSE (meV/\AA$^2$) & 18.0 & 17.4 & 9.0 & 86.5 \\
MAPE (\%) & 13.3 & 14.9 & 5.2 & 71.1 \\
R$^2$ & 0.711 & 0.798 & -2.562 & 0.961 \\
\bottomrule
\end{tabular}
\end{table}


\begin{table}[htbp]
\centering
\small
\caption{Outlier analysis for MLIP models (outliers defined as MAPE > 100\%)}
\label{tab:outlier_analysis}
\begin{tabular}{lcccccc}
\toprule
Model & Viz. & MAPE & \% & Mean MAPE & Max MAPE & Median Error \\
 & Outliers & >100\% & Outliers & of Outliers (\%) & (\%) & Outliers (meV/\AA$^2$) \\
\midrule
eSEN-30M-MP & 150 & 27 & 0.1 & 533 & 4355 & -6.7 \\
eSEN-30M-OAM & 131 & 18 & 0.0 & 878 & 5185 & 35.6 \\
eSEN-30M-OMAT24 & 133 & 140 & 0.4 & 243 & 4337 & -125.9 \\
Eqf.V2-OMAT24 & 130 & 113 & 0.3 & 181 & 1769 & -141.7 \\
Eqf.V2-OMAT24-MP-sAlex & 130 & 24 & 0.1 & 695 & 5274 & 31.3 \\
Eqf.V2-OC20 & 229 & 6641 & 18.1 & 161 & 4340 & -102.8 \\
UMA-m-1p1-OMAT24 & 142 & 145 & 0.4 & 252 & 3778 & -125.7 \\
ORB-v3-OMAT & 130 & 23 & 0.1 & 647 & 4787 & 28.2 \\
ORB-v3-MPA & 130 & 17 & 0.0 & 954 & 5562 & 36.2 \\
ORB-v2 & 146 & 39 & 0.1 & 418 & 4441 & 23.7 \\
SevenNet-MF-OMPA & 133 & 18 & 0.0 & 832 & 5068 & 33.8 \\
GRACE-2L-OMAT & 135 & 15 & 0.0 & 836 & 4007 & 28.2 \\
GRACE-2L-OAM & 130 & 17 & 0.0 & 826 & 4641 & 34.9 \\
GRACE-2L-MP & 178 & 29 & 0.1 & 400 & 2941 & 15.5 \\
MACE-MP-0b3 & 130 & 17 & 0.0 & 495 & 2630 & -6.9 \\
MACE-MPA & 130 & 22 & 0.1 & 801 & 5689 & 32.2 \\
MACE-OMAT & 138 & 15 & 0.0 & 866 & 3931 & 29.6 \\
MatterSim-v1 & 130 & 126 & 0.3 & 207 & 3779 & -108.6 \\
AlphaNet-OMA & 134 & 21 & 0.1 & 814 & 5705 & 33.9 \\
\bottomrule
\end{tabular}
\end{table}


\begin{table}[htbp]
\centering
\caption{MAPE by composition type and symmetry}
\begin{tabular}{lcccccc}
\toprule
 & \multicolumn{3}{c}{Composition} & \multicolumn{2}{c}{Symmetry} \\
\cmidrule(lr){2-4} \cmidrule(lr){5-6}
Model & Unary & Binary & Ternary & Symmetric & Asymmetric \\
\midrule
eSEN-30M-MP & 25.9 & 15.5 & 15.6 & 16.2 & 15.5 \\
eSEN-30M-OAM & 9.9 & 6.3 & 6.3 & 6.4 & 6.3 \\
eSEN-30M-OMAT24 & 9.5 & 7.2 & 5.0 & 6.1 & 5.6 \\
EquiformerV2-153M-OMAT24 & 9.2 & 7.7 & 5.8 & 6.9 & 6.2 \\
EquiformerV2-MP-sAlex & 6.5 & 5.0 & 5.3 & 5.1 & 5.3 \\
EquiformerV2-OC20 & 273.0 & 70.3 & 69.6 & 79.9 & 69.2 \\
UMA-m-1p1-OMAT24 & 9.8 & 6.6 & 5.4 & 6.9 & 5.5 \\
ORB-v3-OMAT & 15.2 & 8.1 & 7.6 & 8.4 & 7.7 \\
ORB-v3-MPA & 11.0 & 7.2 & 7.4 & 7.5 & 7.3 \\
ORB-v2 & 12.3 & 9.3 & 9.1 & 9.7 & 9.1 \\
SevenNet-MF-OMPA & 11.2 & 7.2 & 7.1 & 7.4 & 7.1 \\
GRACE-2L-OMAT & 15.1 & 8.7 & 8.2 & 8.9 & 8.3 \\
GRACE-2L-OAM & 12.8 & 8.3 & 8.3 & 8.8 & 8.2 \\
GRACE-2L-MP & 33.4 & 14.8 & 15.3 & 16.0 & 15.1 \\
MACE-MP-0b3 & 33.2 & 25.6 & 24.9 & 26.7 & 24.9 \\
MACE-MPA & 13.0 & 9.9 & 9.6 & 10.5 & 9.5 \\
MACE-OMAT & 16.3 & 8.4 & 7.4 & 8.4 & 7.6 \\
MatterSim-v1 & 31.8 & 19.1 & 19.3 & 21.3 & 18.9 \\
AlphaNet-OMA & 19.2 & 11.3 & 11.0 & 12.0 & 10.9 \\
\bottomrule
\end{tabular}
\end{table}


\begin{table}[htbp]
\centering
\small
\caption{MAPE by crystal system for each model}
\begin{tabular}{lccccccc}
\toprule
Model & Cubic & Hexagonal & Monoclinic & Orthorhombic & Tetragonal & Trigonal & Triclinic \\
\midrule
eSEN-30M-MP & 17.2 & 17.2 & 11.9 & 15.1 & 13.9 & 21.3 & 16.2 \\
eSEN-30M-OAM & 7.2 & 6.0 & 5.4 & 5.4 & 5.7 & 17.2 & 11.7 \\
eSEN-30M-OMAT24 & 10.1 & 5.4 & 3.9 & 3.8 & 4.8 & 15.1 & 12.1 \\
EquiformerV2-153M-OMAT24 & 8.4 & 5.8 & 5.5 & 5.9 & 6.0 & 11.7 & 11.9 \\
EquiformerV2-MP-sAlex & 5.6 & 4.6 & 4.6 & 4.2 & 5.0 & 17.3 & 9.2 \\
EquiformerV2-OC20 & 88.9 & 78.8 & 88.7 & 65.6 & 61.0 & 77.4 & 83.7 \\
UMA-m-1p1-OMAT24 & 8.3 & 5.9 & 3.8 & 4.1 & 4.9 & 15.0 & 11.5 \\
ORB-v3-OMAT & 9.5 & 7.5 & 7.1 & 6.6 & 7.0 & 18.8 & 14.5 \\
ORB-v3-MPA & 7.7 & 7.0 & 6.9 & 6.5 & 6.7 & 19.4 & 13.9 \\
ORB-v2 & 10.0 & 8.2 & 9.0 & 8.7 & 9.0 & 18.7 & 16.8 \\
SevenNet-MF-OMPA & 7.7 & 6.6 & 6.4 & 6.4 & 6.6 & 18.5 & 14.3 \\
GRACE-2L-OMAT & 9.2 & 8.2 & 7.9 & 7.5 & 7.6 & 19.4 & 20.2 \\
GRACE-2L-OAM & 8.8 & 8.0 & 8.1 & 7.9 & 7.5 & 18.9 & 17.2 \\
GRACE-2L-MP & 16.9 & 16.3 & 15.1 & 13.5 & 13.9 & 22.7 & 20.7 \\
MACE-MP-0b3 & 27.7 & 27.2 & 24.9 & 21.8 & 23.1 & 32.4 & 32.5 \\
MACE-MPA & 10.4 & 9.4 & 9.7 & 9.2 & 8.7 & 23.7 & 16.4 \\
MACE-OMAT & 8.6 & 7.5 & 7.5 & 7.1 & 6.7 & 18.9 & 20.1 \\
MatterSim-v1 & 20.4 & 20.7 & 15.7 & 17.7 & 17.7 & 25.8 & 28.6 \\
AlphaNet-OMA & 11.5 & 11.6 & 11.8 & 11.4 & 9.3 & 22.5 & 22.0 \\
\bottomrule
\end{tabular}
\end{table}


\begin{table}[htbp]
\centering
\caption{Mean errors for most common elements across all models}
\begin{tabular}{lccc}
\toprule
Element & Total Count & Mean MAPE (\%) & Mean MAE (meV/\AA$^2$) \\
\midrule
Si & 1426520 & 10.2 & 12.4 \\
Ge & 1328746 & 11.6 & 11.4 \\
Ni & 1044107 & 9.4 & 10.0 \\
Pd & 948974 & 11.3 & 9.2 \\
Cu & 806474 & 11.1 & 9.7 \\
In & 704463 & 12.6 & 9.6 \\
Au & 646817 & 14.0 & 10.8 \\
Pt & 643112 & 10.3 & 10.9 \\
Al & 549727 & 10.6 & 10.4 \\
Sn & 545224 & 12.1 & 9.1 \\
\bottomrule
\end{tabular}
\end{table}

\begin{table*}[htbp]
\centering
\caption{\textbf{Thickness dependency analysis of MAPE across all models.} Values show how prediction error varies with surface thickness.}
\label{tab:thickness_dependency}
\footnotesize
\begin{tabular}{lccccc}
\toprule
\multirow{2}{*}{Model} & \multicolumn{5}{c}{Mean MAPE by Thickness (\%)} \\
\cmidrule(lr){2-6}
 & $<10$Å & 10-15Å & 15-20Å & 20-25Å & $\geq 25$Å \\
\midrule
\textbf{Best Performers} \\
Eqf.V2-OMAT24-MP-sAlex & 4.5 & 5.4 & 5.2 & 4.2 & 12.2 \\
eSEN-30M-OAM & 6.1 & 6.5 & 6.2 & 5.3 & 13.0 \\
eSEN-30M-OMAT24 & 6.1 & 6.1 & 5.0 & 4.5 & 10.8 \\
UMA-m-1p1-OMAT24 & 6.2 & 6.1 & 5.3 & 4.8 & 10.3 \\
Eqf.V2-OMAT24 & 4.3 & 6.5 & 6.2 & 5.6 & 10.8 \\
\midrule
\textbf{Moderate Performers} \\
SevenNet-MF-OMPA & 6.2 & 7.3 & 7.2 & 5.9 & 15.4 \\
ORB-v3-MPA & 7.0 & 7.6 & 7.2 & 6.0 & 14.3 \\
ORB-v3-OMAT & 7.6 & 8.2 & 7.4 & 6.4 & 14.2 \\
MACE-OMAT & 7.2 & 7.9 & 7.8 & 6.5 & 17.8 \\
GRACE-2L-OAM & 8.1 & 8.5 & 8.3 & 7.3 & 17.5 \\
GRACE-2L-OMAT & 7.8 & 8.6 & 8.3 & 7.5 & 15.4 \\
ORB-v2 & 7.9 & 9.4 & 9.2 & 7.7 & 13.5 \\
MACE-MPA & 9.5 & 10.0 & 9.5 & 7.9 & 19.2 \\
AlphaNet-OMA & 11.2 & 11.6 & 10.7 & 8.8 & 21.8 \\
\midrule
\textbf{Poor Performers} \\
GRACE-2L-MP & 13.9 & 15.5 & 15.1 & 13.8 & 23.9 \\
eSEN-30M-MP & 16.6 & 16.1 & 15.1 & 13.7 & 19.2 \\
MACE-MP-0b3 & 15.9 & 17.5 & 18.2 & 16.8 & 25.3 \\
MatterSim-v1 & 21.7 & 19.6 & 18.9 & 17.9 & 26.4 \\
Eqf.V2-OC20 & 95.0 & 74.1 & 68.0 & 55.5 & 76.5 \\
\bottomrule
\end{tabular}
\end{table*}

\begin{table*}[htbp]
\centering
\caption{\textbf{Mean baseline performance analysis for cleavage energy prediction.} The mean baseline uses the database average cleavage energy (100.04 meV/\AA$^2$) as a constant predictor for all 36,718 structures. Panel A shows overall performance metrics. Panel B presents error distribution across cleavage energy ranges, demonstrating the mathematical amplification of relative errors for low-energy surfaces. Panel C compares the baseline against selected uMLIPs, quantifying the improvement factor achieved through machine learning.}
\label{tab:mean_baseline_analysis}
\begin{tabular}{@{}lccc@{}}
\toprule
\multicolumn{4}{l}{\textbf{A. Mean Baseline Performance Metrics}} \\
\midrule
Metric & Value & \multicolumn{2}{l}{Description} \\
\midrule
Database mean & 100.04 meV/\AA$^2$ & \multicolumn{2}{l}{Constant predictor value} \\
Standard deviation & 45.84 meV/\AA$^2$ & \multicolumn{2}{l}{Database variance} \\
MAE & 33.83 meV/\AA$^2$ & \multicolumn{2}{l}{Mean absolute error} \\
MAPE & 48.70\% & \multicolumn{2}{l}{Mean absolute percentage error} \\
RMSE & 45.84 meV/\AA$^2$ & \multicolumn{2}{l}{Root mean square error} \\
R$^2$ & 0.000 & \multicolumn{2}{l}{Coefficient of determination} \\
APE median & 28.60\% & \multicolumn{2}{l}{Median absolute percentage error} \\
Outliers ($>$100\% error) & 2,816 (7.67\%) & \multicolumn{2}{l}{Structures with MAPE $>$ 100\%} \\
\midrule
\multicolumn{4}{l}{\textbf{B. Performance by Cleavage Energy Range}} \\
\midrule
Energy Range (meV/\AA$^2$) & Structure Count & MAPE (\%) & MAE (meV/\AA$^2$) \\
\midrule
0--50 & 2,813 & 275.92 & 63.16 \\
50--100 & 19,643 & 32.91 & 22.57 \\
100--150 & 9,553 & 15.72 & 20.21 \\
150--200 & 3,404 & 41.08 & 70.84 \\
200--250 & 895 & 54.54 & 120.95 \\
250--300 & 280 & 62.66 & 168.58 \\
300--350 & 82 & 69.32 & 226.62 \\
350--400 & 48 & 72.79 & 268.06 \\
\midrule
\multicolumn{4}{l}{\textbf{C. Comparison with Top-Performing uMLIPs}} \\
\midrule
Model & MAPE (\%) & \multicolumn{2}{l}{Improvement Factor$^a$} \\
\midrule
Mean Baseline & 48.70 & \multicolumn{2}{l}{1.0$\times$ (reference)} \\
UMA-m-1p1-OMAT24 & 5.80 & \multicolumn{2}{l}{8.4$\times$} \\
EquiformerV2-153M-OAM & 5.20 & \multicolumn{2}{l}{9.4$\times$} \\
eSEN-30M-OMAT24 & 5.70 & \multicolumn{2}{l}{8.5$\times$} \\
GRACE-2L-OMAT & 8.40 & \multicolumn{2}{l}{5.8$\times$} \\
SevenNet-MF-OMPA & 7.10 & \multicolumn{2}{l}{6.9$\times$} \\
\bottomrule
\end{tabular}
\begin{flushleft}
\footnotesize{$^a$Improvement factor calculated as baseline MAPE divided by model MAPE.}
\end{flushleft}
\end{table*}


\begin{figure*}[t]
  \centering
  \includegraphics[width=\linewidth]{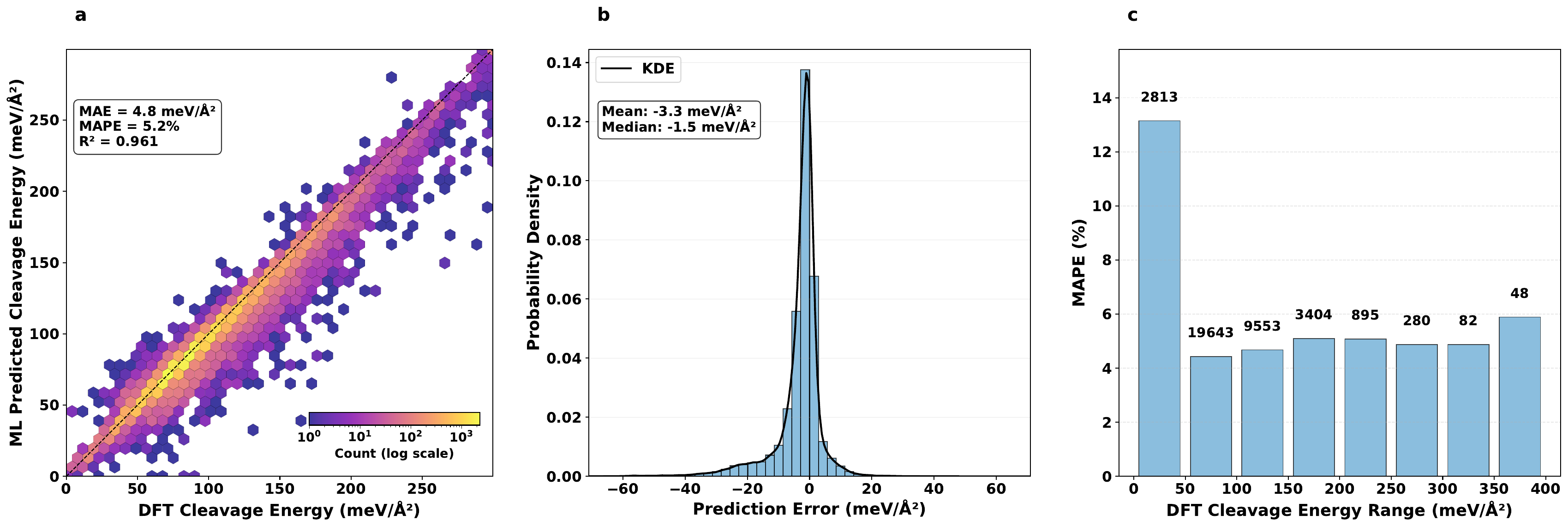}
  \caption{\textbf{Detailed performance analysis of Eqf.V2--OMAT24--MP--sAlex for cleavage energy prediction.}
  \textbf{(a)} Hexagonal density parity plot comparing MLIP-predicted versus DFT-calculated cleavage energies. Note that the color bar is logarithmic.
  \textbf{(b)} Probability density distribution of prediction errors (MLIP $-$ DFT), showing a kernel density estimate (KDE) curve calculated using a Gaussian kernel with Scott's rule bandwidth selection. The mean and median error values are displayed in the text box.
  \textbf{(c)} MAPE as a function of DFT cleavage energy bins. Numbers above bars indicate the sample count in each bin.}
  \label{fig:eqfv2_omat24_mp_salex_analysis}
\end{figure*}

\begin{figure*}[t]
  \centering
  \includegraphics[width=\linewidth]{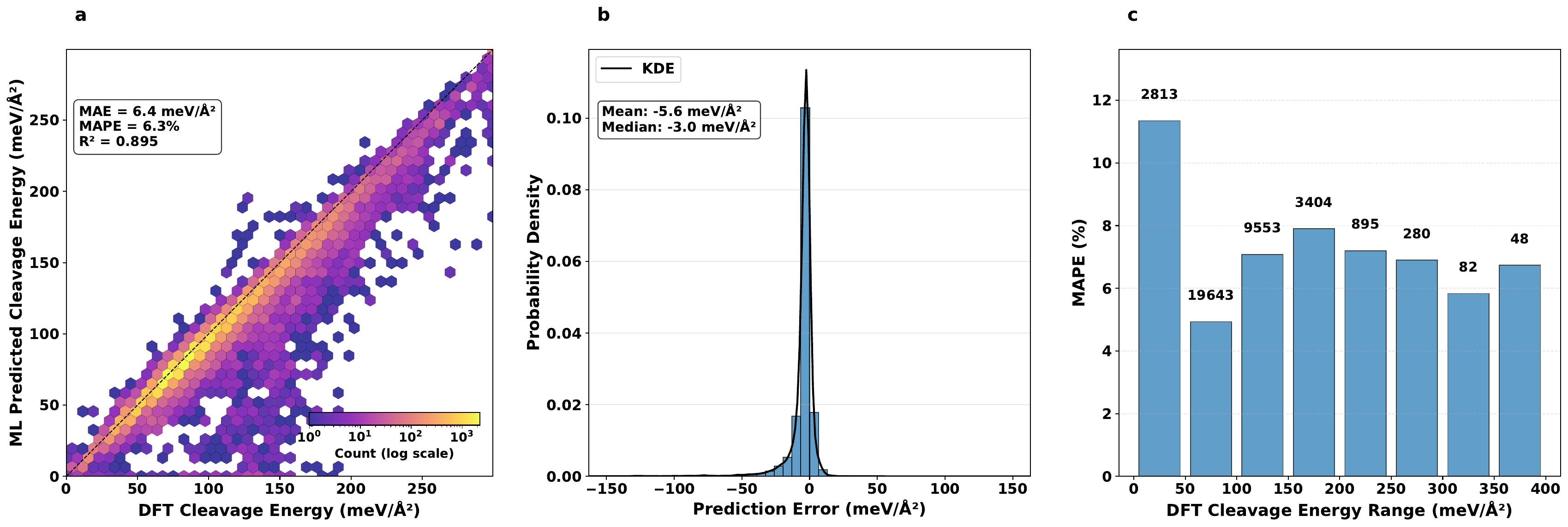}
  \caption{\textbf{Detailed performance analysis of Eqf.V2--OMAT24 for cleavage energy prediction.}
  \textbf{(a)} Hexagonal density parity plot comparing MLIP-predicted versus DFT-calculated cleavage energies. Note that the color bar is logarithmic.
  \textbf{(b)} Probability density distribution of prediction errors (MLIP $-$ DFT), showing a kernel density estimate (KDE) curve calculated using a Gaussian kernel with Scott's rule bandwidth selection. The mean and median error values are displayed in the text box.
  \textbf{(c)} MAPE as a function of DFT cleavage energy bins. Numbers above bars indicate the sample count in each bin.}
  \label{fig:eqfv2_omat24_analysis}
\end{figure*}

\begin{figure*}[t]
  \centering
  \includegraphics[width=\linewidth]{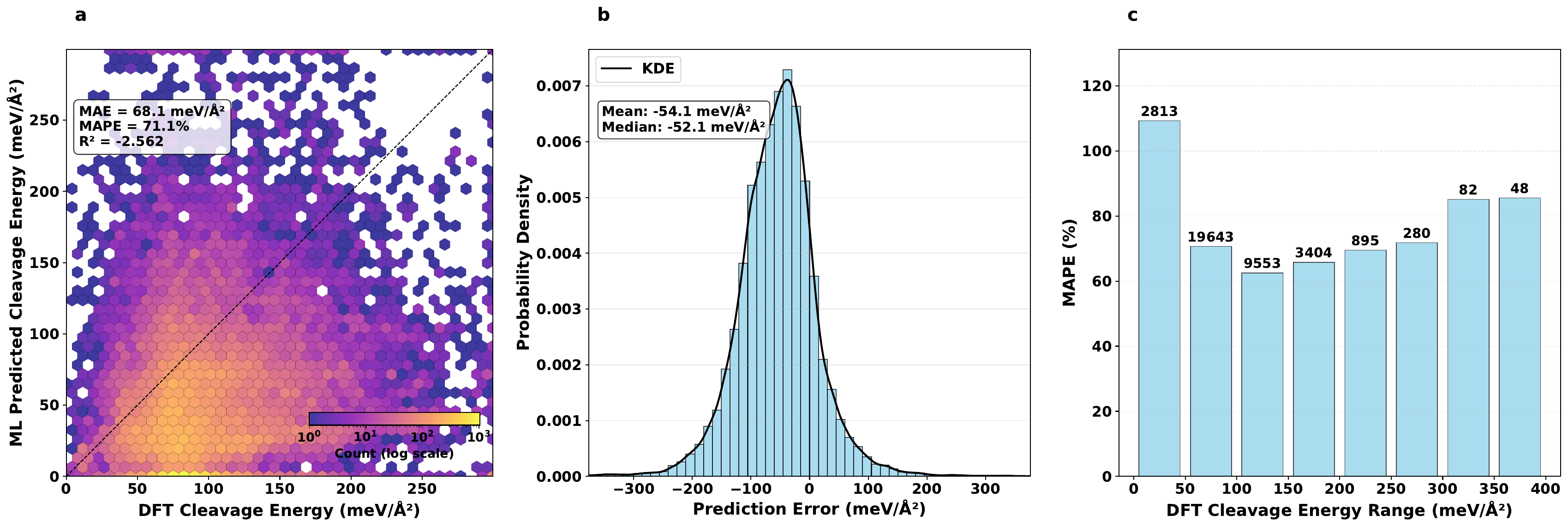}
  \caption{\textbf{Detailed performance analysis of Eqf.V2--OC20 for cleavage energy prediction.}
  \textbf{(a)} Hexagonal density parity plot comparing MLIP-predicted versus DFT-calculated cleavage energies. Note that the color bar is logarithmic.
  \textbf{(b)} Probability density distribution of prediction errors (MLIP $-$ DFT), showing a kernel density estimate (KDE) curve calculated using a Gaussian kernel with Scott's rule bandwidth selection. The mean and median error values are displayed in the text box.
  \textbf{(c)} MAPE as a function of DFT cleavage energy bins. Numbers above bars indicate the sample count in each bin.}
  \label{fig:eqfv2_oc20_analysis}
\end{figure*}

\begin{figure*}[t]
  \centering
  \includegraphics[width=\linewidth]{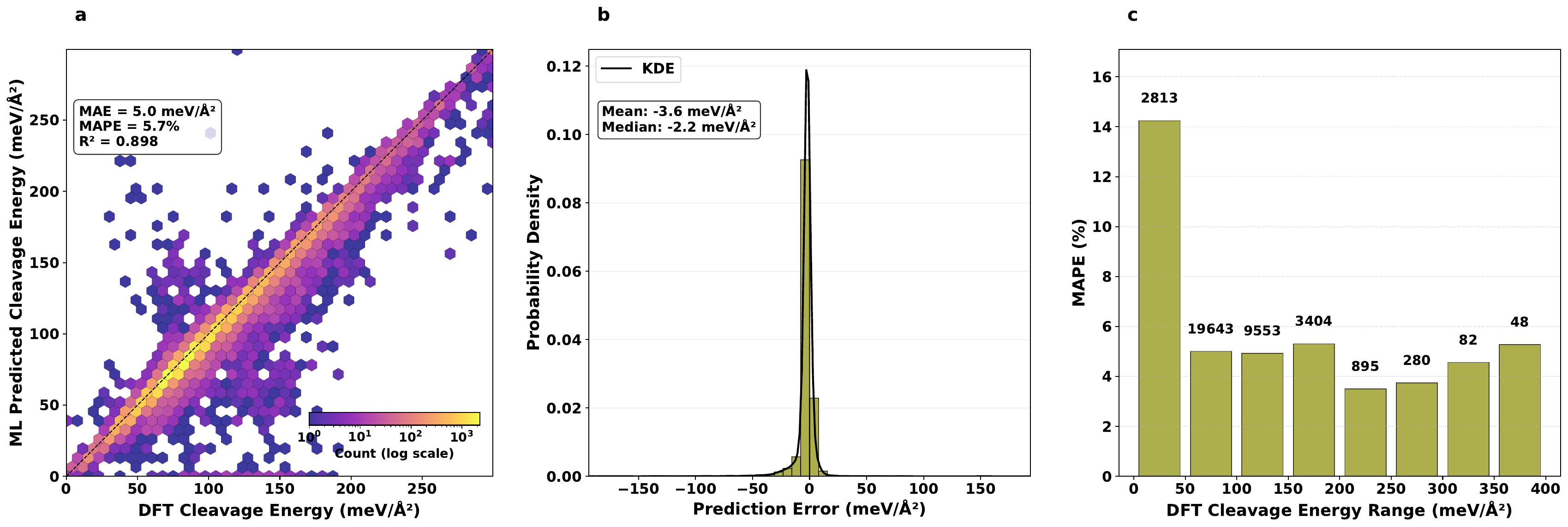}
  \caption{\textbf{Detailed performance analysis of eSEN--30M--OMAT24 for cleavage energy prediction.}
  \textbf{(a)} Hexagonal density parity plot comparing MLIP-predicted versus DFT-calculated cleavage energies. Note that the color bar is logarithmic.
  \textbf{(b)} Probability density distribution of prediction errors (MLIP $-$ DFT), showing a kernel density estimate (KDE) curve calculated using a Gaussian kernel with Scott's rule bandwidth selection. The mean and median error values are displayed in the text box.
  \textbf{(c)} MAPE as a function of DFT cleavage energy bins. Numbers above bars indicate the sample count in each bin.}
  \label{fig:esen_30m_omat24_analysis}
\end{figure*}

\begin{figure*}[t]
  \centering
  \includegraphics[width=\linewidth]{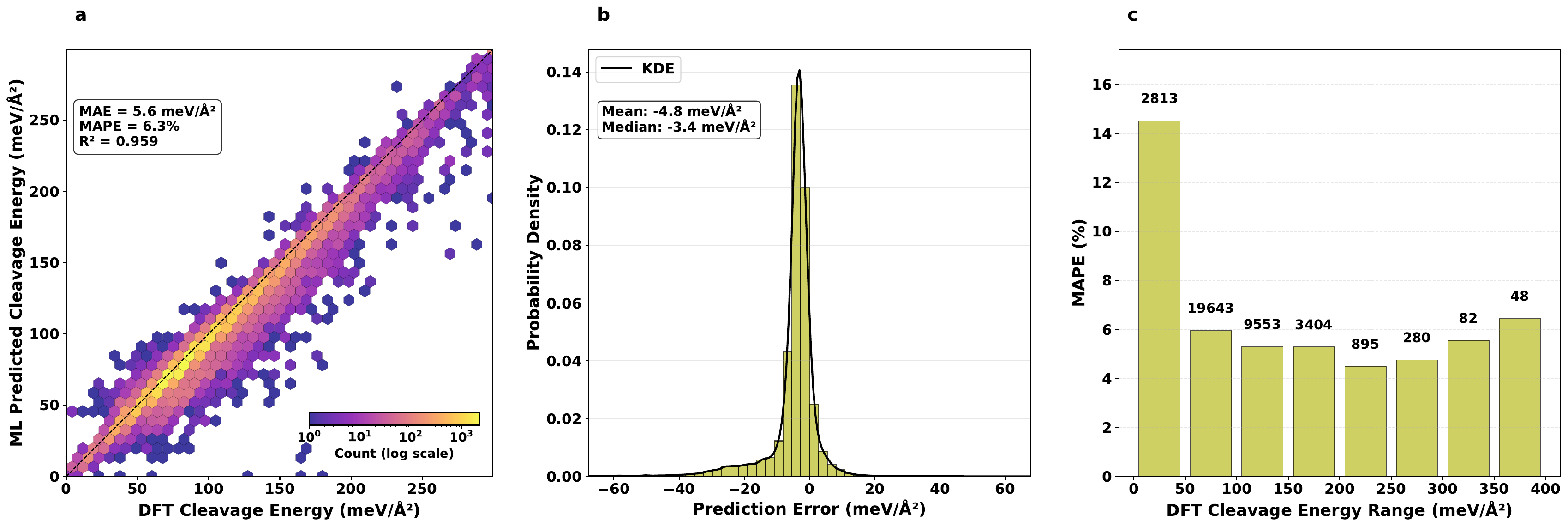}
  \caption{\textbf{Detailed performance analysis of eSEN--30M--OAM for cleavage energy prediction.}
  \textbf{(a)} Hexagonal density parity plot comparing MLIP-predicted versus DFT-calculated cleavage energies. Note that the color bar is logarithmic.
  \textbf{(b)} Probability density distribution of prediction errors (MLIP $-$ DFT), showing a kernel density estimate (KDE) curve calculated using a Gaussian kernel with Scott's rule bandwidth selection. The mean and median error values are displayed in the text box.
  \textbf{(c)} MAPE as a function of DFT cleavage energy bins. Numbers above bars indicate the sample count in each bin.}
  \label{fig:esen_30m_oam_analysis}
\end{figure*}

\begin{figure*}[t]
  \centering
  \includegraphics[width=\linewidth]{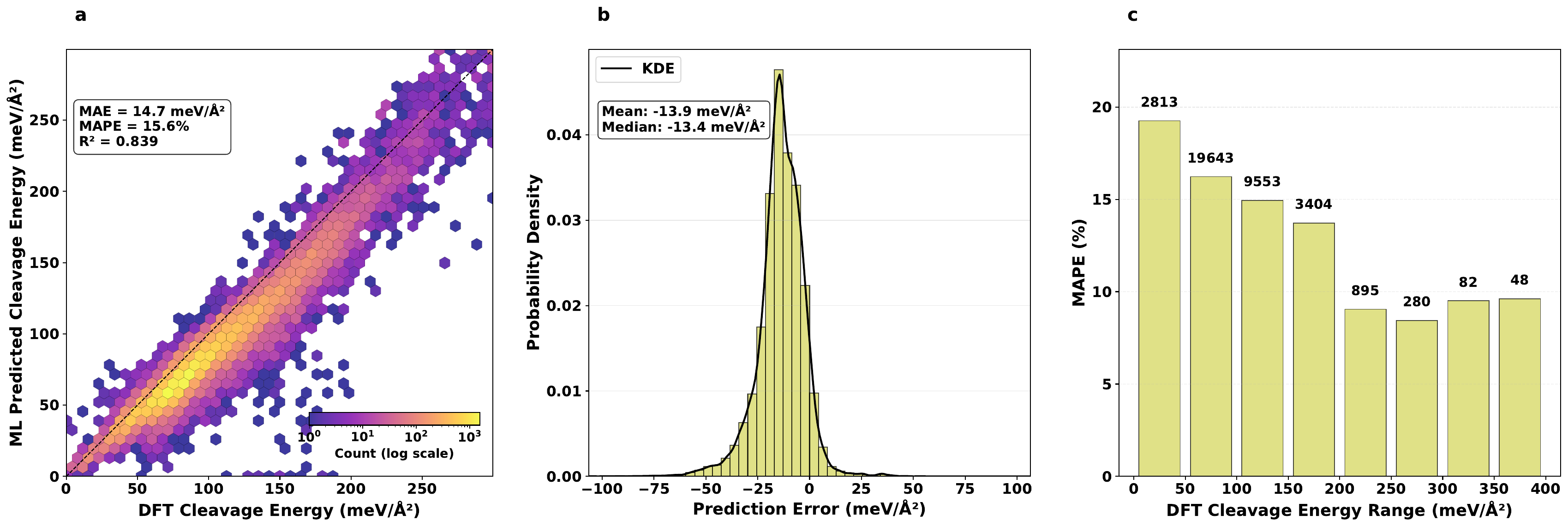}
  \caption{\textbf{Detailed performance analysis of eSEN--30M--MP for cleavage energy prediction.}
  \textbf{(a)} Hexagonal density parity plot comparing MLIP-predicted versus DFT-calculated cleavage energies. Note that the color bar is logarithmic.
  \textbf{(b)} Probability density distribution of prediction errors (MLIP $-$ DFT), showing a kernel density estimate (KDE) curve calculated using a Gaussian kernel with Scott's rule bandwidth selection. The mean and median error values are displayed in the text box.
  \textbf{(c)} MAPE as a function of DFT cleavage energy bins. Numbers above bars indicate the sample count in each bin.}
  \label{fig:esen_30m_mp_analysis}
\end{figure*}

\begin{figure*}[t]
  \centering
  \includegraphics[width=\linewidth]{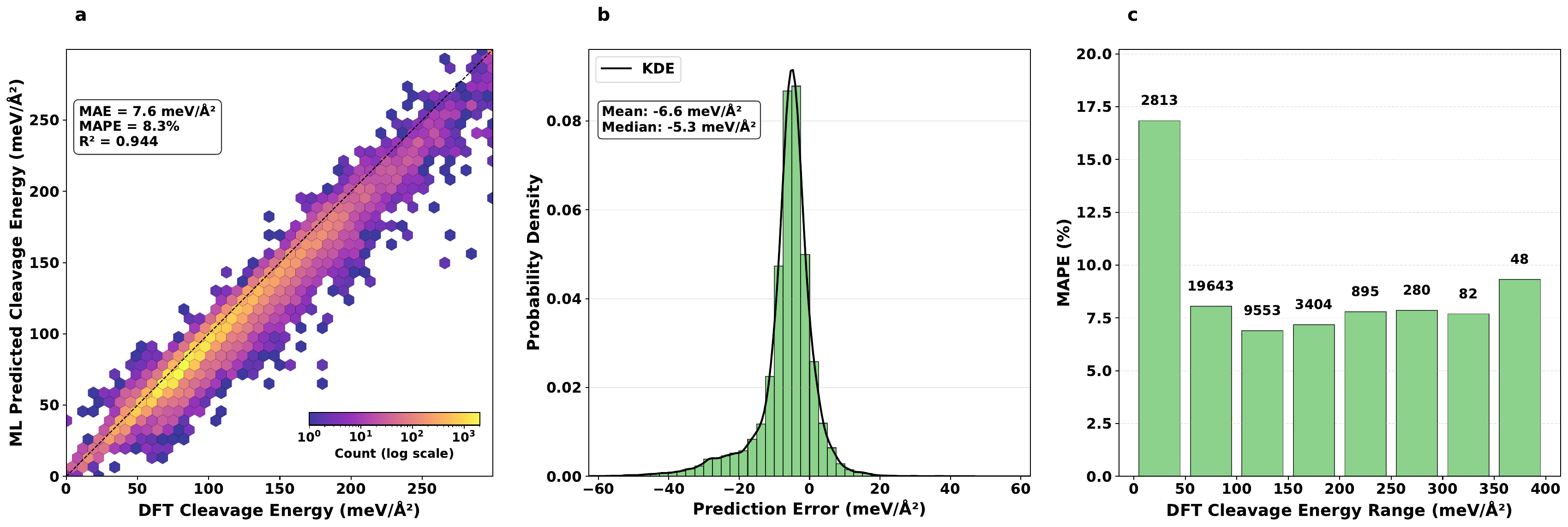}
  \caption{\textbf{Detailed performance analysis of GRACE--2L--OAM for cleavage energy prediction.}
  \textbf{(a)} Hexagonal density parity plot comparing MLIP-predicted versus DFT-calculated cleavage energies. Note that the color bar is logarithmic.
  \textbf{(b)} Probability density distribution of prediction errors (MLIP $-$ DFT), showing a kernel density estimate (KDE) curve calculated using a Gaussian kernel with Scott's rule bandwidth selection. The mean and median error values are displayed in the text box.
  \textbf{(c)} MAPE as a function of DFT cleavage energy bins. Numbers above bars indicate the sample count in each bin.}
  \label{fig:grace_2l_oam_analysis}
\end{figure*}

\begin{figure*}[t]
  \centering
  \includegraphics[width=\linewidth]{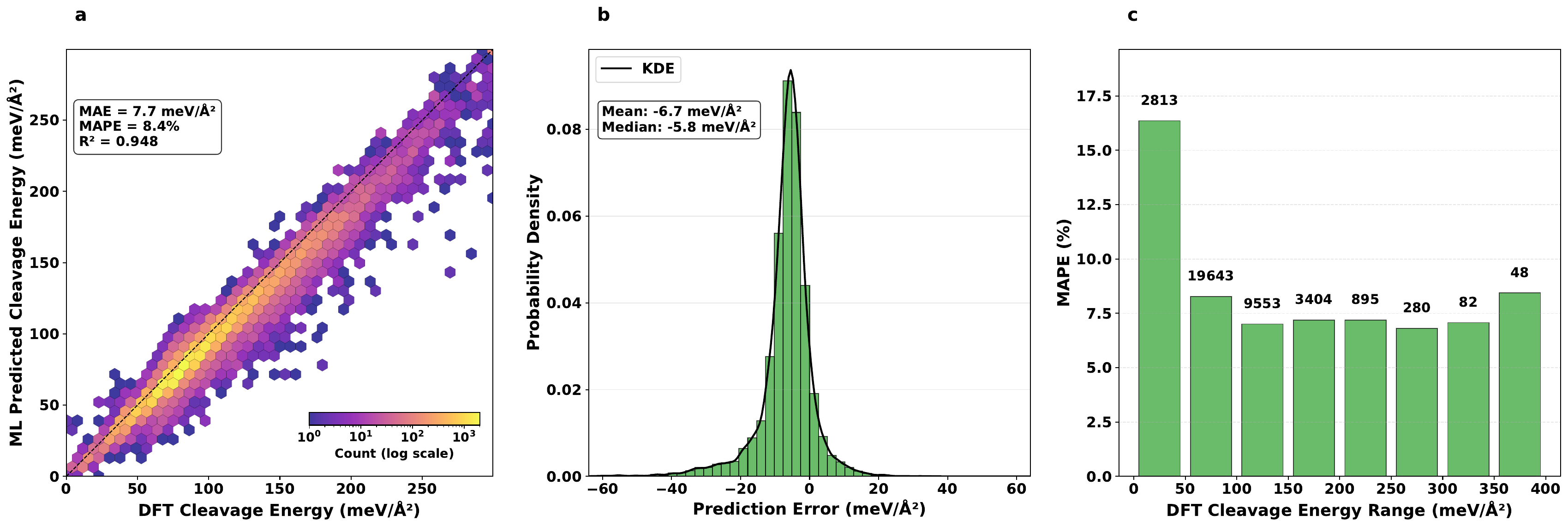}
  \caption{\textbf{Detailed performance analysis of GRACE--2L--OMAT for cleavage energy prediction.}
  \textbf{(a)} Hexagonal density parity plot comparing MLIP-predicted versus DFT-calculated cleavage energies. Note that the color bar is logarithmic.
  \textbf{(b)} Probability density distribution of prediction errors (MLIP $-$ DFT), showing a kernel density estimate (KDE) curve calculated using a Gaussian kernel with Scott's rule bandwidth selection. The mean and median error values are displayed in the text box.
  \textbf{(c)} MAPE as a function of DFT cleavage energy bins. Numbers above bars indicate the sample count in each bin.}
  \label{fig:grace_2l_omat_analysis}
\end{figure*}

\begin{figure*}[t]
  \centering
  \includegraphics[width=\linewidth]{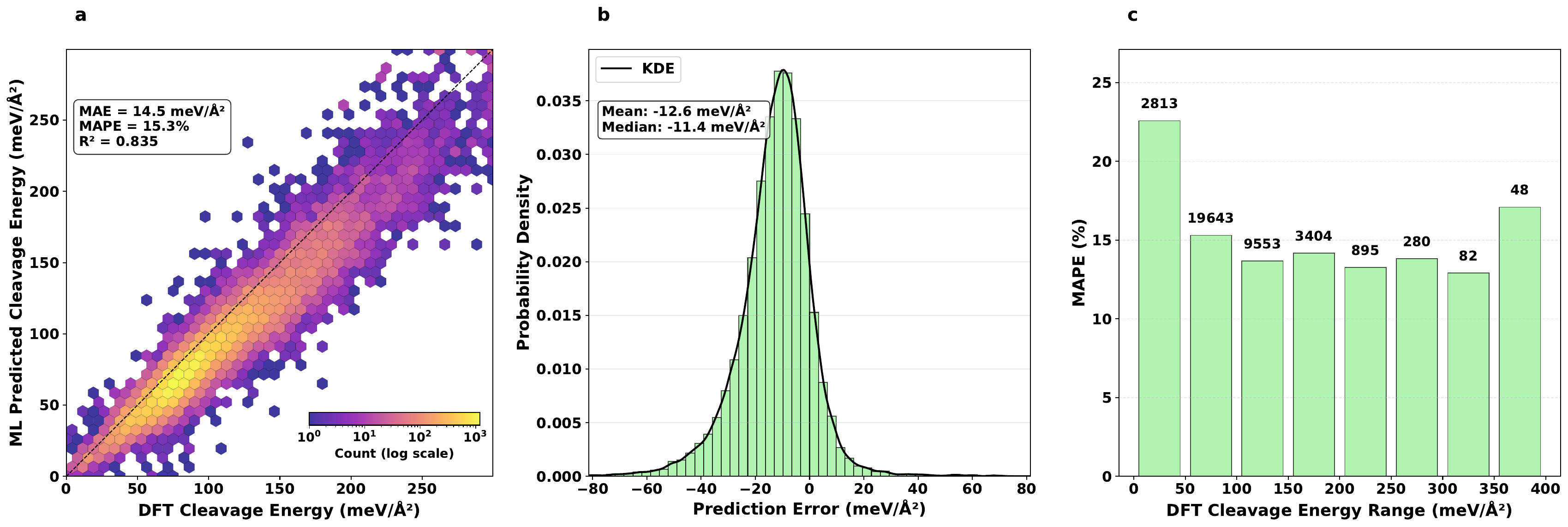}
  \caption{\textbf{Detailed performance analysis of GRACE--2L--MP for cleavage energy prediction.}
  \textbf{(a)} Hexagonal density parity plot comparing MLIP-predicted versus DFT-calculated cleavage energies. Note that the color bar is logarithmic.
  \textbf{(b)} Probability density distribution of prediction errors (MLIP $-$ DFT), showing a kernel density estimate (KDE) curve calculated using a Gaussian kernel with Scott's rule bandwidth selection. The mean and median error values are displayed in the text box.
  \textbf{(c)} MAPE as a function of DFT cleavage energy bins. Numbers above bars indicate the sample count in each bin.}
  \label{fig:grace_2l_mp_analysis}
\end{figure*}

\begin{figure*}[t]
  \centering
  \includegraphics[width=\linewidth]{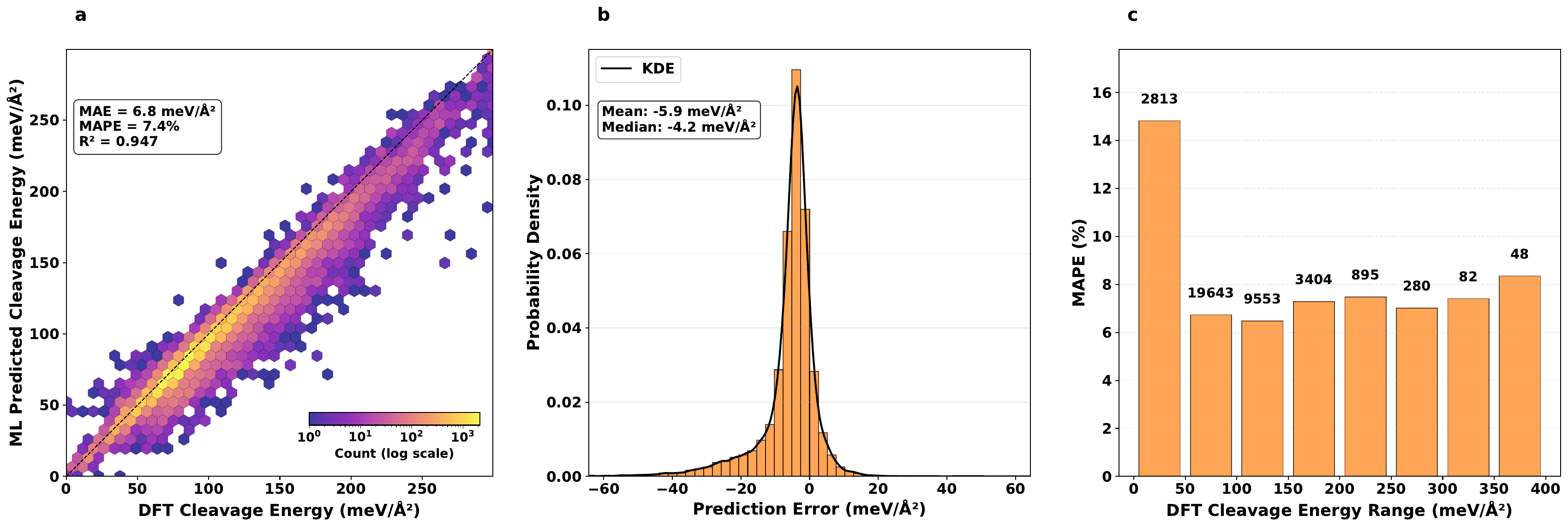}
  \caption{\textbf{Detailed performance analysis of ORB--v3--MPA for cleavage energy prediction.}
  \textbf{(a)} Hexagonal density parity plot comparing MLIP-predicted versus DFT-calculated cleavage energies. Note that the color bar is logarithmic.
  \textbf{(b)} Probability density distribution of prediction errors (MLIP $-$ DFT), showing a kernel density estimate (KDE) curve calculated using a Gaussian kernel with Scott's rule bandwidth selection. The mean and median error values are displayed in the text box.
  \textbf{(c)} MAPE as a function of DFT cleavage energy bins. Numbers above bars indicate the sample count in each bin.}
  \label{fig:orb_v3_mpa_analysis}
\end{figure*}

\begin{figure*}[t]
  \centering
  \includegraphics[width=\linewidth]{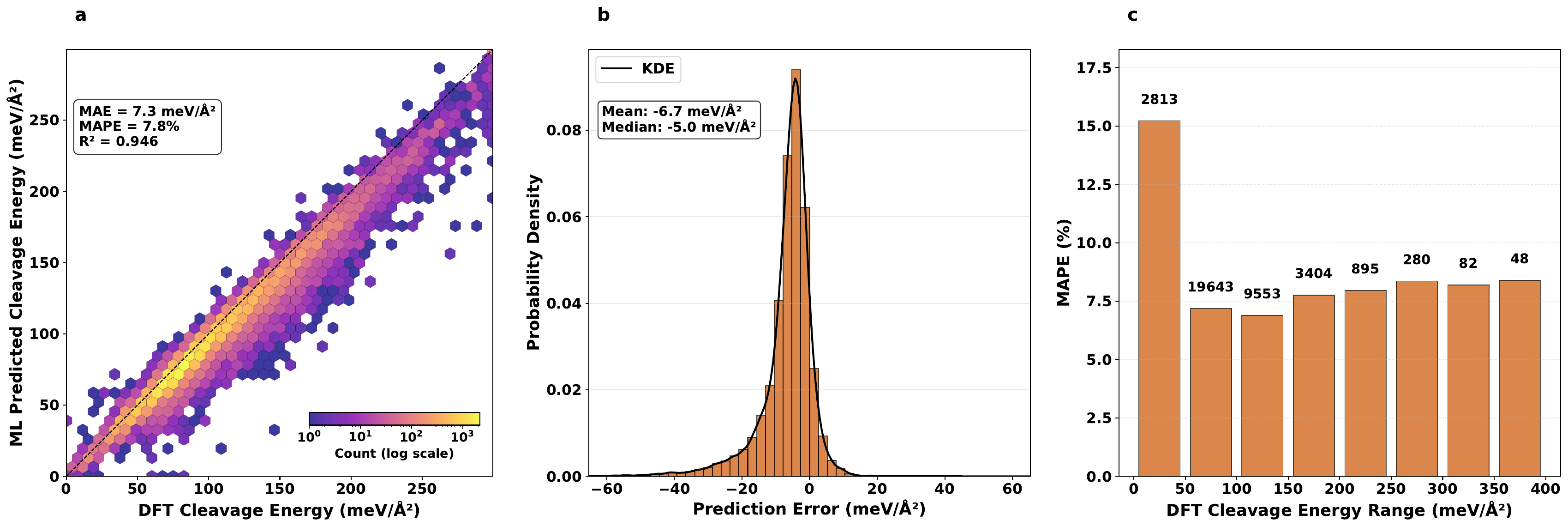}
  \caption{\textbf{Detailed performance analysis of ORB--v3--OMAT for cleavage energy prediction.}
  \textbf{(a)} Hexagonal density parity plot comparing MLIP-predicted versus DFT-calculated cleavage energies. Note that the color bar is logarithmic.
  \textbf{(b)} Probability density distribution of prediction errors (MLIP $-$ DFT), showing a kernel density estimate (KDE) curve calculated using a Gaussian kernel with Scott's rule bandwidth selection. The mean and median error values are displayed in the text box.
  \textbf{(c)} MAPE as a function of DFT cleavage energy bins. Numbers above bars indicate the sample count in each bin.}
  \label{fig:orb_v3_omat_analysis}
\end{figure*}

\begin{figure*}[t]
  \centering
  \includegraphics[width=\linewidth]{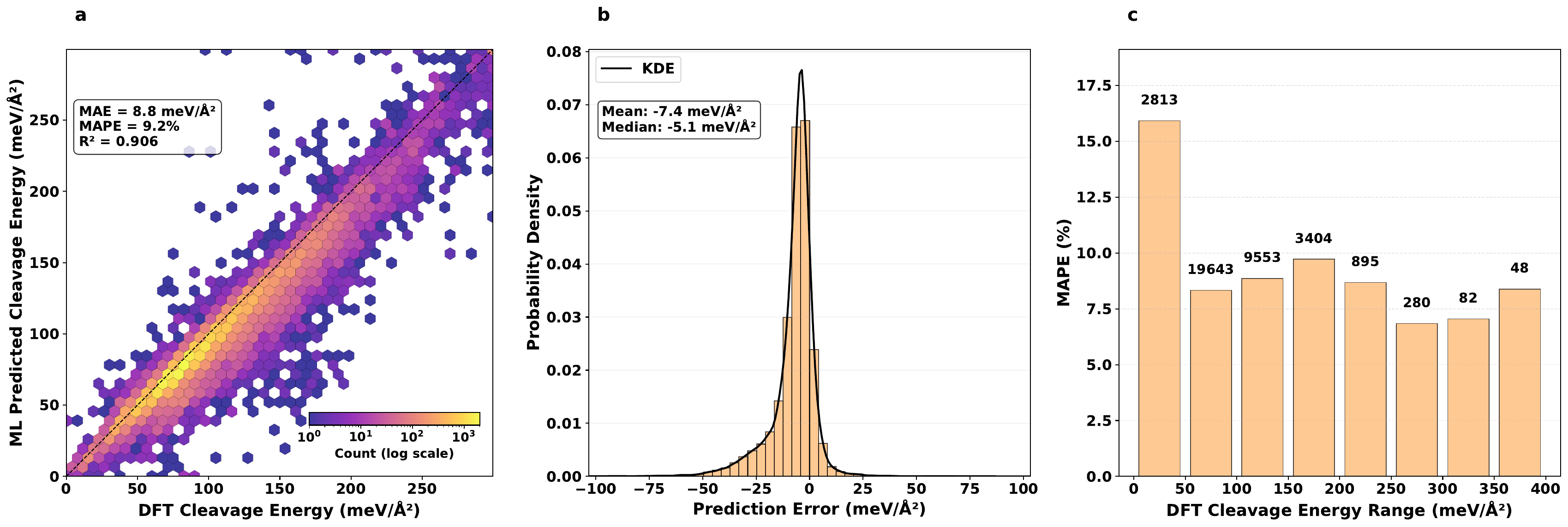}
  \caption{\textbf{Detailed performance analysis of ORB--v2 for cleavage energy prediction.}
  \textbf{(a)} Hexagonal density parity plot comparing MLIP-predicted versus DFT-calculated cleavage energies. Note that the color bar is logarithmic.
  \textbf{(b)} Probability density distribution of prediction errors (MLIP $-$ DFT), showing a kernel density estimate (KDE) curve calculated using a Gaussian kernel with Scott's rule bandwidth selection. The mean and median error values are displayed in the text box.
  \textbf{(c)} MAPE as a function of DFT cleavage energy bins. Numbers above bars indicate the sample count in each bin.}
  \label{fig:orb_v2_analysis}
\end{figure*}

\begin{figure*}[t]
  \centering
  \includegraphics[width=\linewidth]{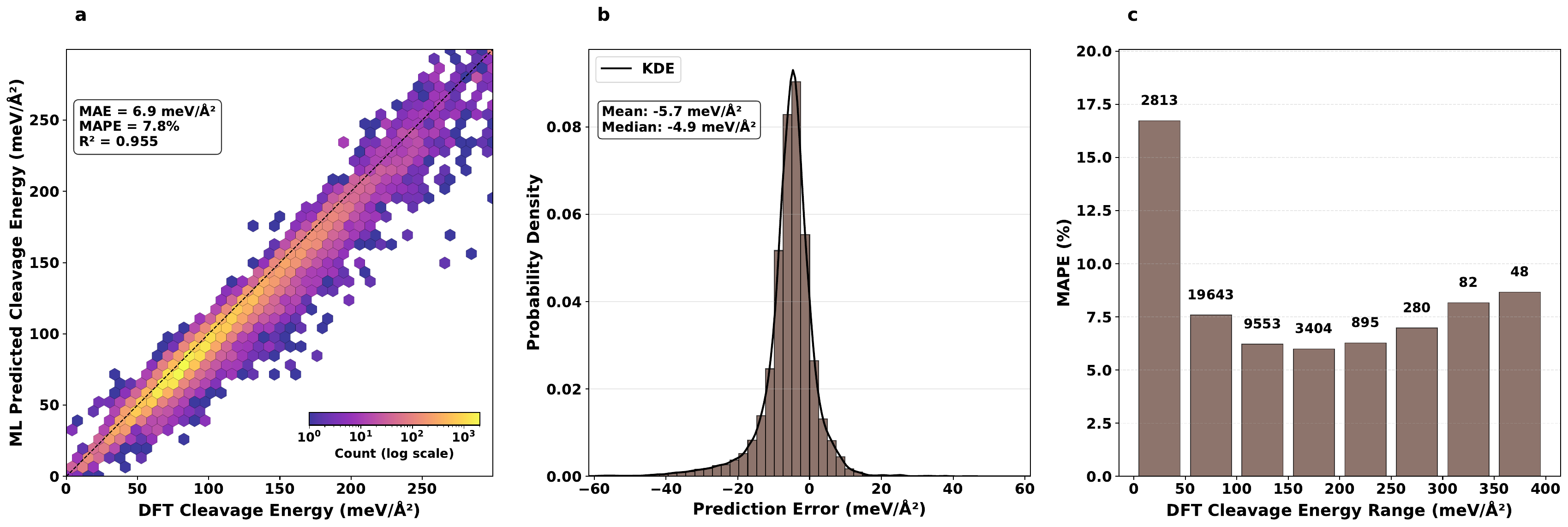}
  \caption{\textbf{Detailed performance analysis of MACE--OMAT for cleavage energy prediction.}
  \textbf{(a)} Hexagonal density parity plot comparing MLIP-predicted versus DFT-calculated cleavage energies. Note that the color bar is logarithmic.
  \textbf{(b)} Probability density distribution of prediction errors (MLIP $-$ DFT), showing a kernel density estimate (KDE) curve calculated using a Gaussian kernel with Scott's rule bandwidth selection. The mean and median error values are displayed in the text box.
  \textbf{(c)} MAPE as a function of DFT cleavage energy bins. Numbers above bars indicate the sample count in each bin.}
  \label{fig:mace_omat_analysis}
\end{figure*}

\begin{figure*}[t]
  \centering
  \includegraphics[width=\linewidth]{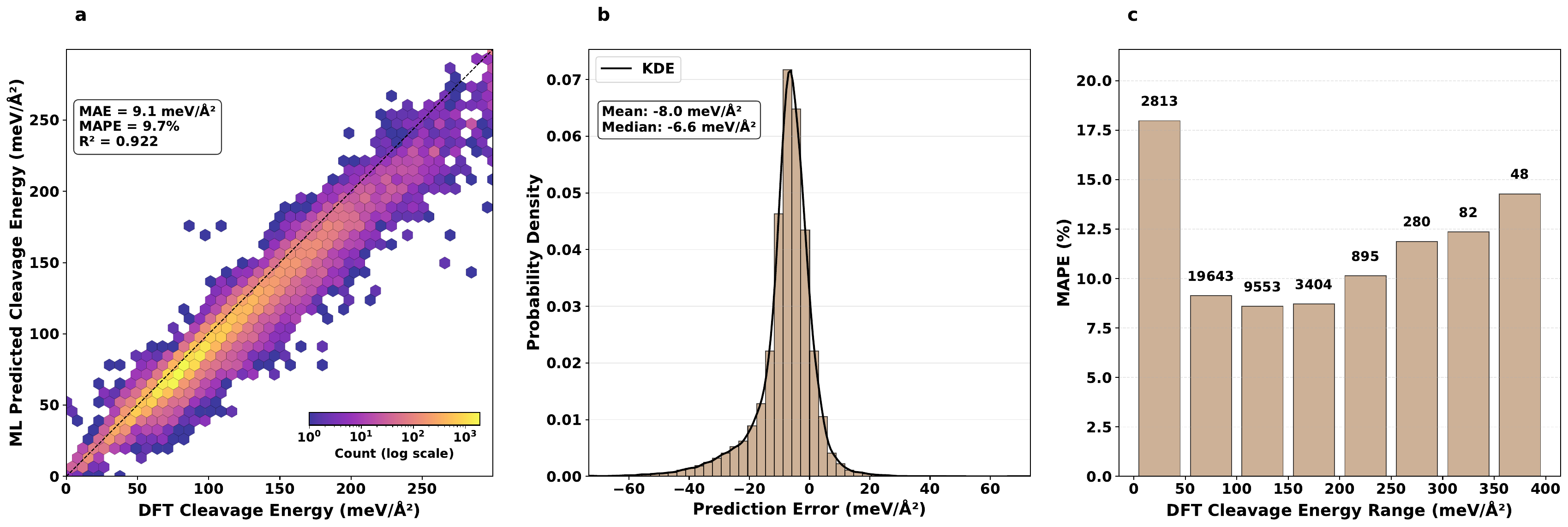}
  \caption{\textbf{Detailed performance analysis of MACE--MPA for cleavage energy prediction.}
  \textbf{(a)} Hexagonal density parity plot comparing MLIP-predicted versus DFT-calculated cleavage energies. Note that the color bar is logarithmic.
  \textbf{(b)} Probability density distribution of prediction errors (MLIP $-$ DFT), showing a kernel density estimate (KDE) curve calculated using a Gaussian kernel with Scott's rule bandwidth selection. The mean and median error values are displayed in the text box.
  \textbf{(c)} MAPE as a function of DFT cleavage energy bins. Numbers above bars indicate the sample count in each bin.}
  \label{fig:mace_mpa_analysis}
\end{figure*}

\begin{figure*}[t]
  \centering
  \includegraphics[width=\linewidth]{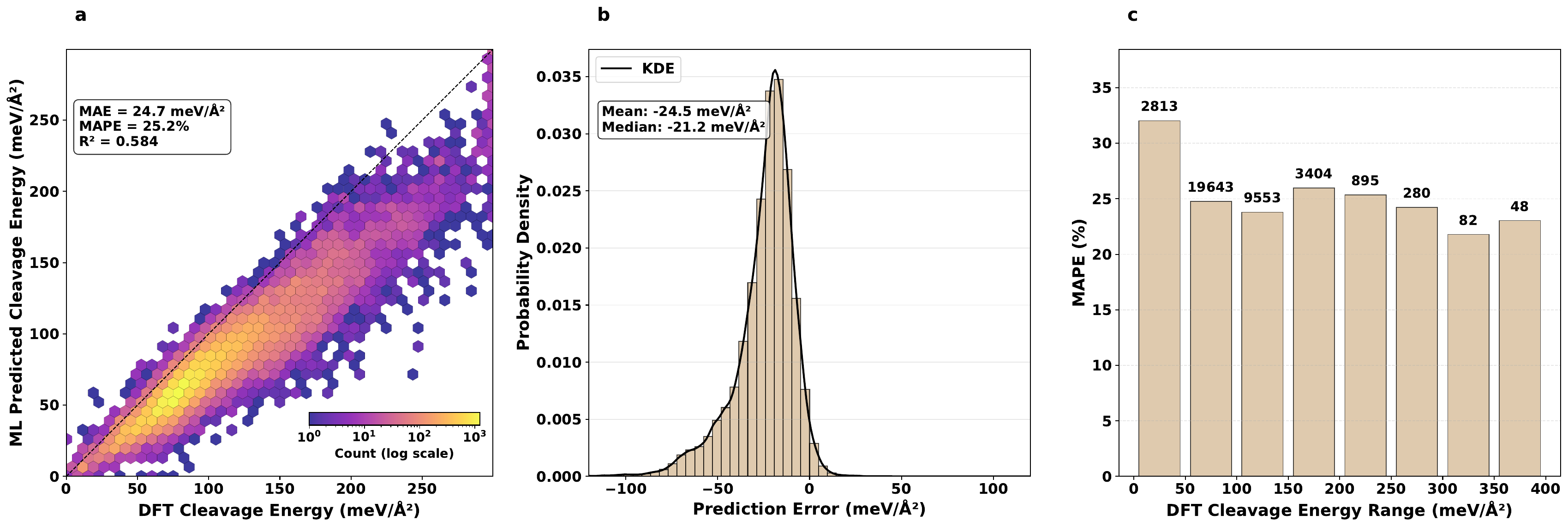}
  \caption{\textbf{Detailed performance analysis of MACE--MP--0b3 for cleavage energy prediction.}
  \textbf{(a)} Hexagonal density parity plot comparing MLIP-predicted versus DFT-calculated cleavage energies. Note that the color bar is logarithmic.
  \textbf{(b)} Probability density distribution of prediction errors (MLIP $-$ DFT), showing a kernel density estimate (KDE) curve calculated using a Gaussian kernel with Scott's rule bandwidth selection. The mean and median error values are displayed in the text box.
  \textbf{(c)} MAPE as a function of DFT cleavage energy bins. Numbers above bars indicate the sample count in each bin.}
  \label{fig:mace_mp_0b3_analysis}
\end{figure*}

\begin{figure*}[t]
  \centering
  \includegraphics[width=\linewidth]{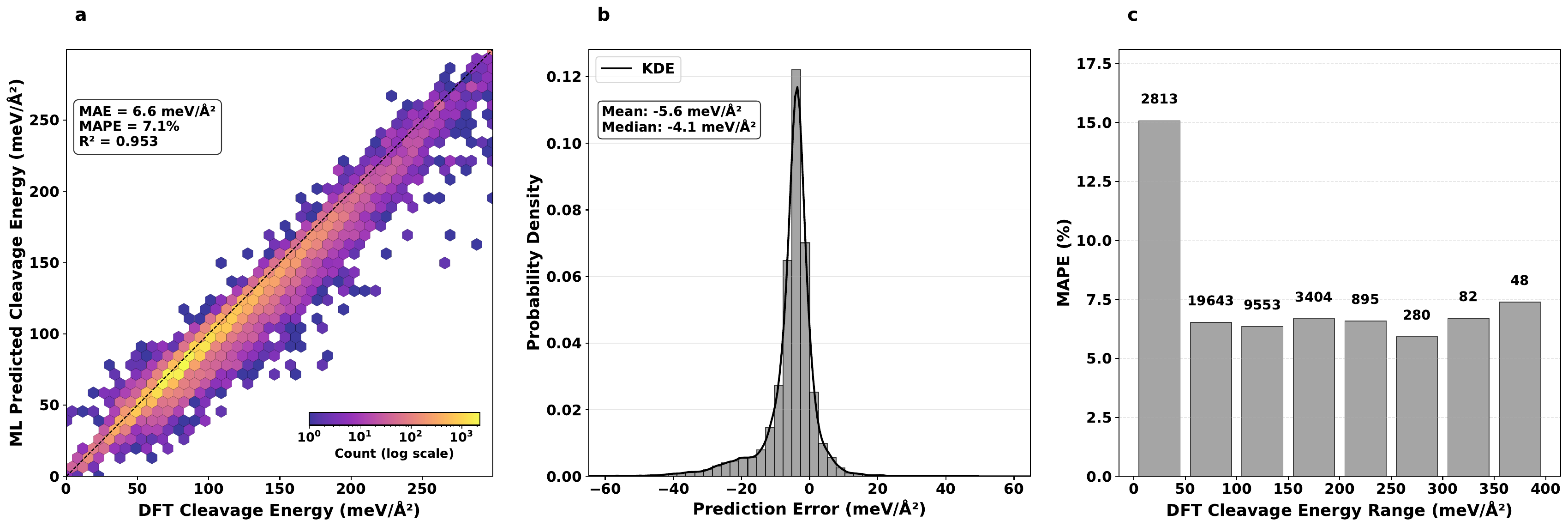}
  \caption{\textbf{Detailed performance analysis of SevenNet--MF--OMPA for cleavage energy prediction.}
  \textbf{(a)} Hexagonal density parity plot comparing MLIP-predicted versus DFT-calculated cleavage energies. Note that the color bar is logarithmic.
  \textbf{(b)} Probability density distribution of prediction errors (MLIP $-$ DFT), showing a kernel density estimate (KDE) curve calculated using a Gaussian kernel with Scott's rule bandwidth selection. The mean and median error values are displayed in the text box.
  \textbf{(c)} MAPE as a function of DFT cleavage energy bins. Numbers above bars indicate the sample count in each bin.}
  \label{fig:sevennet_mf_ompa_analysis}
\end{figure*}

\begin{figure*}[t]
  \centering
  \includegraphics[width=\linewidth]{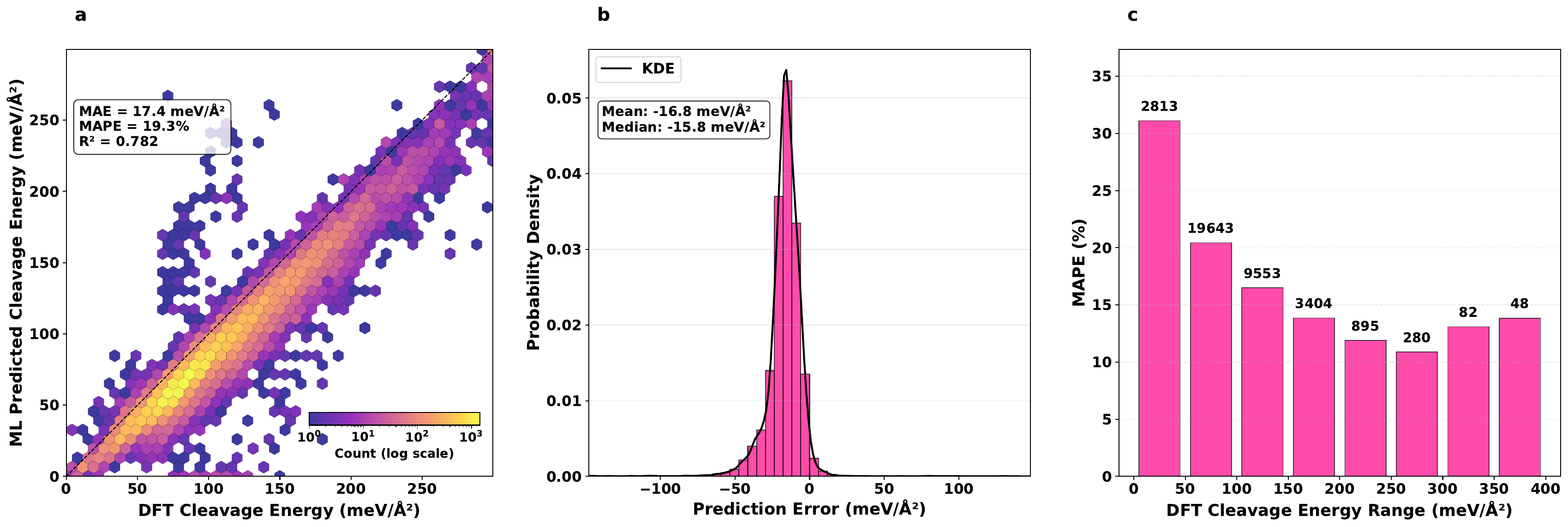}
  \caption{\textbf{Detailed performance analysis of MatterSim--v1 for cleavage energy prediction.}
  \textbf{(a)} Hexagonal density parity plot comparing MLIP-predicted versus DFT-calculated cleavage energies. Note that the color bar is logarithmic.
  \textbf{(b)} Probability density distribution of prediction errors (MLIP $-$ DFT), showing a kernel density estimate (KDE) curve calculated using a Gaussian kernel with Scott's rule bandwidth selection. The mean and median error values are displayed in the text box.
  \textbf{(c)} MAPE as a function of DFT cleavage energy bins. Numbers above bars indicate the sample count in each bin.}
  \label{fig:mattersim_v1_analysis}
\end{figure*}

\begin{figure*}[t]
  \centering
  \includegraphics[width=\linewidth]{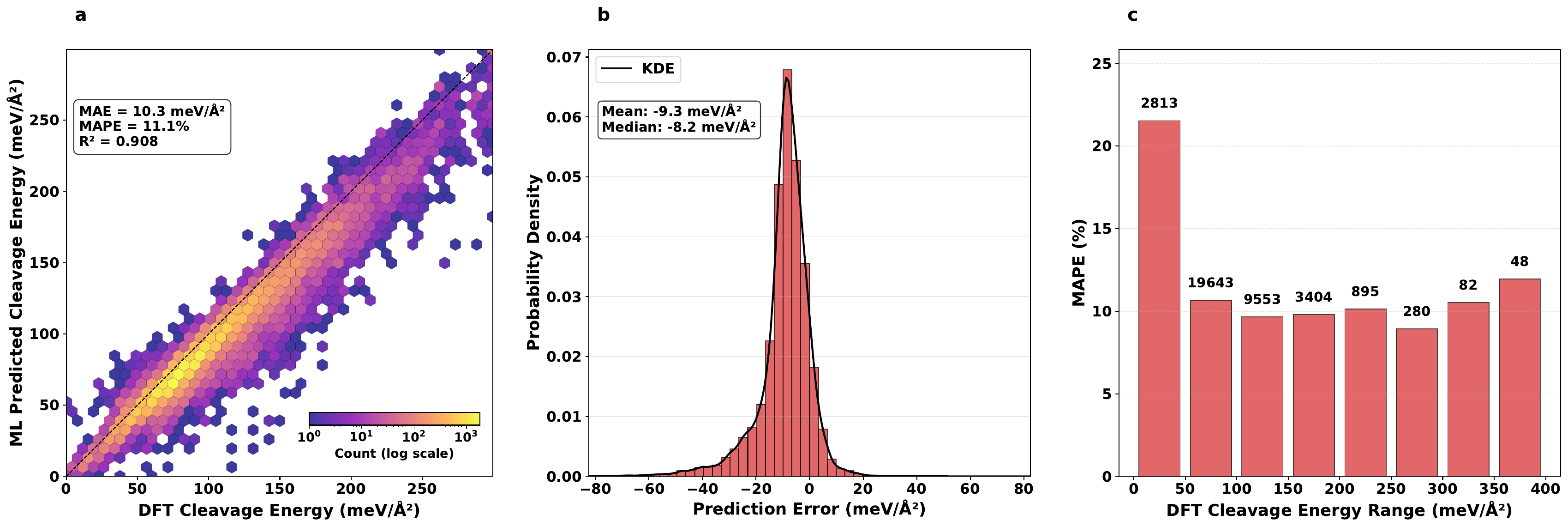}
  \caption{\textbf{Detailed performance analysis of AlphaNet--OMA for cleavage energy prediction.}
  \textbf{(a)} Hexagonal density parity plot comparing MLIP-predicted versus DFT-calculated cleavage energies. Note that the color bar is logarithmic.
  \textbf{(b)} Probability density distribution of prediction errors (MLIP $-$ DFT), showing a kernel density estimate (KDE) curve calculated using a Gaussian kernel with Scott's rule bandwidth selection. The mean and median error values are displayed in the text box.
  \textbf{(c)} MAPE as a function of DFT cleavage energy bins. Numbers above bars indicate the sample count in each bin.}
  \label{fig:alphanet_oma_analysis}
\end{figure*}


\begin{figure}[ht]
\centering
\includegraphics[width=\columnwidth]{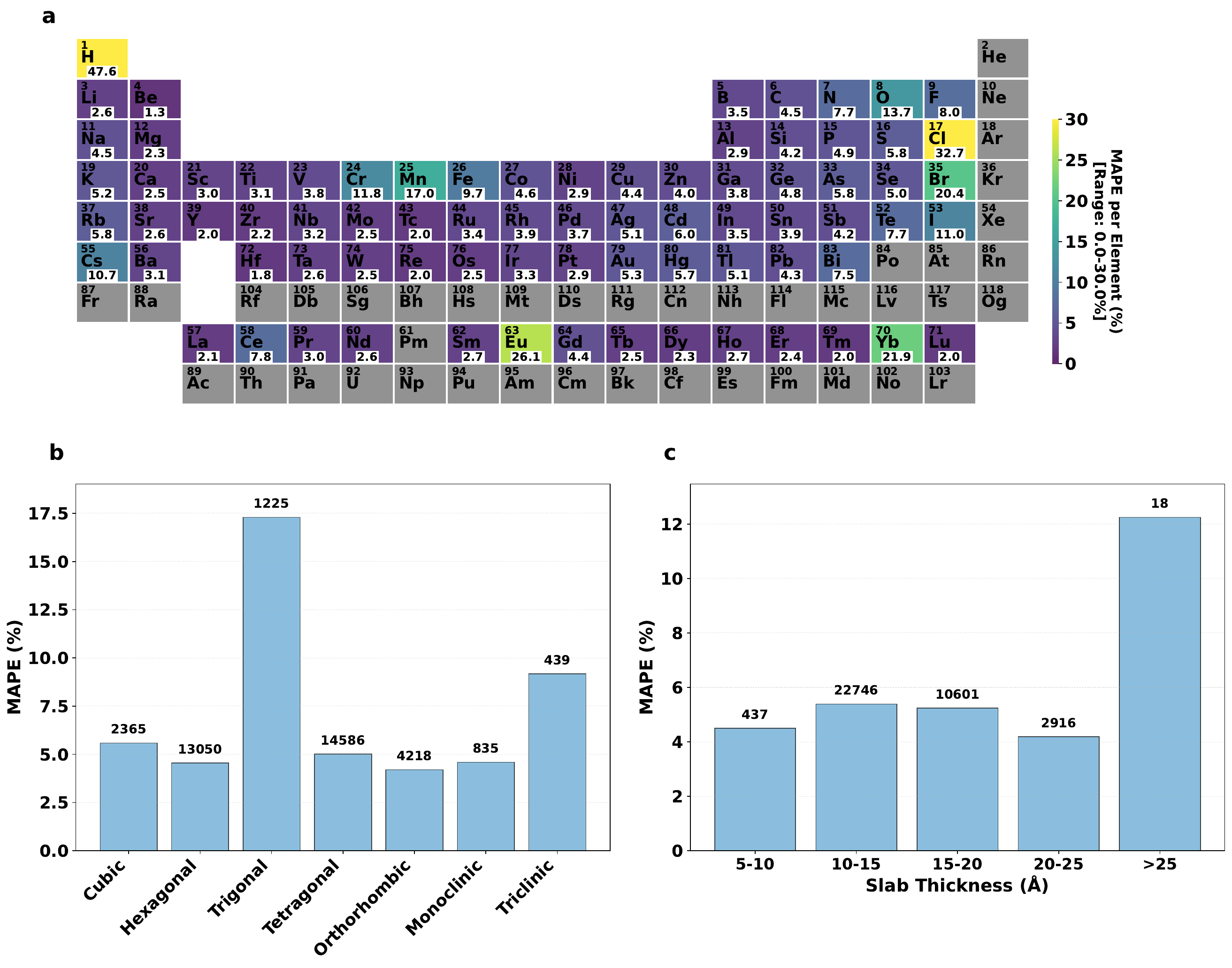}
\caption{\textbf{Decomposition of Eqf.V2--OMAT24--MP--sAlex prediction errors by chemical, structural, and geometric features.} 
\textbf{(a)} Periodic table heat map showing MAPE for surfaces containing each element, with viridis colormap normalized to 0--30\% range. 
\textbf{(b)} MAPE distribution across seven crystal systems ordered by symmetry, with sample counts indicated above bars. 
\textbf{(c)} MAPE dependency on slab thickness, binned into 5~Å intervals showing consistent accuracy across different slab thicknesses with sample counts per bin.}
\label{fig:eqfv2_omat24_mp_salex_error_decomposition}
\end{figure}

\begin{figure}[ht]
\centering
\includegraphics[width=\columnwidth]{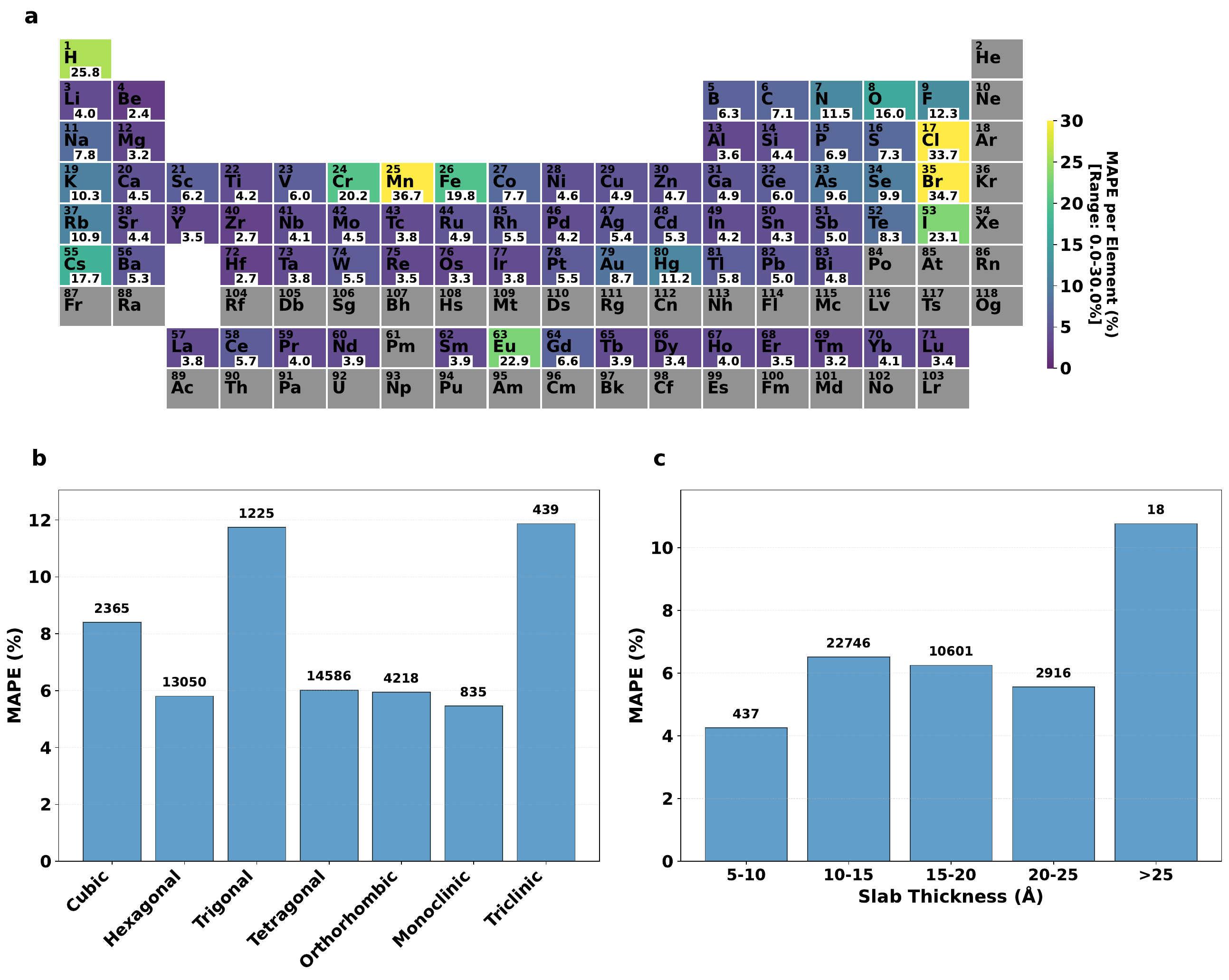}
\caption{\textbf{Decomposition of Eqf.V2--OMAT24 prediction errors by chemical, structural, and geometric features.} 
\textbf{(a)} Periodic table heat map showing MAPE for surfaces containing each element, with viridis colormap normalized to 0--30\% range. 
\textbf{(b)} MAPE distribution across seven crystal systems ordered by symmetry, with sample counts indicated above bars. 
\textbf{(c)} MAPE dependency on slab thickness, binned into 5~Å intervals showing consistent accuracy across different slab thicknesses with sample counts per bin.}
\label{fig:eqfv2_omat24_error_decomposition}
\end{figure}

\begin{figure}[ht]
\centering
\includegraphics[width=\columnwidth]{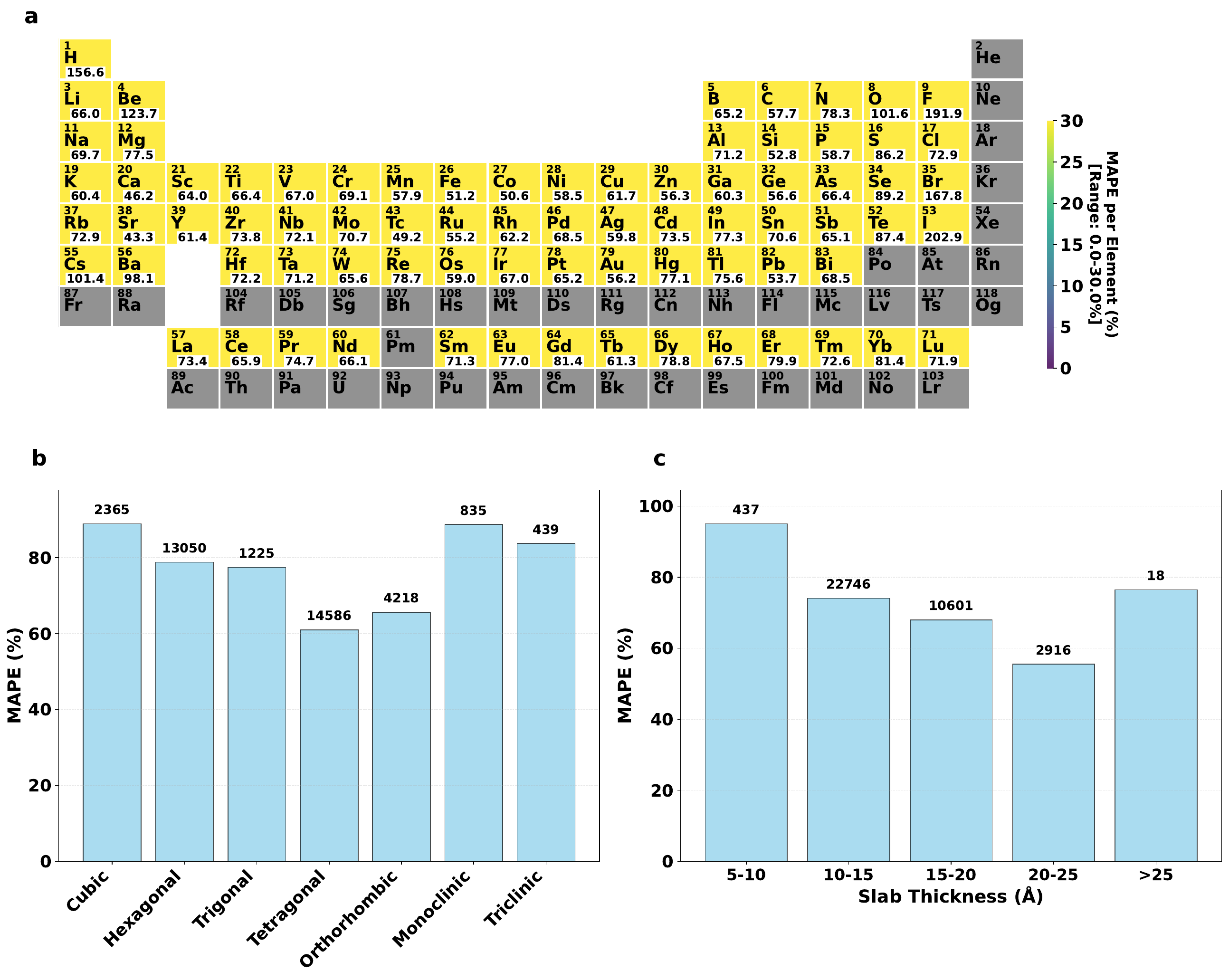}
\caption{\textbf{Decomposition of Eqf.V2--OC20 prediction errors by chemical, structural, and geometric features.} 
\textbf{(a)} Periodic table heat map showing MAPE for surfaces containing each element, with viridis colormap normalized to 0--30\% range. 
\textbf{(b)} MAPE distribution across seven crystal systems ordered by symmetry, with sample counts indicated above bars. 
\textbf{(c)} MAPE dependency on slab thickness, binned into 5~Å intervals showing consistent accuracy across different slab thicknesses with sample counts per bin.}
\label{fig:eqfv2_oc20_error_decomposition}
\end{figure}

\begin{figure}[ht]
\centering
\includegraphics[width=\columnwidth]{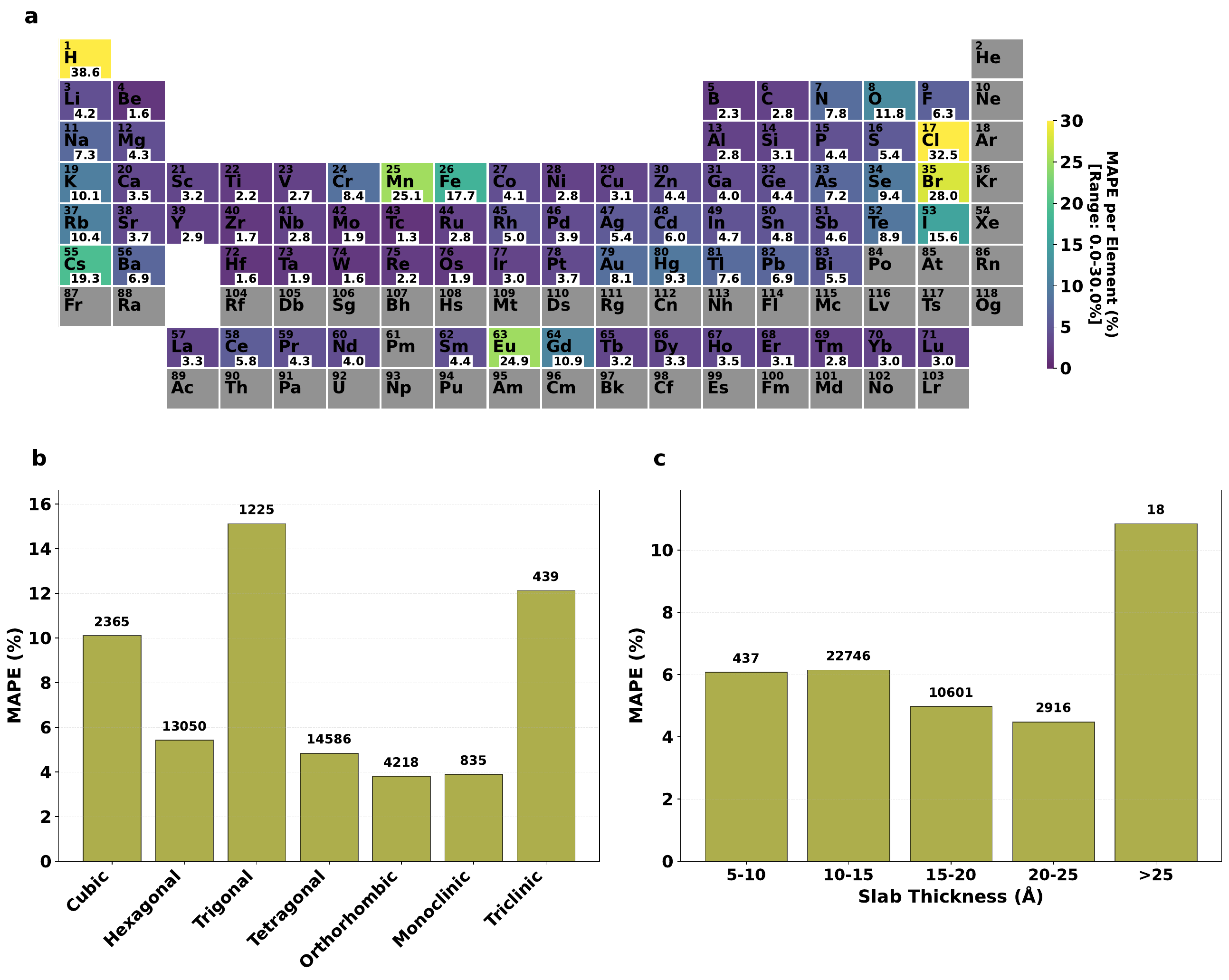}
\caption{\textbf{Decomposition of eSEN--30M--OMAT24 prediction errors by chemical, structural, and geometric features.} 
\textbf{(a)} Periodic table heat map showing MAPE for surfaces containing each element, with viridis colormap normalized to 0--30\% range. 
\textbf{(b)} MAPE distribution across seven crystal systems ordered by symmetry, with sample counts indicated above bars. 
\textbf{(c)} MAPE dependency on slab thickness, binned into 5~Å intervals showing consistent accuracy across different slab thicknesses with sample counts per bin.}
\label{fig:esen_30m_omat24_error_decomposition}
\end{figure}

\begin{figure}[ht]
\centering
\includegraphics[width=\columnwidth]{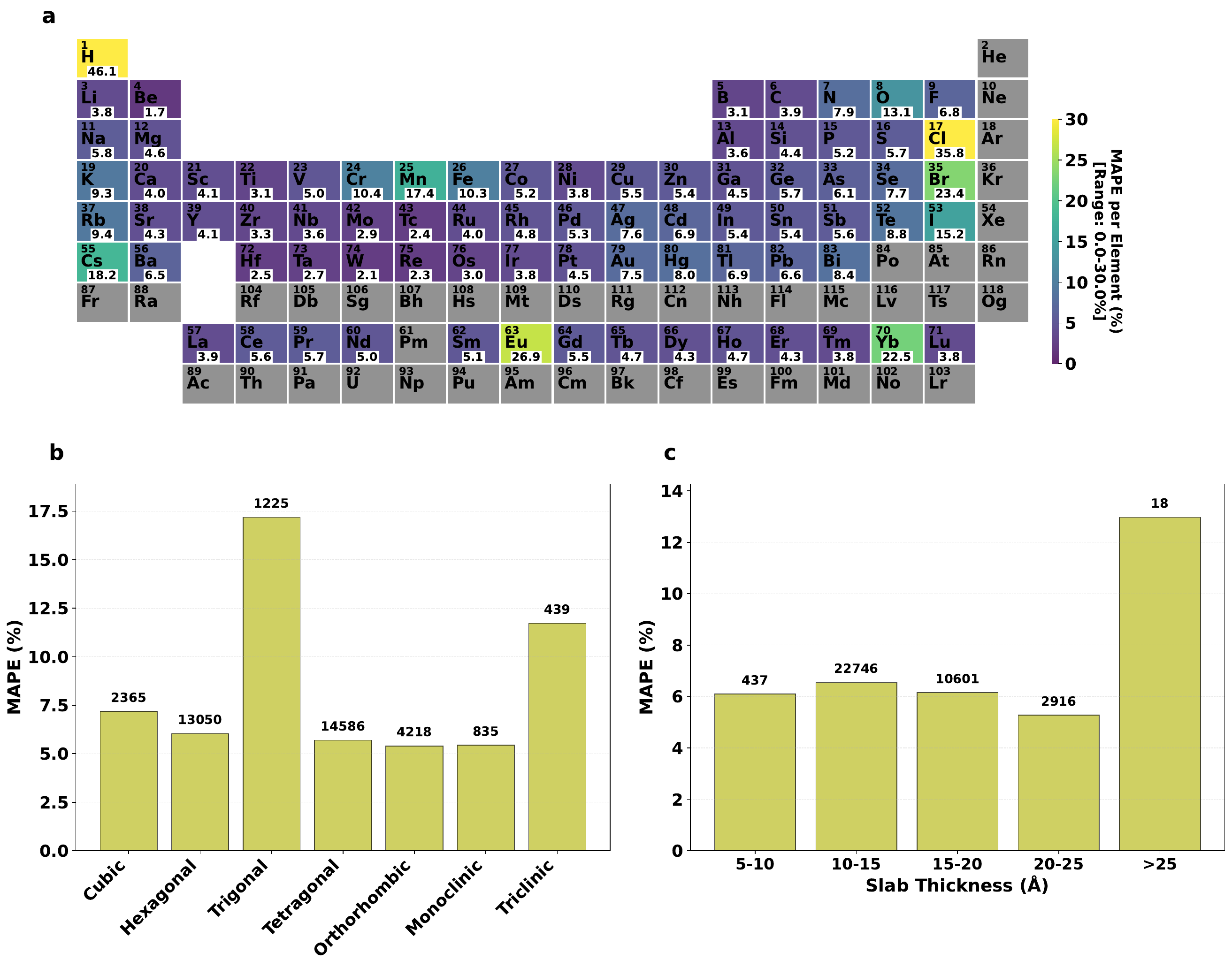}
\caption{\textbf{Decomposition of eSEN--30M--OAM prediction errors by chemical, structural, and geometric features.} 
\textbf{(a)} Periodic table heat map showing MAPE for surfaces containing each element, with viridis colormap normalized to 0--30\% range. 
\textbf{(b)} MAPE distribution across seven crystal systems ordered by symmetry, with sample counts indicated above bars. 
\textbf{(c)} MAPE dependency on slab thickness, binned into 5~Å intervals showing consistent accuracy across different slab thicknesses with sample counts per bin.}
\label{fig:esen_30m_oam_error_decomposition}
\end{figure}

\begin{figure}[ht]
\centering
\includegraphics[width=\columnwidth]{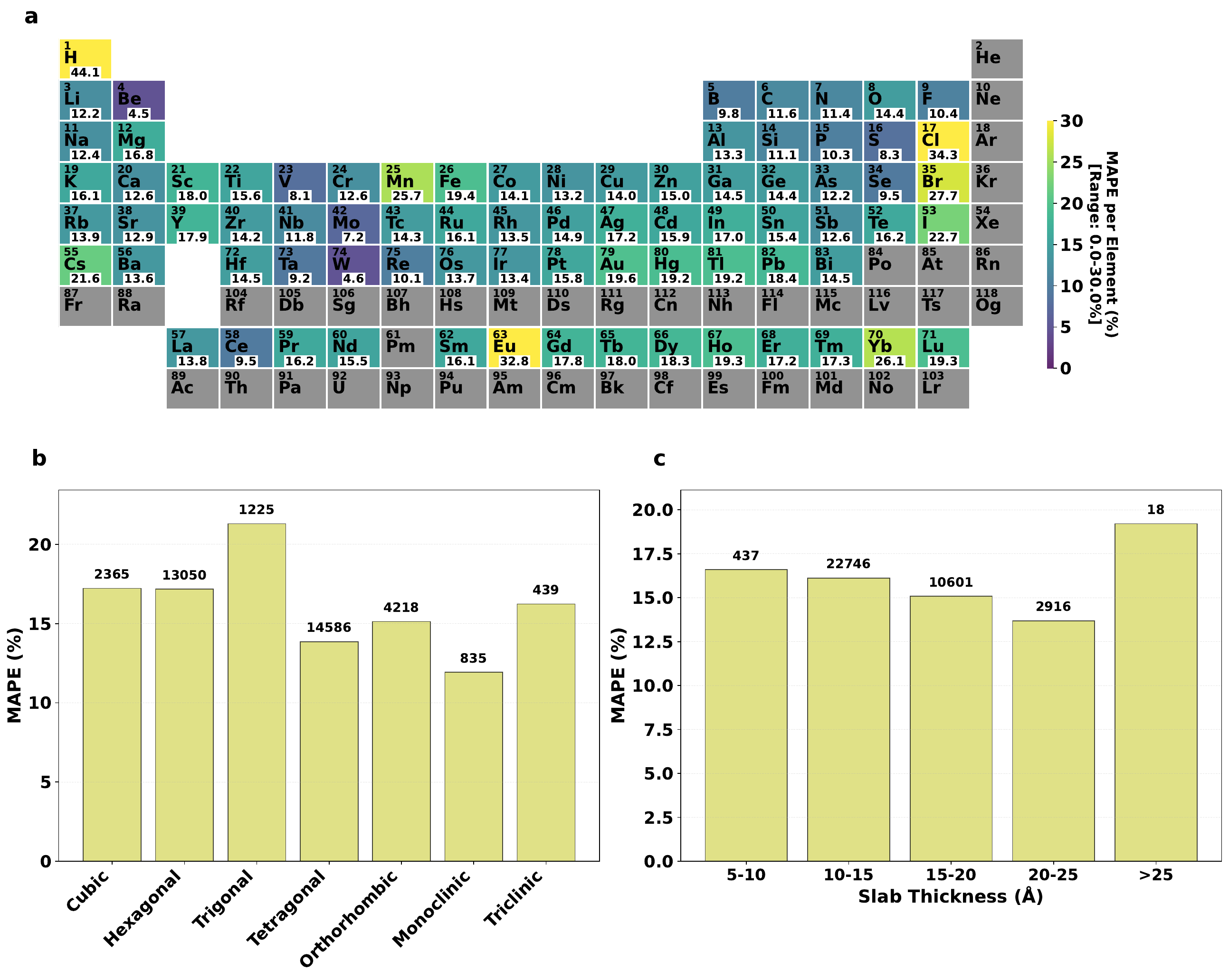}
\caption{\textbf{Decomposition of eSEN--30M--MP prediction errors by chemical, structural, and geometric features.} 
\textbf{(a)} Periodic table heat map showing MAPE for surfaces containing each element, with viridis colormap normalized to 0--30\% range. 
\textbf{(b)} MAPE distribution across seven crystal systems ordered by symmetry, with sample counts indicated above bars. 
\textbf{(c)} MAPE dependency on slab thickness, binned into 5~Å intervals showing consistent accuracy across different slab thicknesses with sample counts per bin.}
\label{fig:esen_30m_mp_error_decomposition}
\end{figure}

\begin{figure}[ht]
\centering
\includegraphics[width=\columnwidth]{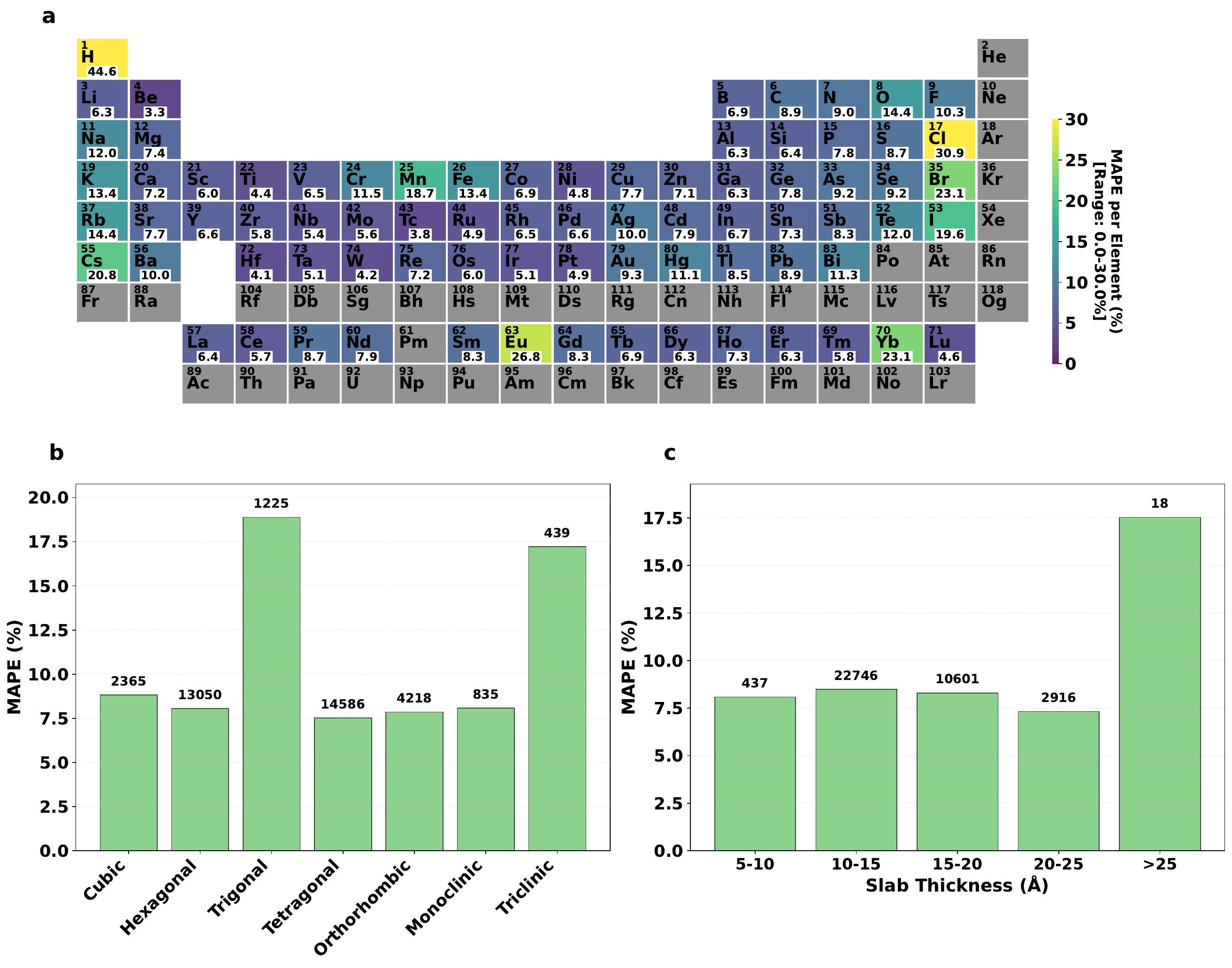}
\caption{\textbf{Decomposition of GRACE--2L--OAM prediction errors by chemical, structural, and geometric features.} 
\textbf{(a)} Periodic table heat map showing MAPE for surfaces containing each element, with viridis colormap normalized to 0--30\% range. 
\textbf{(b)} MAPE distribution across seven crystal systems ordered by symmetry, with sample counts indicated above bars. 
\textbf{(c)} MAPE dependency on slab thickness, binned into 5~Å intervals showing consistent accuracy across different slab thicknesses with sample counts per bin.}
\label{fig:grace_2l_oam_error_decomposition}
\end{figure}

\begin{figure}[ht]
\centering
\includegraphics[width=\columnwidth]{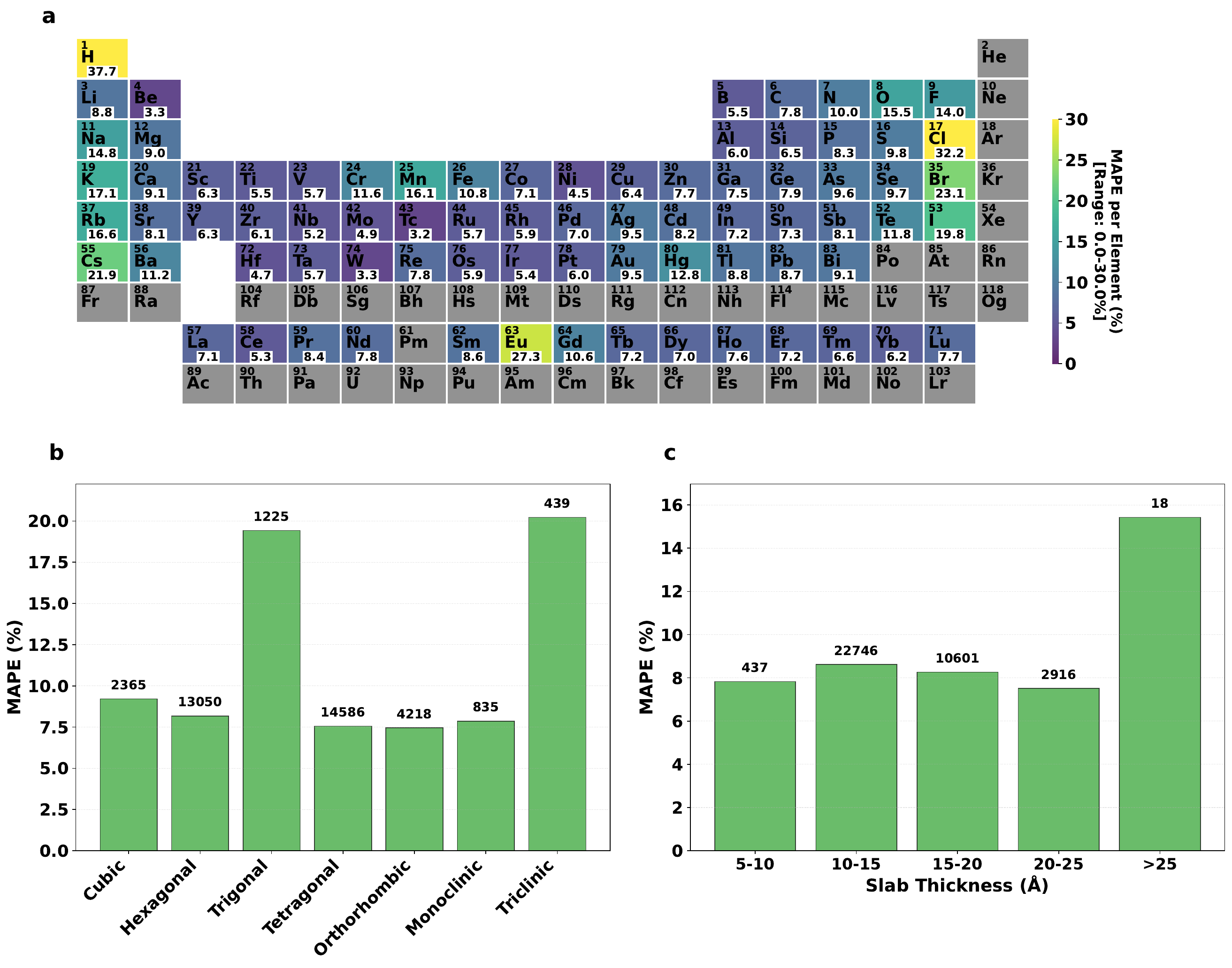}
\caption{\textbf{Decomposition of GRACE--2L--OMAT prediction errors by chemical, structural, and geometric features.} 
\textbf{(a)} Periodic table heat map showing MAPE for surfaces containing each element, with viridis colormap normalized to 0--30\% range. 
\textbf{(b)} MAPE distribution across seven crystal systems ordered by symmetry, with sample counts indicated above bars. 
\textbf{(c)} MAPE dependency on slab thickness, binned into 5~Å intervals showing consistent accuracy across different slab thicknesses with sample counts per bin.}
\label{fig:grace_2l_omat_error_decomposition}
\end{figure}

\begin{figure}[ht]
\centering
\includegraphics[width=\columnwidth]{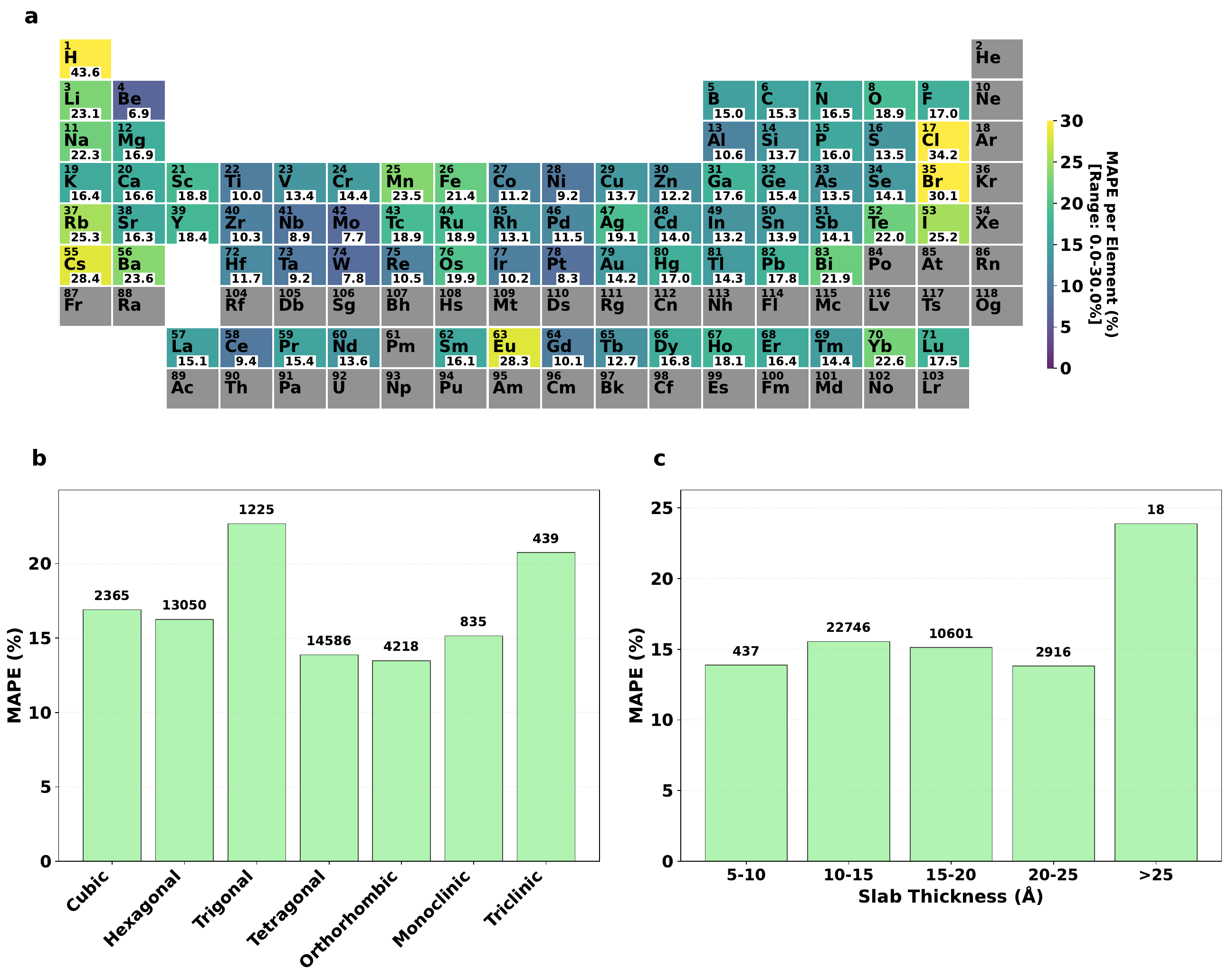}
\caption{\textbf{Decomposition of GRACE--2L--MP prediction errors by chemical, structural, and geometric features.} 
\textbf{(a)} Periodic table heat map showing MAPE for surfaces containing each element, with viridis colormap normalized to 0--30\% range. 
\textbf{(b)} MAPE distribution across seven crystal systems ordered by symmetry, with sample counts indicated above bars. 
\textbf{(c)} MAPE dependency on slab thickness, binned into 5~Å intervals showing consistent accuracy across different slab thicknesses with sample counts per bin.}
\label{fig:grace_2l_mp_error_decomposition}
\end{figure}

\begin{figure}[ht]
\centering
\includegraphics[width=\columnwidth]{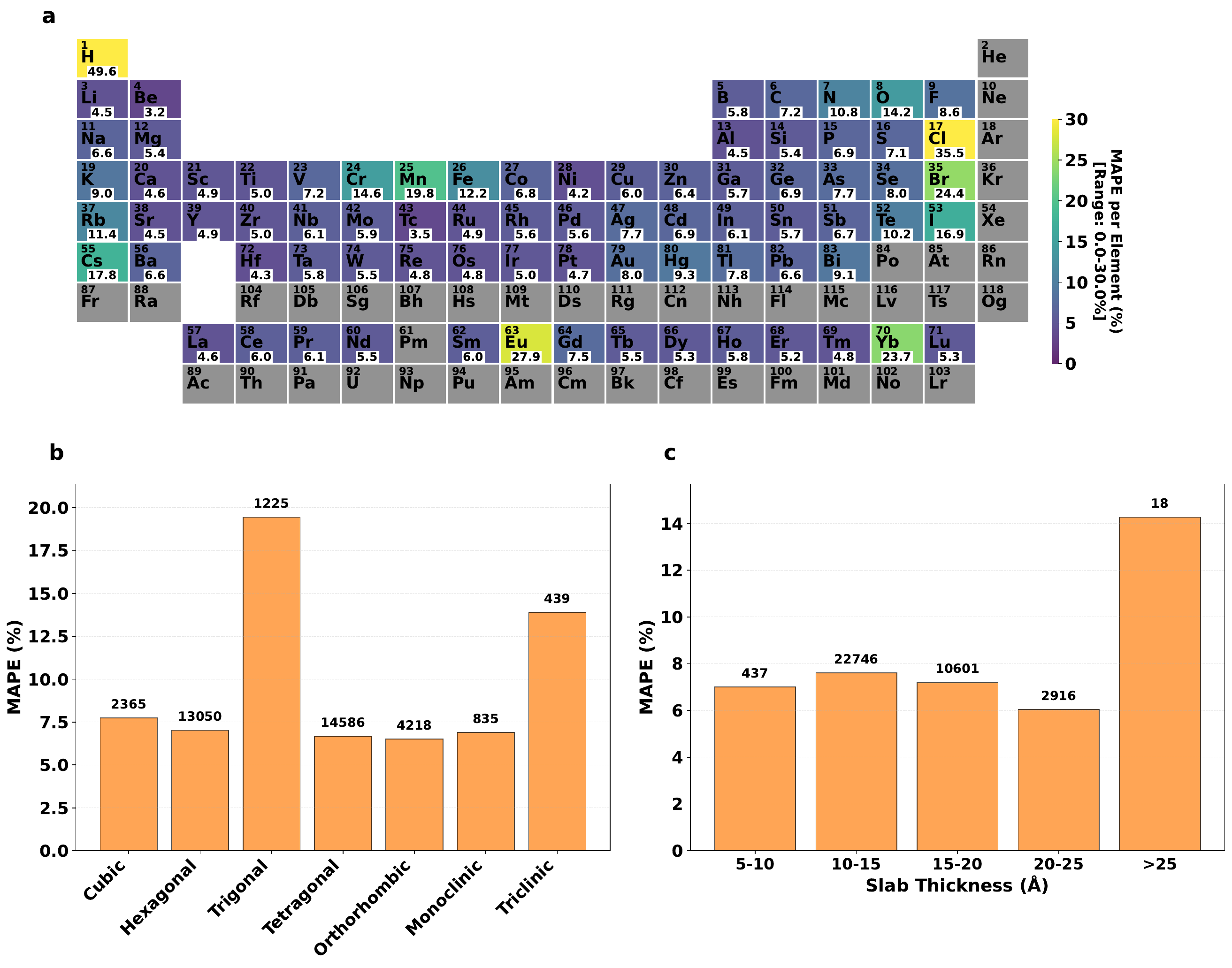}
\caption{\textbf{Decomposition of ORB--v3--MPA prediction errors by chemical, structural, and geometric features.} 
\textbf{(a)} Periodic table heat map showing MAPE for surfaces containing each element, with viridis colormap normalized to 0--30\% range. 
\textbf{(b)} MAPE distribution across seven crystal systems ordered by symmetry, with sample counts indicated above bars. 
\textbf{(c)} MAPE dependency on slab thickness, binned into 5~Å intervals showing consistent accuracy across different slab thicknesses with sample counts per bin.}
\label{fig:orb_v3_mpa_error_decomposition}
\end{figure}

\begin{figure}[ht]
\centering
\includegraphics[width=\columnwidth]{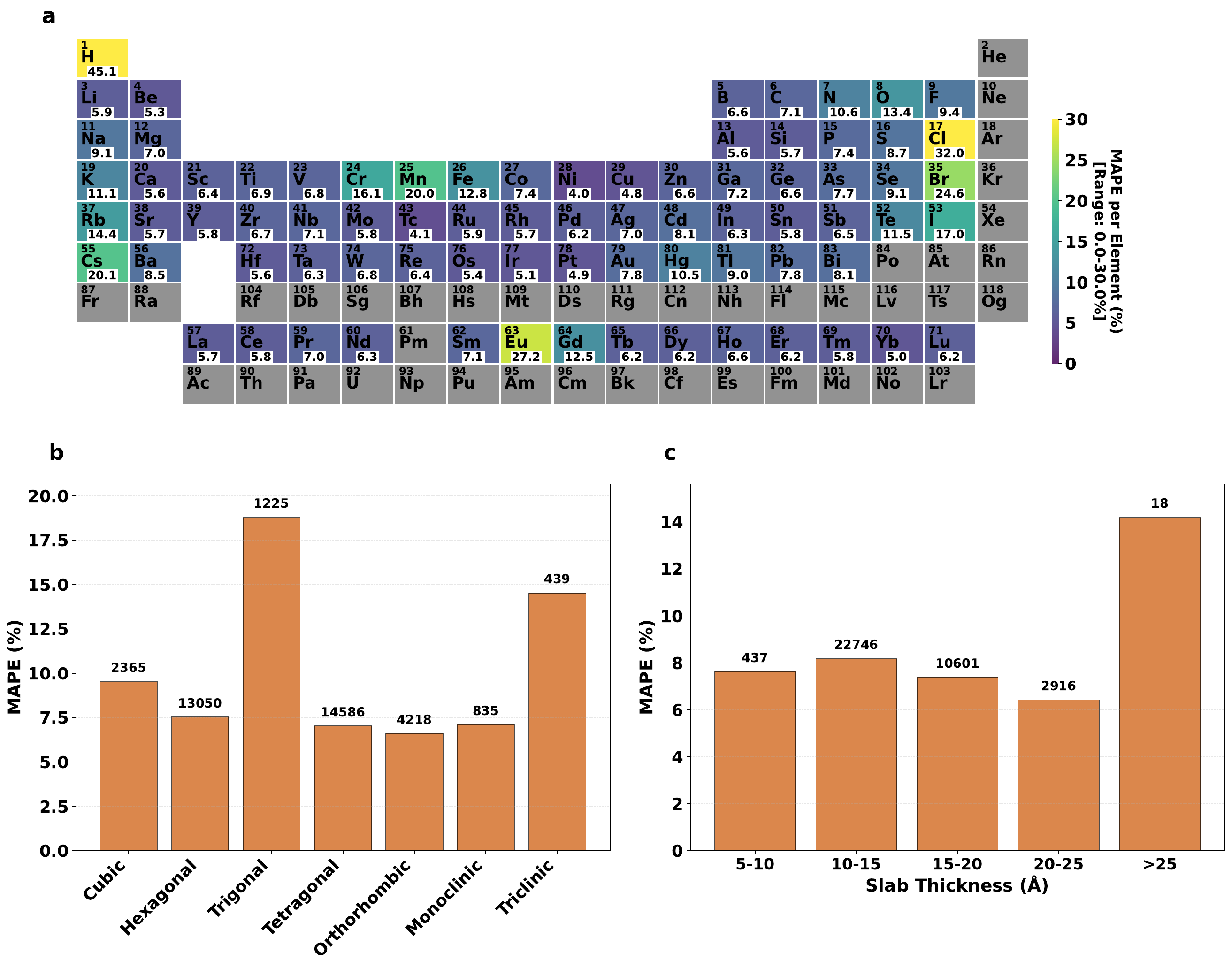}
\caption{\textbf{Decomposition of ORB--v3--OMAT prediction errors by chemical, structural, and geometric features.} 
\textbf{(a)} Periodic table heat map showing MAPE for surfaces containing each element, with viridis colormap normalized to 0--30\% range. 
\textbf{(b)} MAPE distribution across seven crystal systems ordered by symmetry, with sample counts indicated above bars. 
\textbf{(c)} MAPE dependency on slab thickness, binned into 5~Å intervals showing consistent accuracy across different slab thicknesses with sample counts per bin.}
\label{fig:orb_v3_omat_error_decomposition}
\end{figure}

\begin{figure}[ht]
\centering
\includegraphics[width=\columnwidth]{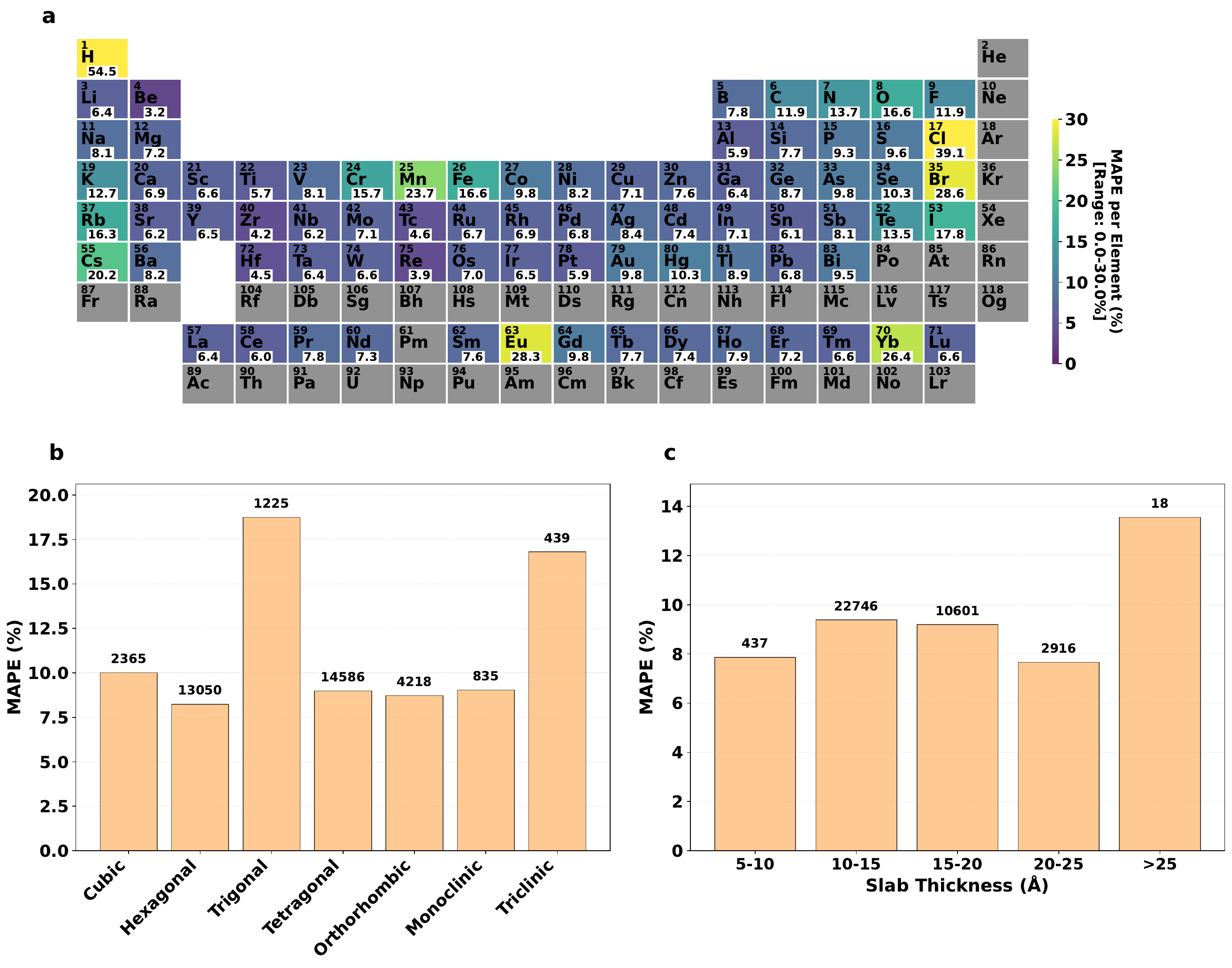}
\caption{\textbf{Decomposition of ORB--v2 prediction errors by chemical, structural, and geometric features.} 
\textbf{(a)} Periodic table heat map showing MAPE for surfaces containing each element, with viridis colormap normalized to 0--30\% range. 
\textbf{(b)} MAPE distribution across seven crystal systems ordered by symmetry, with sample counts indicated above bars. 
\textbf{(c)} MAPE dependency on slab thickness, binned into 5~Å intervals showing consistent accuracy across different slab thicknesses with sample counts per bin.}
\label{fig:orb_v2_error_decomposition}
\end{figure}

\begin{figure}[ht]
\centering
\includegraphics[width=\columnwidth]{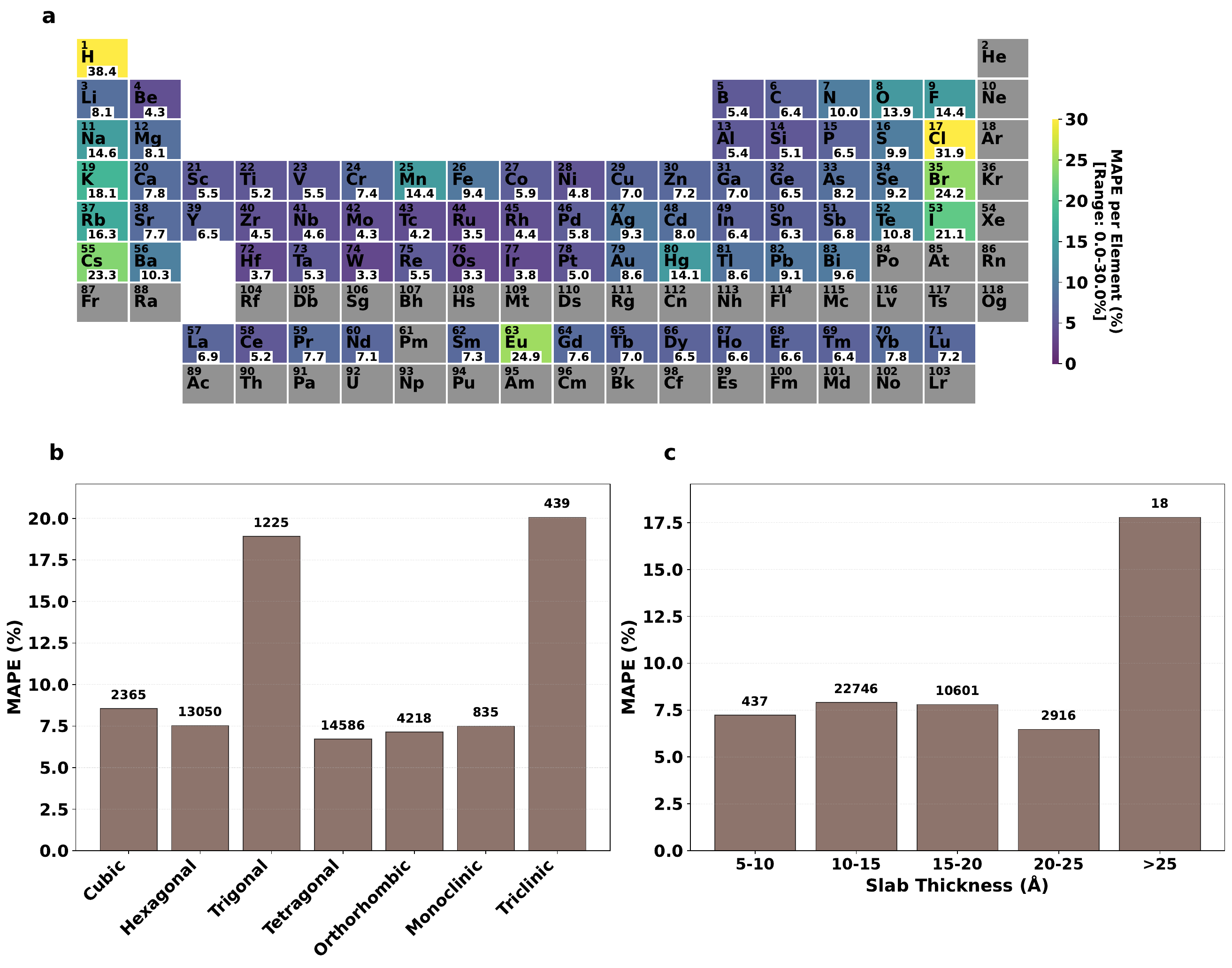}
\caption{\textbf{Decomposition of MACE--OMAT prediction errors by chemical, structural, and geometric features.} 
\textbf{(a)} Periodic table heat map showing MAPE for surfaces containing each element, with viridis colormap normalized to 0--30\% range. 
\textbf{(b)} MAPE distribution across seven crystal systems ordered by symmetry, with sample counts indicated above bars. 
\textbf{(c)} MAPE dependency on slab thickness, binned into 5~Å intervals showing consistent accuracy across different slab thicknesses with sample counts per bin.}
\label{fig:mace_omat_error_decomposition}
\end{figure}

\begin{figure}[ht]
\centering
\includegraphics[width=\columnwidth]{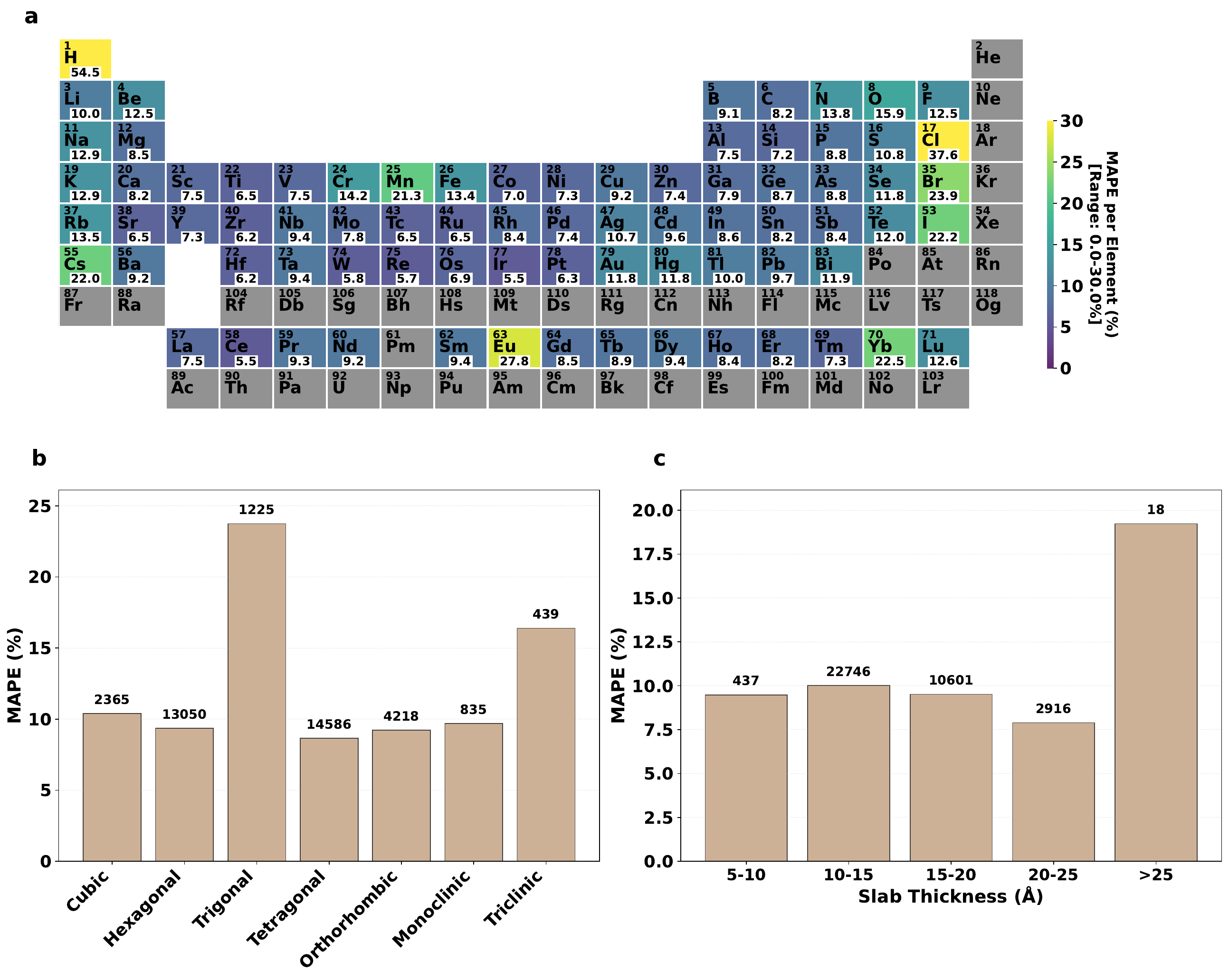}
\caption{\textbf{Decomposition of MACE--MPA prediction errors by chemical, structural, and geometric features.} 
\textbf{(a)} Periodic table heat map showing MAPE for surfaces containing each element, with viridis colormap normalized to 0--30\% range. 
\textbf{(b)} MAPE distribution across seven crystal systems ordered by symmetry, with sample counts indicated above bars. 
\textbf{(c)} MAPE dependency on slab thickness, binned into 5~Å intervals showing consistent accuracy across different slab thicknesses with sample counts per bin.}
\label{fig:mace_mpa_error_decomposition}
\end{figure}

\begin{figure}[ht]
\centering
\includegraphics[width=\columnwidth]{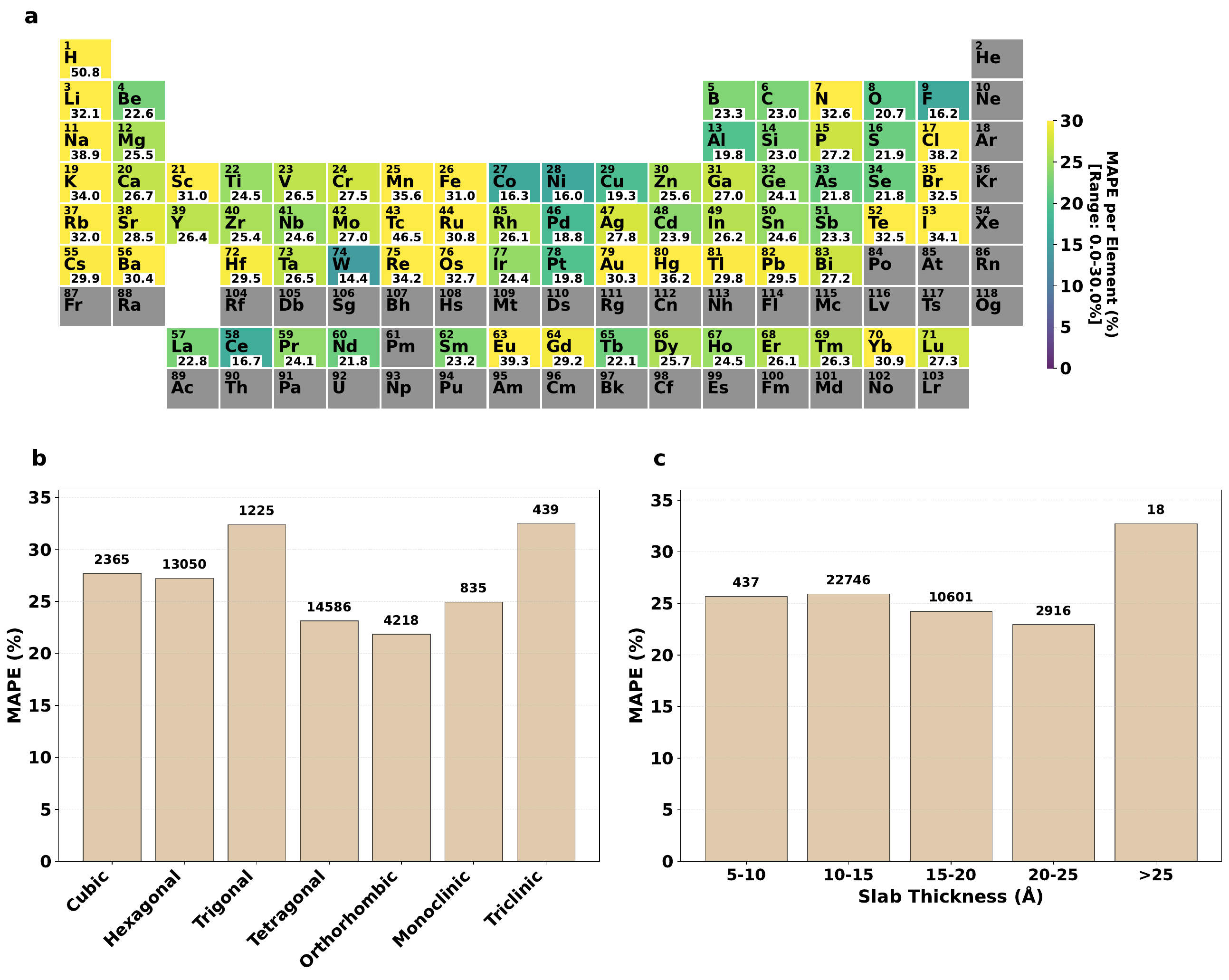}
\caption{\textbf{Decomposition of MACE--MP--0b3 prediction errors by chemical, structural, and geometric features.} 
\textbf{(a)} Periodic table heat map showing MAPE for surfaces containing each element, with viridis colormap normalized to 0--30\% range. 
\textbf{(b)} MAPE distribution across seven crystal systems ordered by symmetry, with sample counts indicated above bars. 
\textbf{(c)} MAPE dependency on slab thickness, binned into 5~Å intervals showing consistent accuracy across different slab thicknesses with sample counts per bin.}
\label{fig:mace_mp_0b3_error_decomposition}
\end{figure}

\begin{figure}[ht]
\centering
\includegraphics[width=\columnwidth]{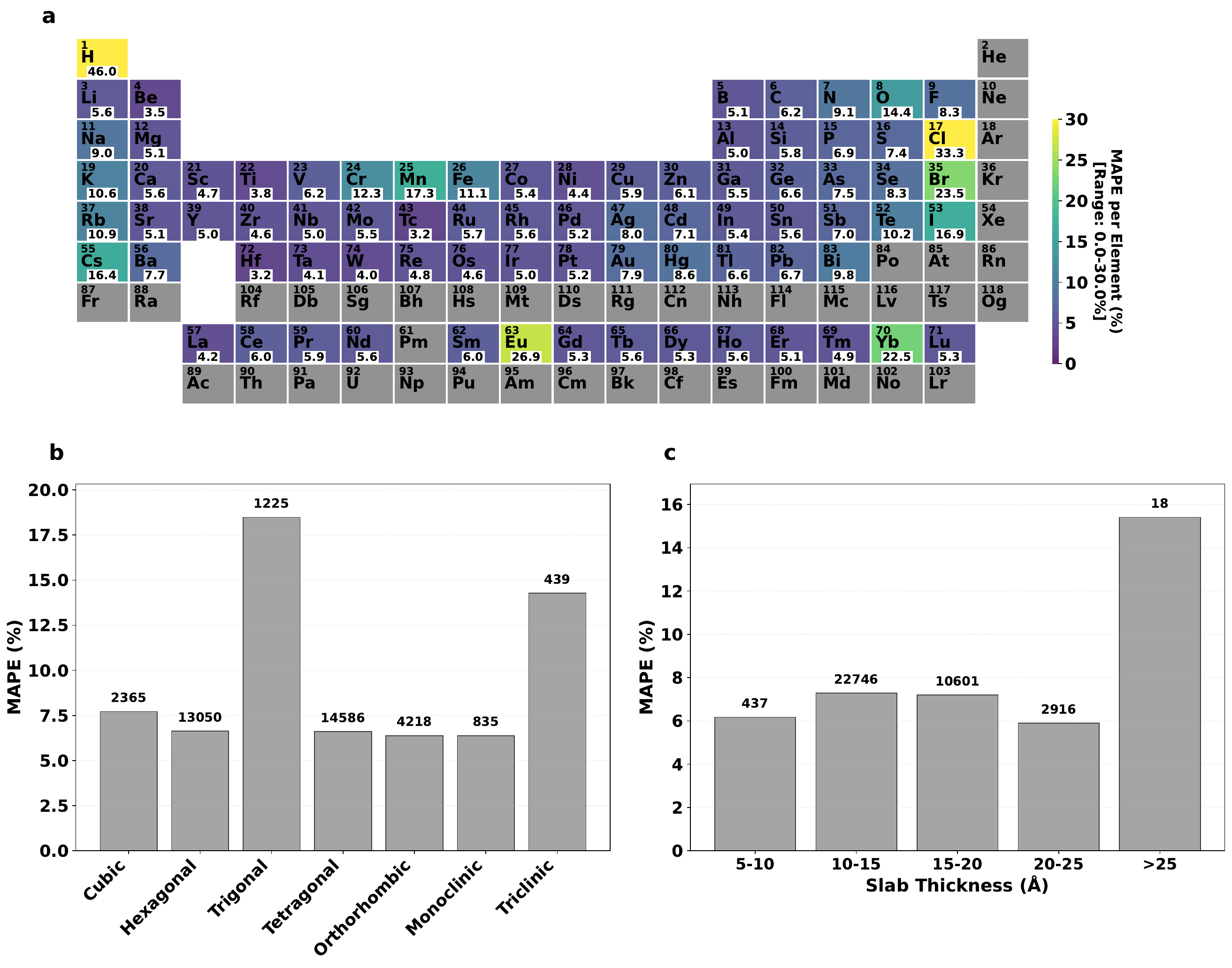}
\caption{\textbf{Decomposition of SevenNet--MF--OMPA prediction errors by chemical, structural, and geometric features.} 
\textbf{(a)} Periodic table heat map showing MAPE for surfaces containing each element, with viridis colormap normalized to 0--30\% range. 
\textbf{(b)} MAPE distribution across seven crystal systems ordered by symmetry, with sample counts indicated above bars. 
\textbf{(c)} MAPE dependency on slab thickness, binned into 5~Å intervals showing consistent accuracy across different slab thicknesses with sample counts per bin.}
\label{fig:sevennet_mf_ompa_error_decomposition}
\end{figure}

\begin{figure}[ht]
\centering
\includegraphics[width=\columnwidth]{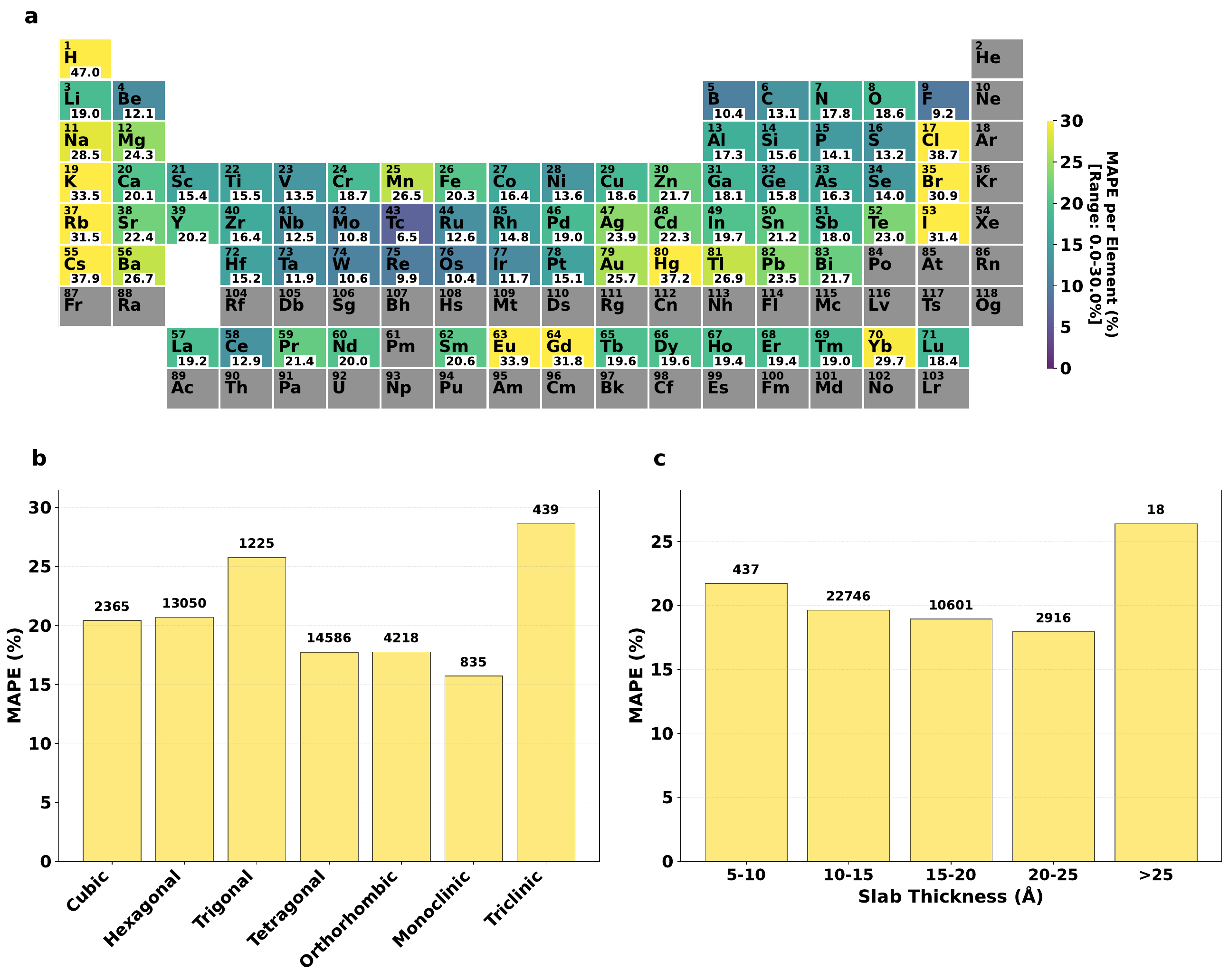}
\caption{\textbf{Decomposition of MatterSim--v1 prediction errors by chemical, structural, and geometric features.} 
\textbf{(a)} Periodic table heat map showing MAPE for surfaces containing each element, with viridis colormap normalized to 0--30\% range. 
\textbf{(b)} MAPE distribution across seven crystal systems ordered by symmetry, with sample counts indicated above bars. 
\textbf{(c)} MAPE dependency on slab thickness, binned into 5~Å intervals showing consistent accuracy across different slab thicknesses with sample counts per bin.}
\label{fig:mattersim_v1_error_decomposition}
\end{figure}

\begin{figure}[ht]
\centering
\includegraphics[width=\columnwidth]{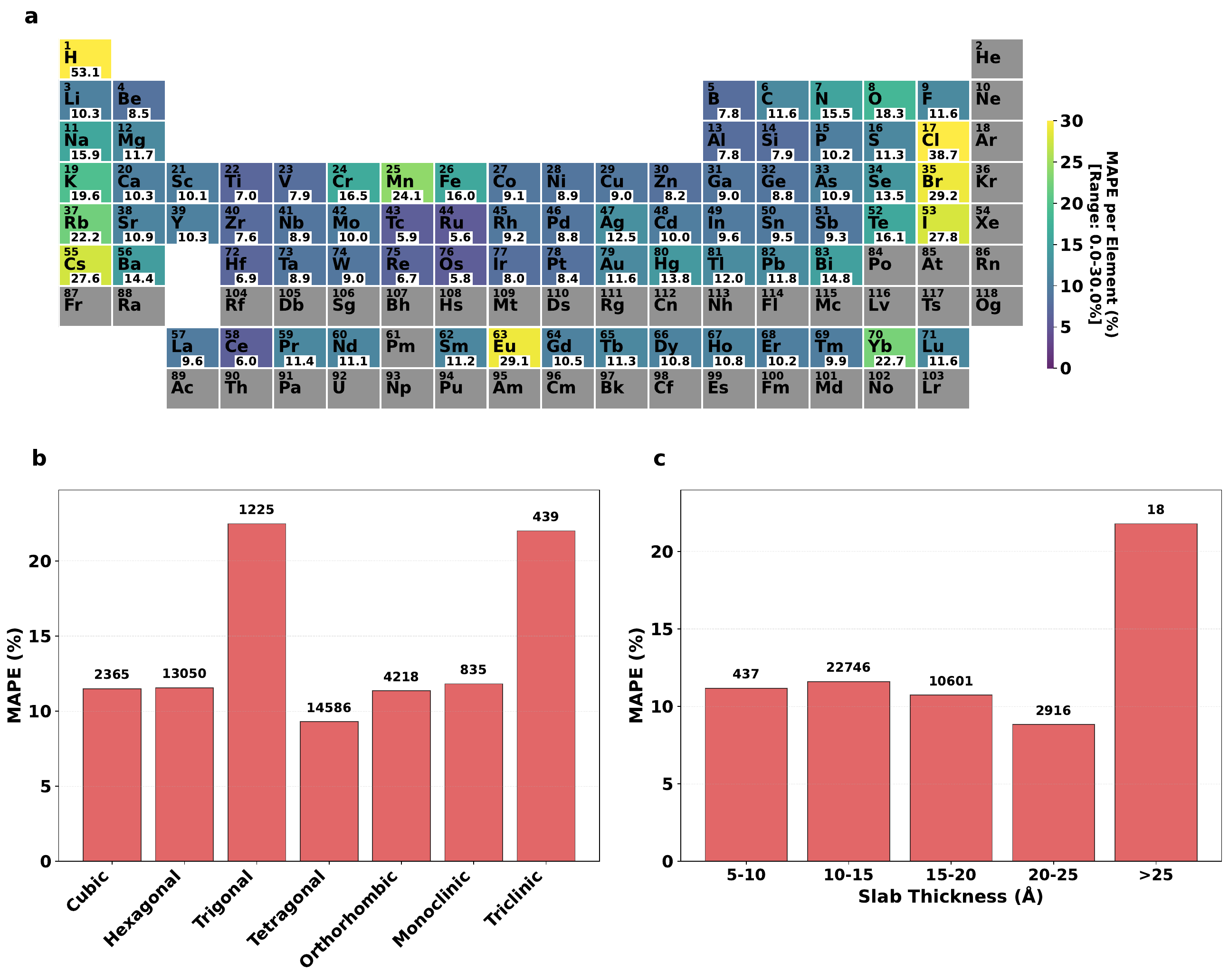}
\caption{\textbf{Decomposition of AlphaNet--OMA prediction errors by chemical, structural, and geometric features.} 
\textbf{(a)} Periodic table heat map showing MAPE for surfaces containing each element, with viridis colormap normalized to 0--30\% range. 
\textbf{(b)} MAPE distribution across seven crystal systems ordered by symmetry, with sample counts indicated above bars. 
\textbf{(c)} MAPE dependency on slab thickness, binned into 5~Å intervals showing consistent accuracy across different slab thicknesses with sample counts per bin.}
\label{fig:alphanet_oma_error_decomposition}
\end{figure}


\begin{figure}[ht]
\centering
\includegraphics[width=0.8\columnwidth]{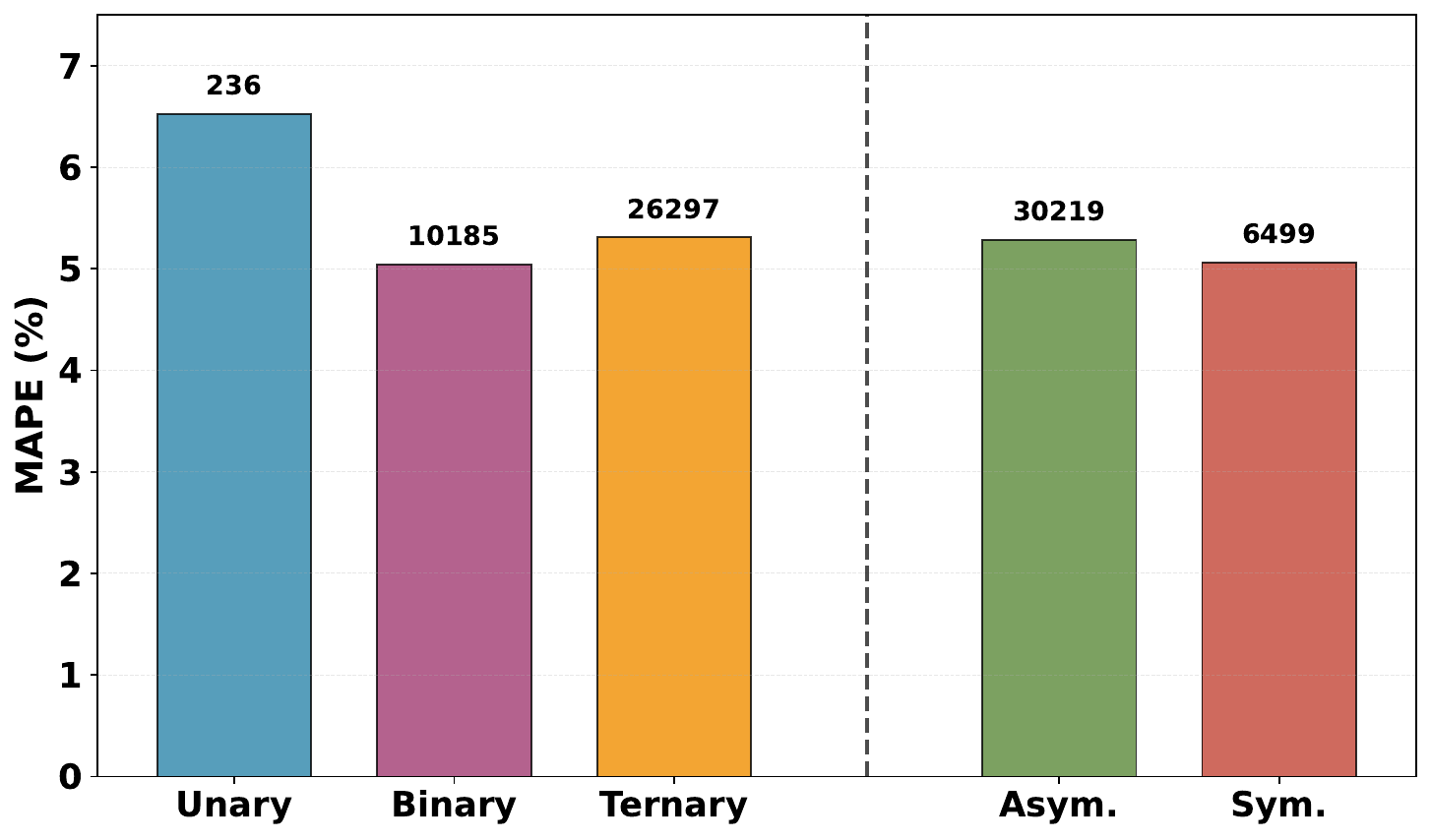}
\caption{\textbf{MAPE analysis by compositional complexity and surface symmetry for Eqf.V2--OMAT24--MP--sAlex.} 
Bar chart showing prediction errors grouped by compositional complexity (unary, binary, ternary) and surface symmetry (asymmetric, symmetric), with sample counts indicated above each bar. The dashed line separates composition-based analysis (left) from symmetry-based analysis (right).}
\label{fig:eqfv2_omat24_mp_salex_composition_symmetry}
\end{figure}

\begin{figure}[ht]
\centering
\includegraphics[width=0.8\columnwidth]{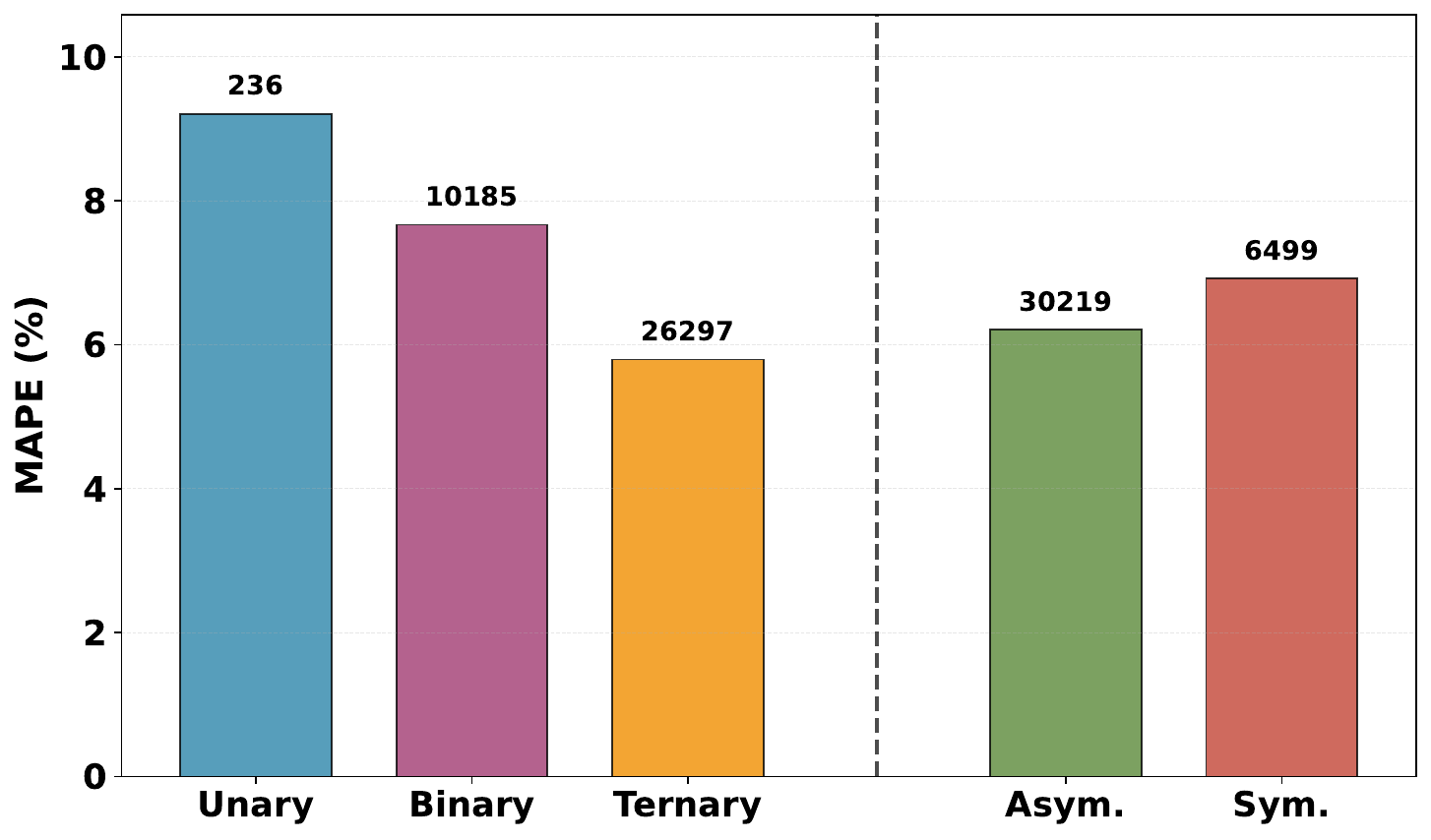}
\caption{\textbf{MAPE analysis by compositional complexity and surface symmetry for Eqf.V2--OMAT24.} 
Bar chart showing prediction errors grouped by compositional complexity (unary, binary, ternary) and surface symmetry (asymmetric, symmetric), with sample counts indicated above each bar. The dashed line separates composition-based analysis (left) from symmetry-based analysis (right).}
\label{fig:eqfv2_omat24_composition_symmetry}
\end{figure}

\begin{figure}[ht]
\centering
\includegraphics[width=0.8\columnwidth]{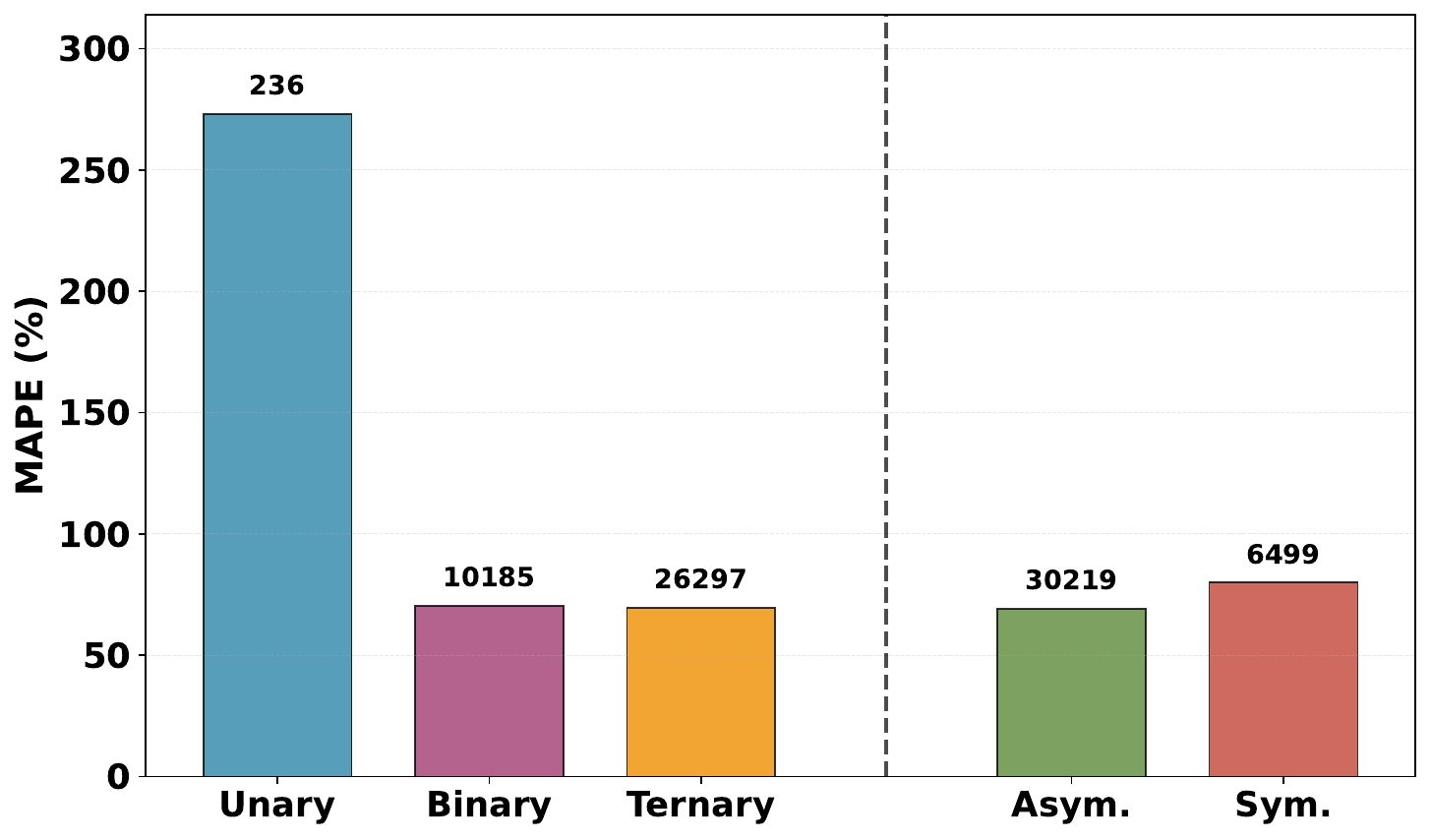}
\caption{\textbf{MAPE analysis by compositional complexity and surface symmetry for Eqf.V2--OC20.} 
Bar chart showing prediction errors grouped by compositional complexity (unary, binary, ternary) and surface symmetry (asymmetric, symmetric), with sample counts indicated above each bar. The dashed line separates composition-based analysis (left) from symmetry-based analysis (right).}
\label{fig:eqfv2_oc20_composition_symmetry}
\end{figure}

\begin{figure}[ht]
\centering
\includegraphics[width=0.8\columnwidth]{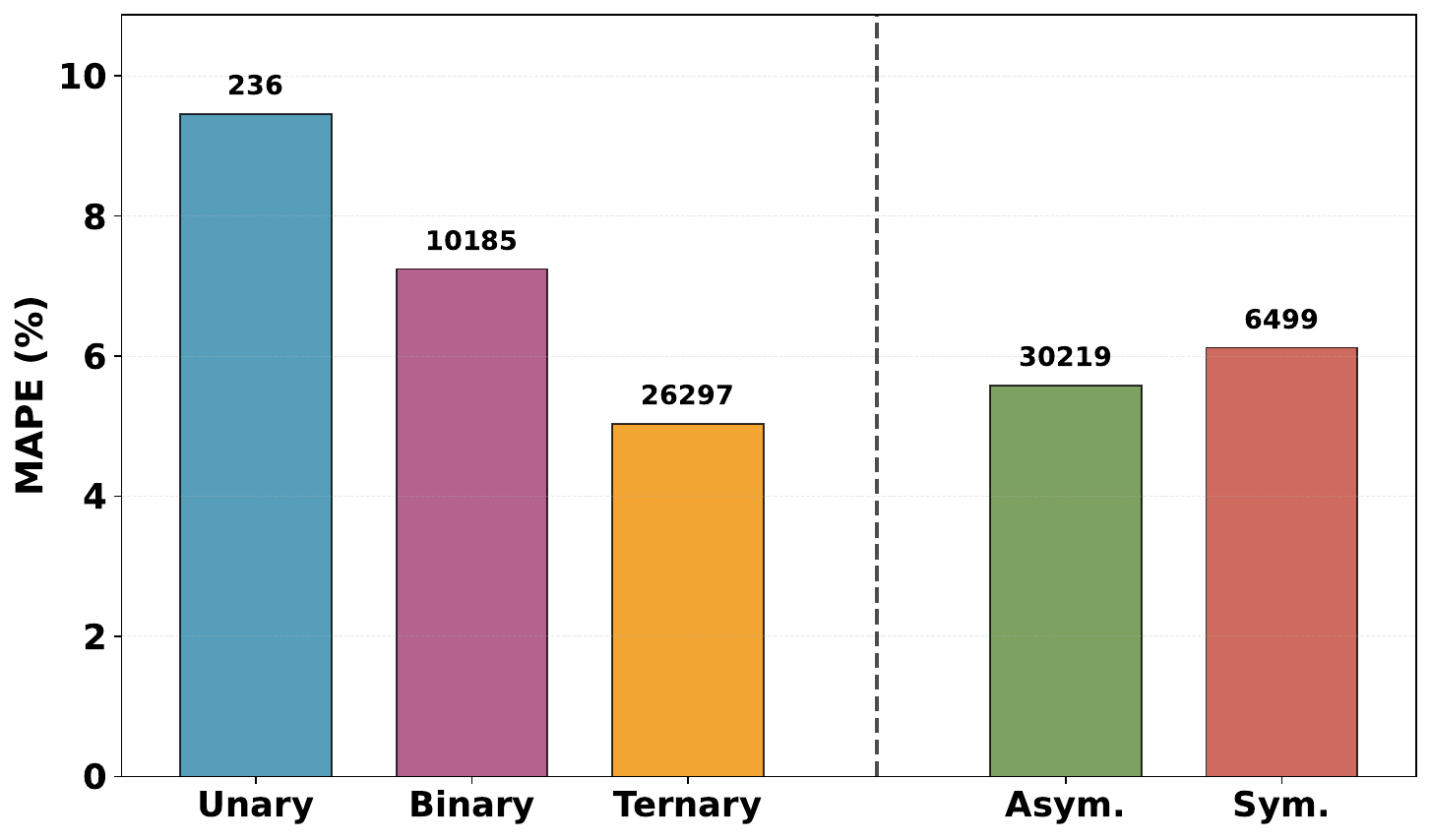}
\caption{\textbf{MAPE analysis by compositional complexity and surface symmetry for eSEN--30M--OMAT24.} 
Bar chart showing prediction errors grouped by compositional complexity (unary, binary, ternary) and surface symmetry (asymmetric, symmetric), with sample counts indicated above each bar. The dashed line separates composition-based analysis (left) from symmetry-based analysis (right).}
\label{fig:esen_30m_omat24_composition_symmetry}
\end{figure}

\begin{figure}[ht]
\centering
\includegraphics[width=0.8\columnwidth]{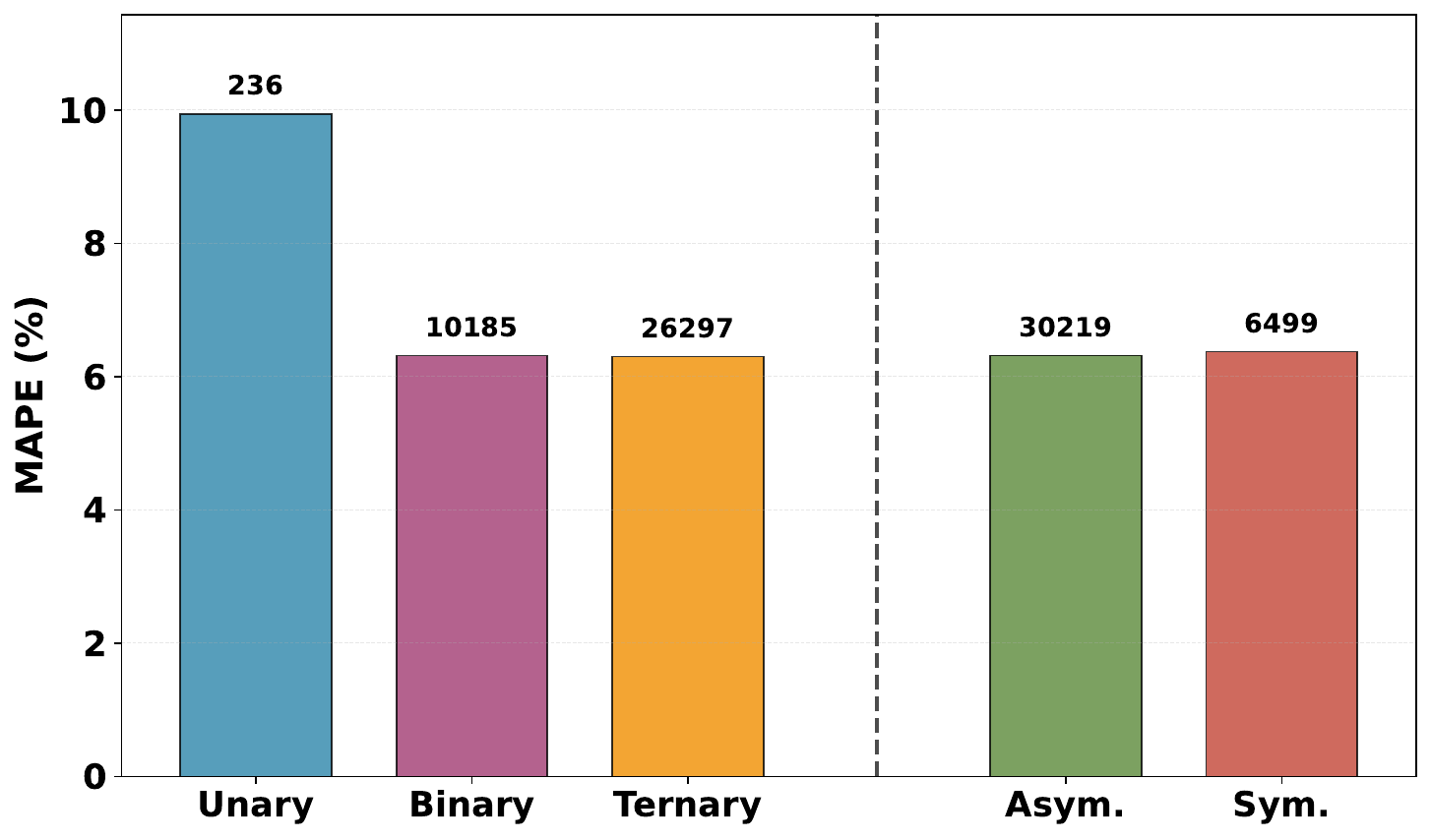}
\caption{\textbf{MAPE analysis by compositional complexity and surface symmetry for eSEN--30M--OAM.} 
Bar chart showing prediction errors grouped by compositional complexity (unary, binary, ternary) and surface symmetry (asymmetric, symmetric), with sample counts indicated above each bar. The dashed line separates composition-based analysis (left) from symmetry-based analysis (right).}
\label{fig:esen_30m_oam_composition_symmetry}
\end{figure}

\begin{figure}[ht]
\centering
\includegraphics[width=0.8\columnwidth]{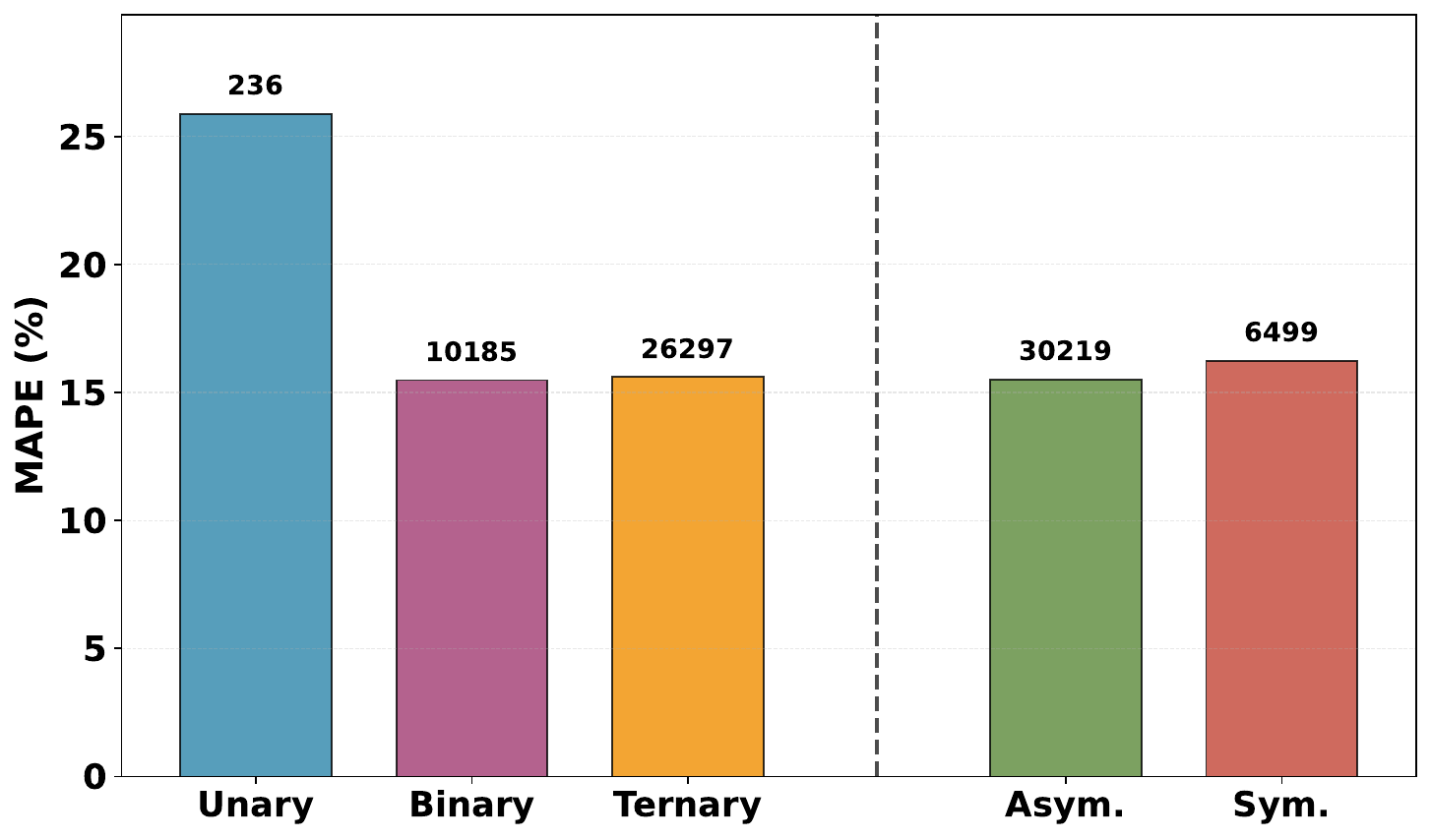}
\caption{\textbf{MAPE analysis by compositional complexity and surface symmetry for eSEN--30M--MP.} 
Bar chart showing prediction errors grouped by compositional complexity (unary, binary, ternary) and surface symmetry (asymmetric, symmetric), with sample counts indicated above each bar. The dashed line separates composition-based analysis (left) from symmetry-based analysis (right).}
\label{fig:esen_30m_mp_composition_symmetry}
\end{figure}

\begin{figure}[ht]
\centering
\includegraphics[width=0.8\columnwidth]{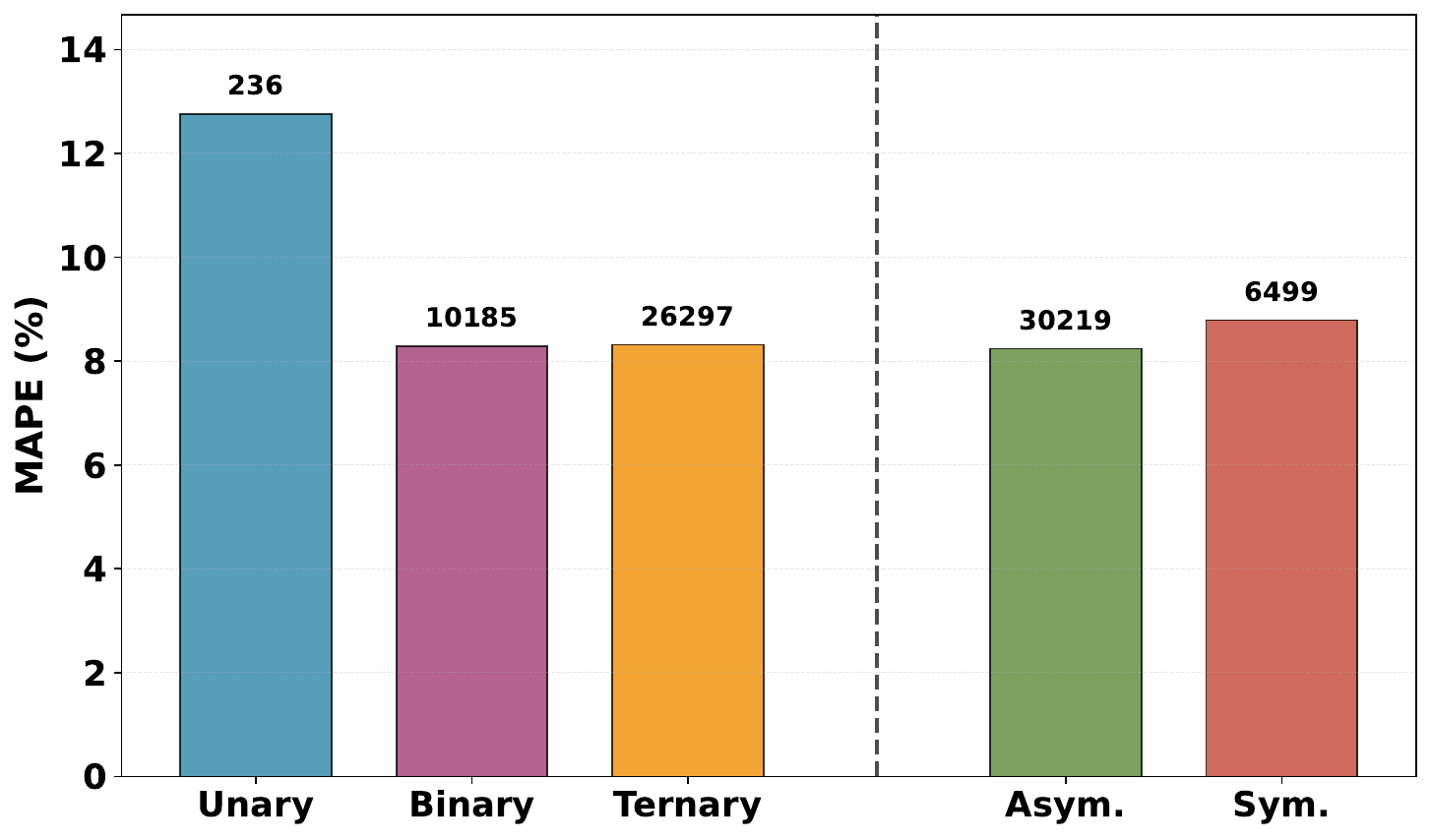}
\caption{\textbf{MAPE analysis by compositional complexity and surface symmetry for GRACE--2L--OAM.} 
Bar chart showing prediction errors grouped by compositional complexity (unary, binary, ternary) and surface symmetry (asymmetric, symmetric), with sample counts indicated above each bar. The dashed line separates composition-based analysis (left) from symmetry-based analysis (right).}
\label{fig:grace_2l_oam_composition_symmetry}
\end{figure}

\clearpage

\begin{figure}[ht]
\centering
\includegraphics[width=0.8\columnwidth]{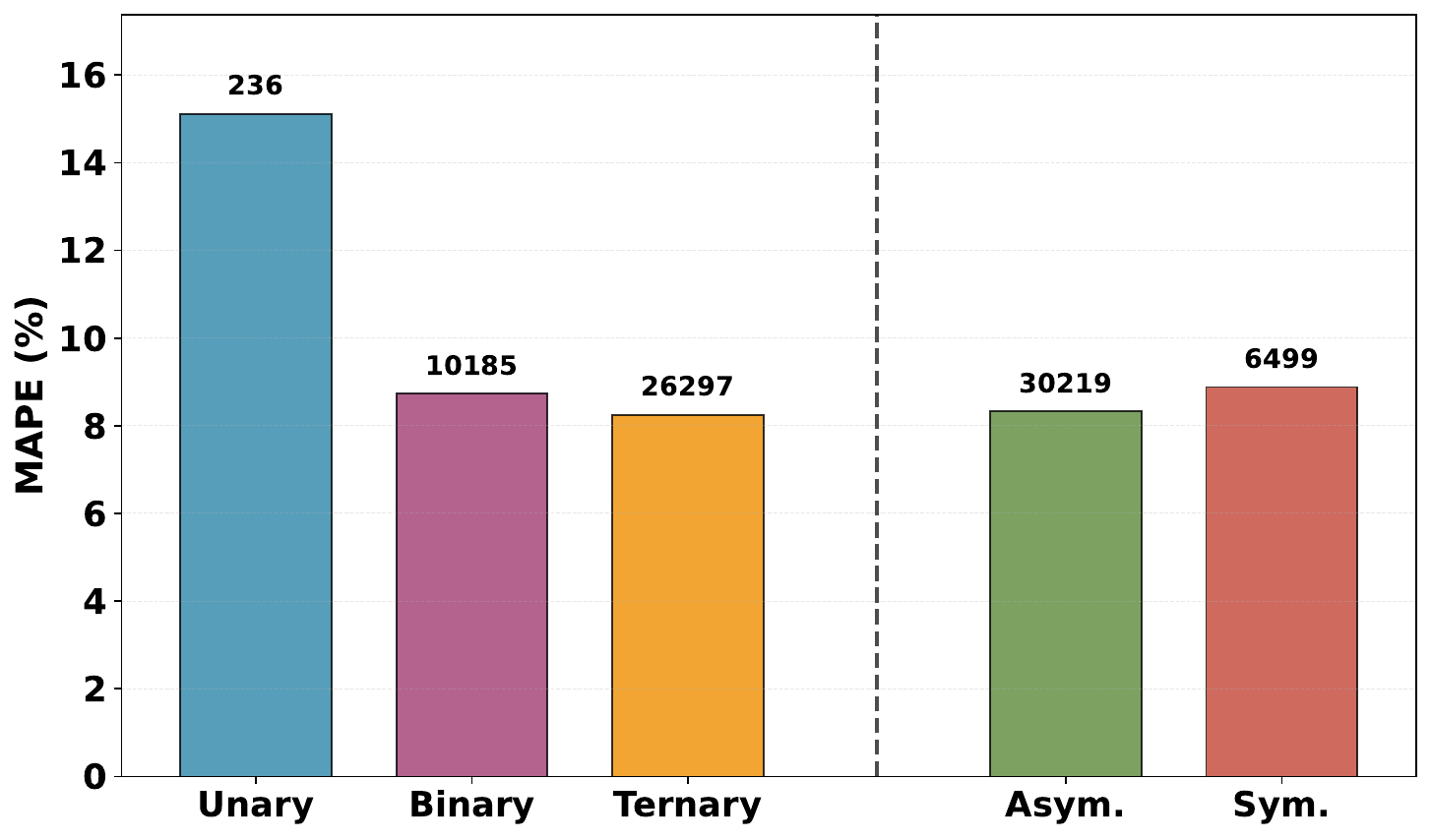}
\caption{\textbf{MAPE analysis by compositional complexity and surface symmetry for GRACE--2L--OMAT.} 
Bar chart showing prediction errors grouped by compositional complexity (unary, binary, ternary) and surface symmetry (asymmetric, symmetric), with sample counts indicated above each bar. The dashed line separates composition-based analysis (left) from symmetry-based analysis (right).}
\label{fig:grace_2l_omat_composition_symmetry}
\end{figure}

\begin{figure}[ht]
\centering
\includegraphics[width=0.8\columnwidth]{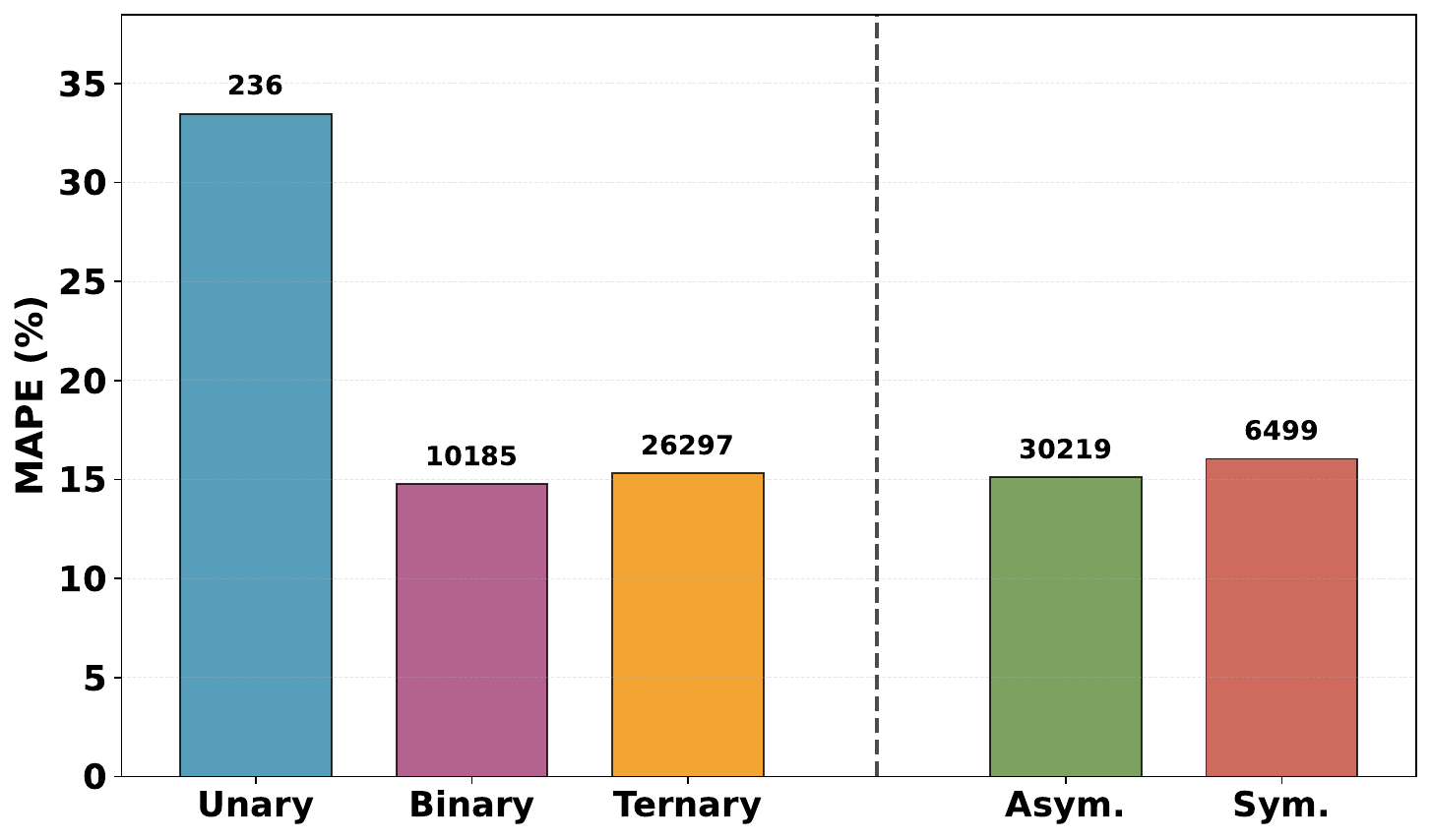}
\caption{\textbf{MAPE analysis by compositional complexity and surface symmetry for GRACE--2L--MP.} 
Bar chart showing prediction errors grouped by compositional complexity (unary, binary, ternary) and surface symmetry (asymmetric, symmetric), with sample counts indicated above each bar. The dashed line separates composition-based analysis (left) from symmetry-based analysis (right).}
\label{fig:grace_2l_mp_composition_symmetry}
\end{figure}

\clearpage

\begin{figure}[ht]
\centering
\includegraphics[width=0.8\columnwidth]{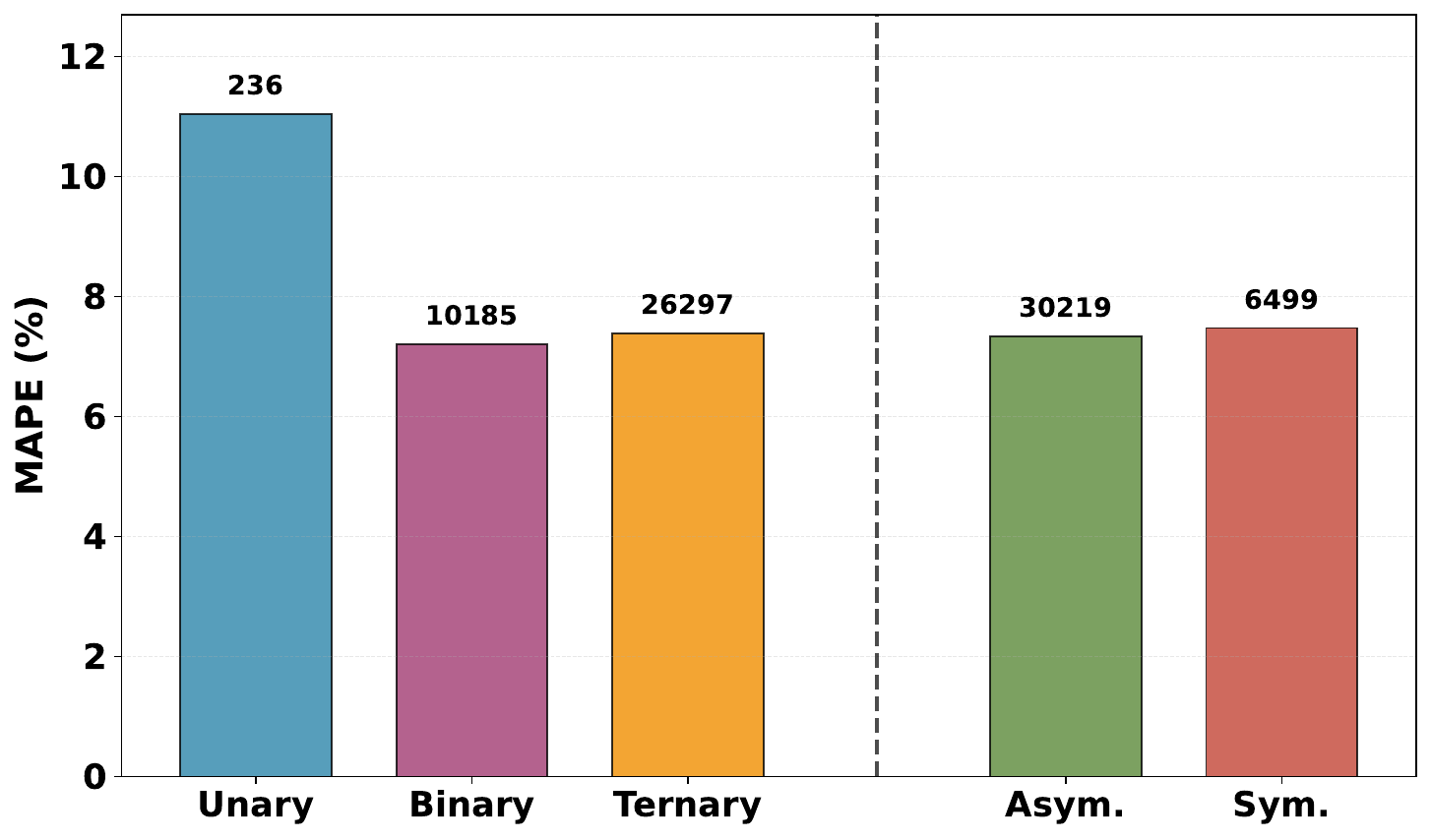}
\caption{\textbf{MAPE analysis by compositional complexity and surface symmetry for ORB--v3--MPA.} 
Bar chart showing prediction errors grouped by compositional complexity (unary, binary, ternary) and surface symmetry (asymmetric, symmetric), with sample counts indicated above each bar. The dashed line separates composition-based analysis (left) from symmetry-based analysis (right).}
\label{fig:orb_v3_mpa_composition_symmetry}
\end{figure}

\begin{figure}[ht]
\centering
\includegraphics[width=0.8\columnwidth]{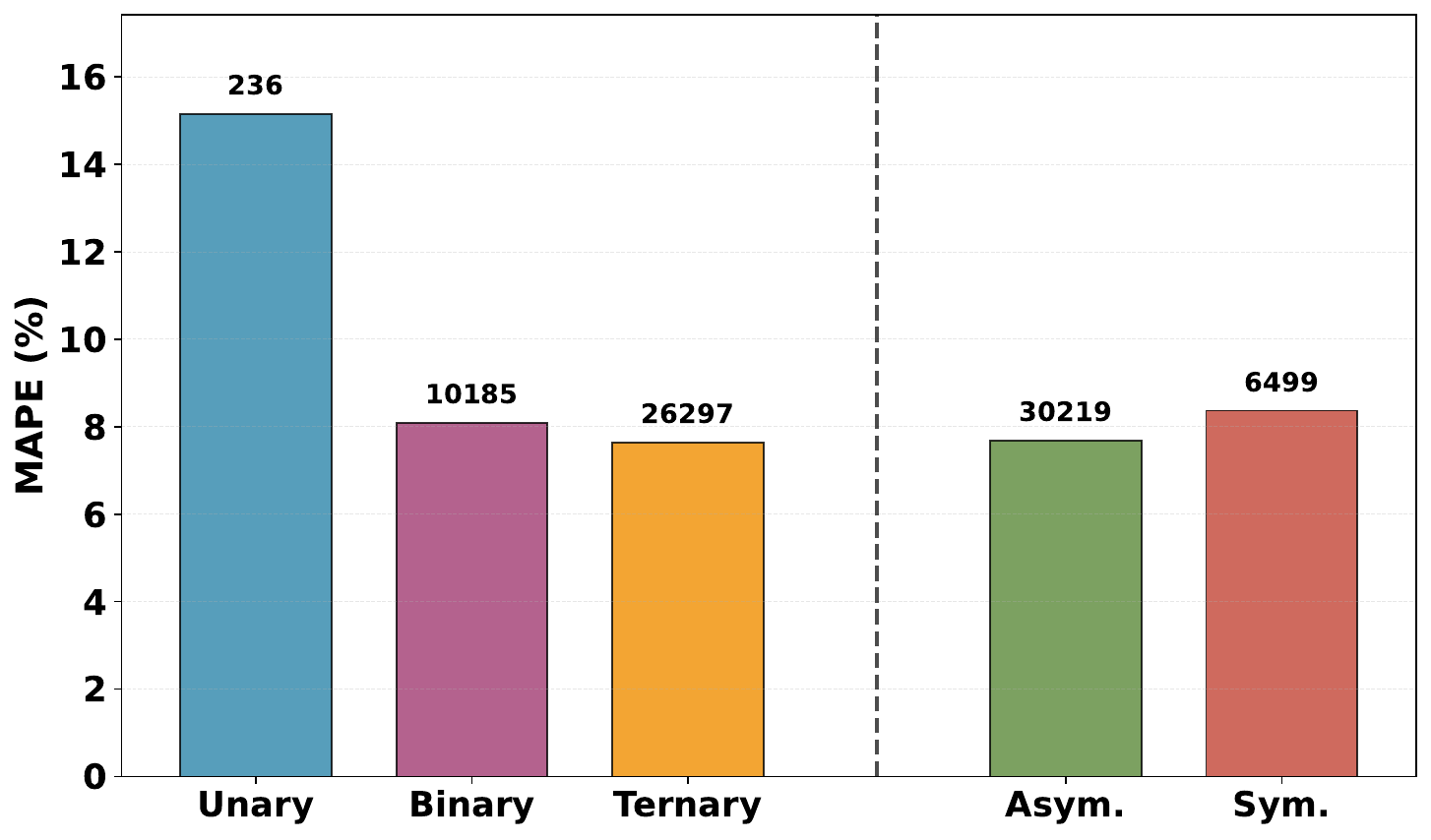}
\caption{\textbf{MAPE analysis by compositional complexity and surface symmetry for ORB--v3--OMAT.} 
Bar chart showing prediction errors grouped by compositional complexity (unary, binary, ternary) and surface symmetry (asymmetric, symmetric), with sample counts indicated above each bar. The dashed line separates composition-based analysis (left) from symmetry-based analysis (right).}
\label{fig:orb_v3_omat_composition_symmetry}
\end{figure}

\begin{figure}[ht]
\centering
\includegraphics[width=0.8\columnwidth]{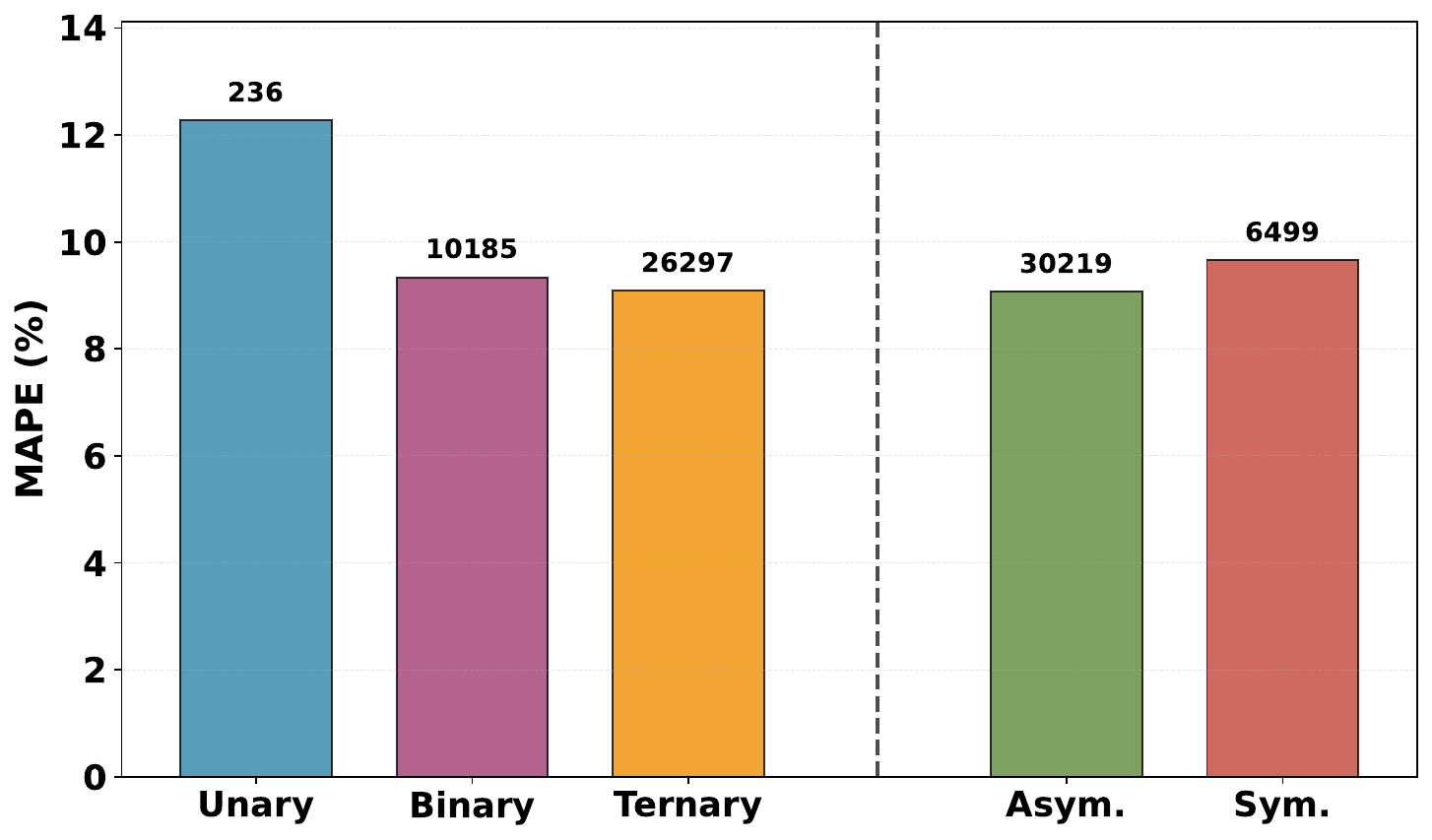}
\caption{\textbf{MAPE analysis by compositional complexity and surface symmetry for ORB--v2.} 
Bar chart showing prediction errors grouped by compositional complexity (unary, binary, ternary) and surface symmetry (asymmetric, symmetric), with sample counts indicated above each bar. The dashed line separates composition-based analysis (left) from symmetry-based analysis (right).}
\label{fig:orb_v2_composition_symmetry}
\end{figure}

\begin{figure}[ht]
\centering
\includegraphics[width=0.8\columnwidth]{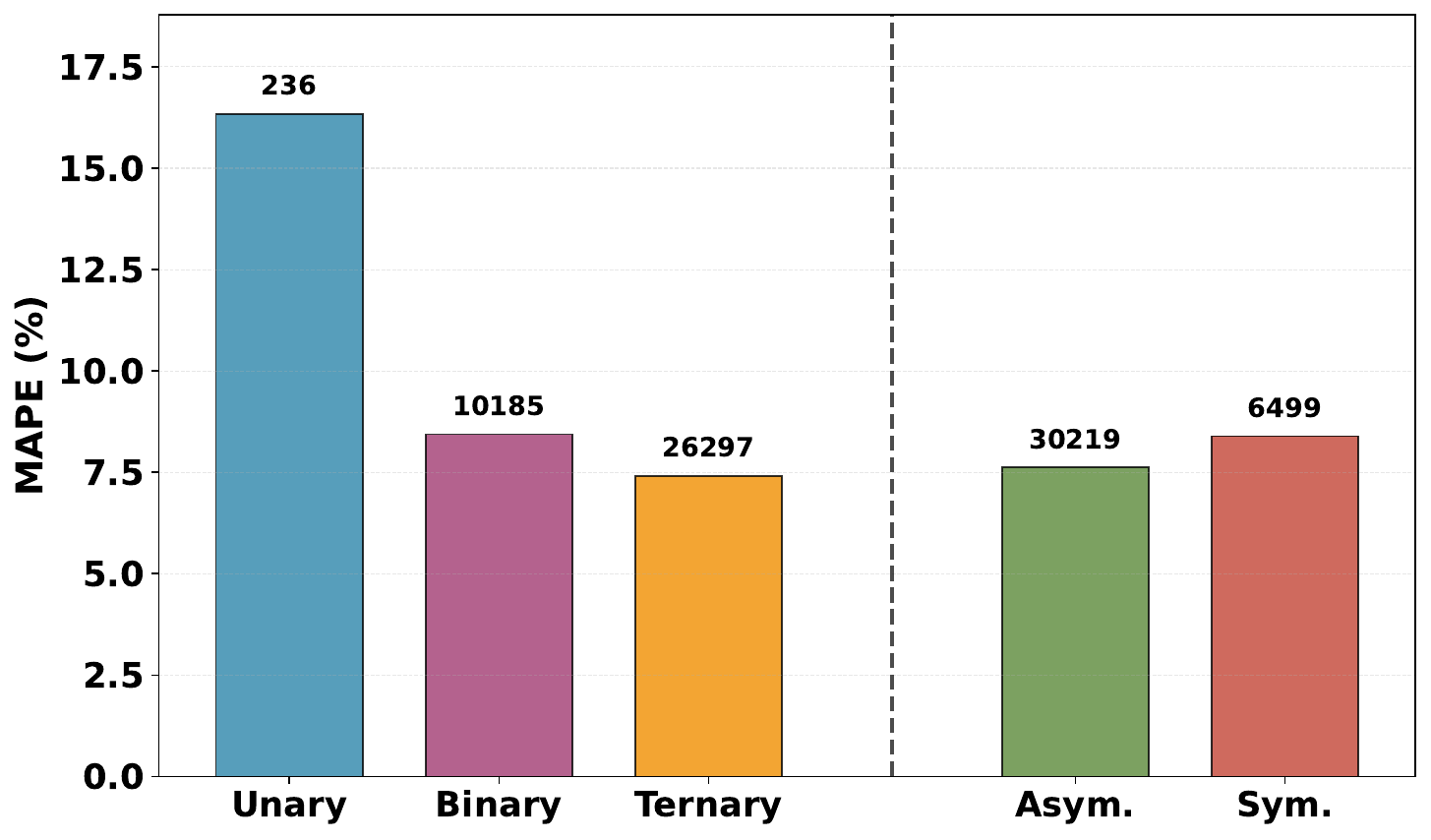}
\caption{\textbf{MAPE analysis by compositional complexity and surface symmetry for MACE--OMAT.} 
Bar chart showing prediction errors grouped by compositional complexity (unary, binary, ternary) and surface symmetry (asymmetric, symmetric), with sample counts indicated above each bar. The dashed line separates composition-based analysis (left) from symmetry-based analysis (right).}
\label{fig:mace_omat_composition_symmetry}
\end{figure}

\begin{figure}[ht]
\centering
\includegraphics[width=0.8\columnwidth]{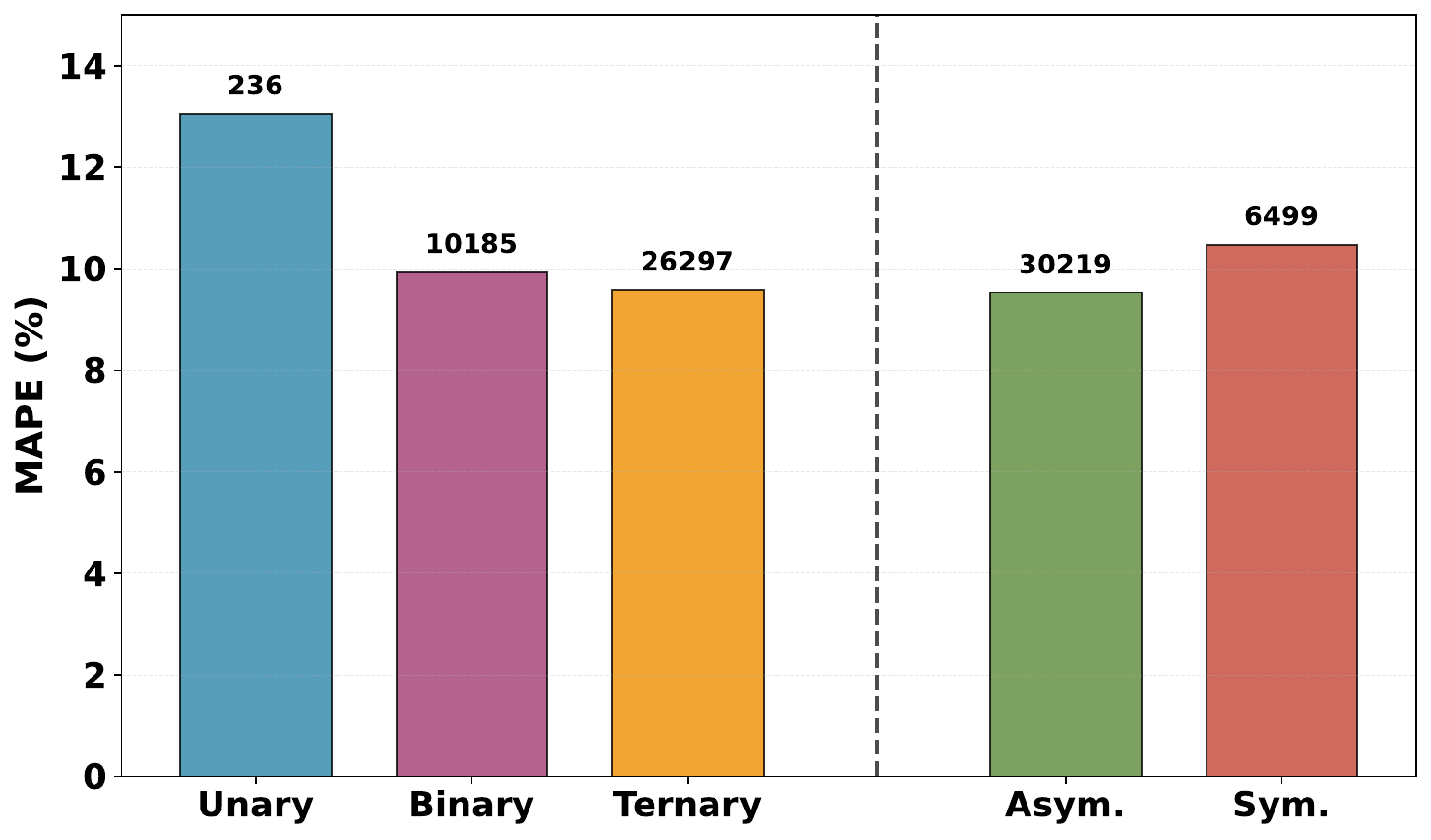}
\caption{\textbf{MAPE analysis by compositional complexity and surface symmetry for MACE--MPA.} 
Bar chart showing prediction errors grouped by compositional complexity (unary, binary, ternary) and surface symmetry (asymmetric, symmetric), with sample counts indicated above each bar. The dashed line separates composition-based analysis (left) from symmetry-based analysis (right).}
\label{fig:mace_mpa_composition_symmetry}
\end{figure}

\begin{figure}[ht]
\centering
\includegraphics[width=0.8\columnwidth]{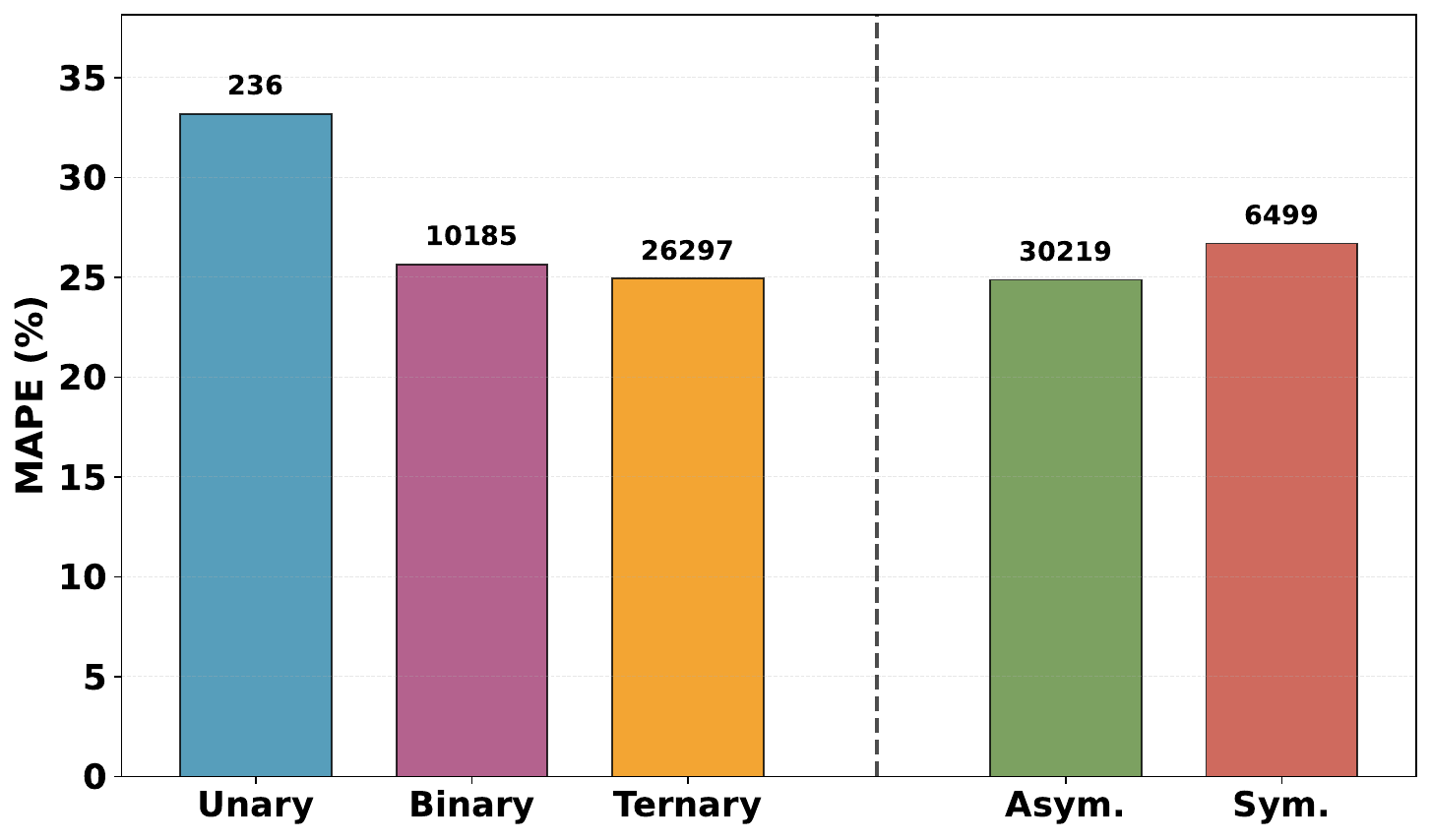}
\caption{\textbf{MAPE analysis by compositional complexity and surface symmetry for MACE--MP--0b3.} 
Bar chart showing prediction errors grouped by compositional complexity (unary, binary, ternary) and surface symmetry (asymmetric, symmetric), with sample counts indicated above each bar. The dashed line separates composition-based analysis (left) from symmetry-based analysis (right).}
\label{fig:mace_mp_0b3_composition_symmetry}
\end{figure}

\begin{figure}[ht]
\centering
\includegraphics[width=0.8\columnwidth]{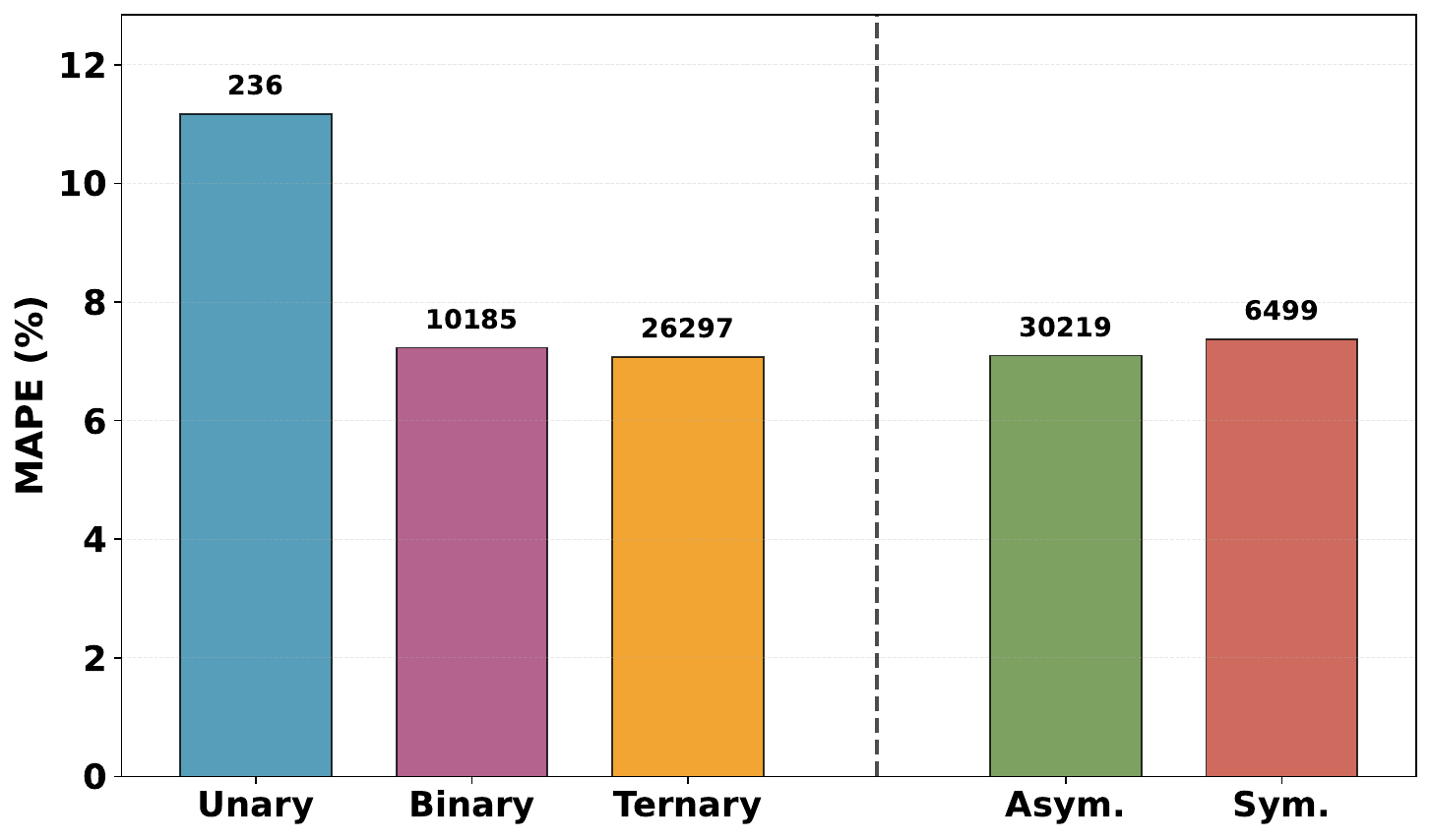}
\caption{\textbf{MAPE analysis by compositional complexity and surface symmetry for SevenNet--MF--OMPA.} 
Bar chart showing prediction errors grouped by compositional complexity (unary, binary, ternary) and surface symmetry (asymmetric, symmetric), with sample counts indicated above each bar. The dashed line separates composition-based analysis (left) from symmetry-based analysis (right).}
\label{fig:sevennet_mf_ompa_composition_symmetry}
\end{figure}

\begin{figure}[ht]
\centering
\includegraphics[width=0.8\columnwidth]{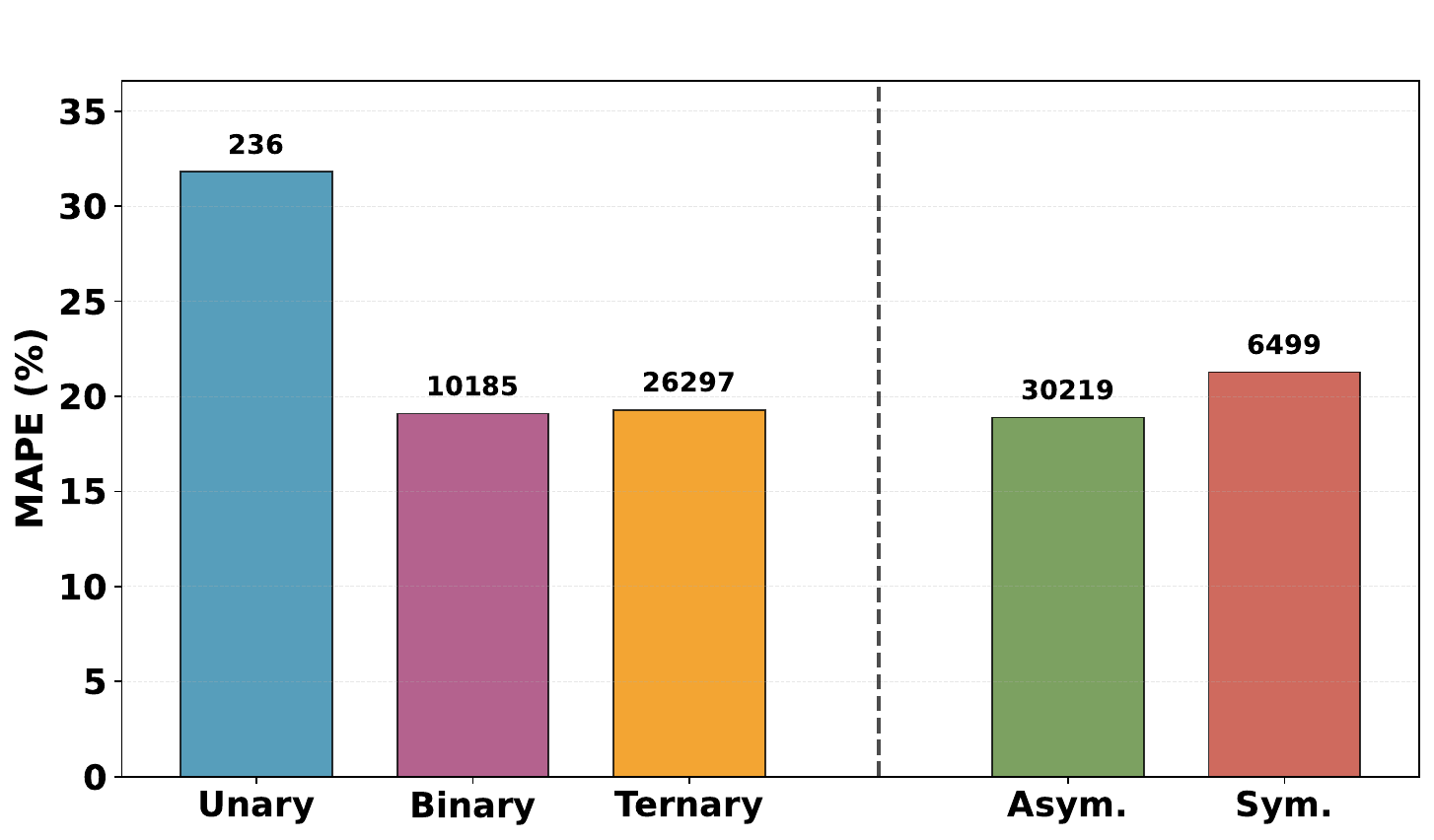}
\caption{\textbf{MAPE analysis by compositional complexity and surface symmetry for MatterSim--v1.} 
Bar chart showing prediction errors grouped by compositional complexity (unary, binary, ternary) and surface symmetry (asymmetric, symmetric), with sample counts indicated above each bar. The dashed line separates composition-based analysis (left) from symmetry-based analysis (right).}
\label{fig:mattersim_v1_composition_symmetry}
\end{figure}

\begin{figure}[ht]
\centering
\includegraphics[width=0.8\columnwidth]{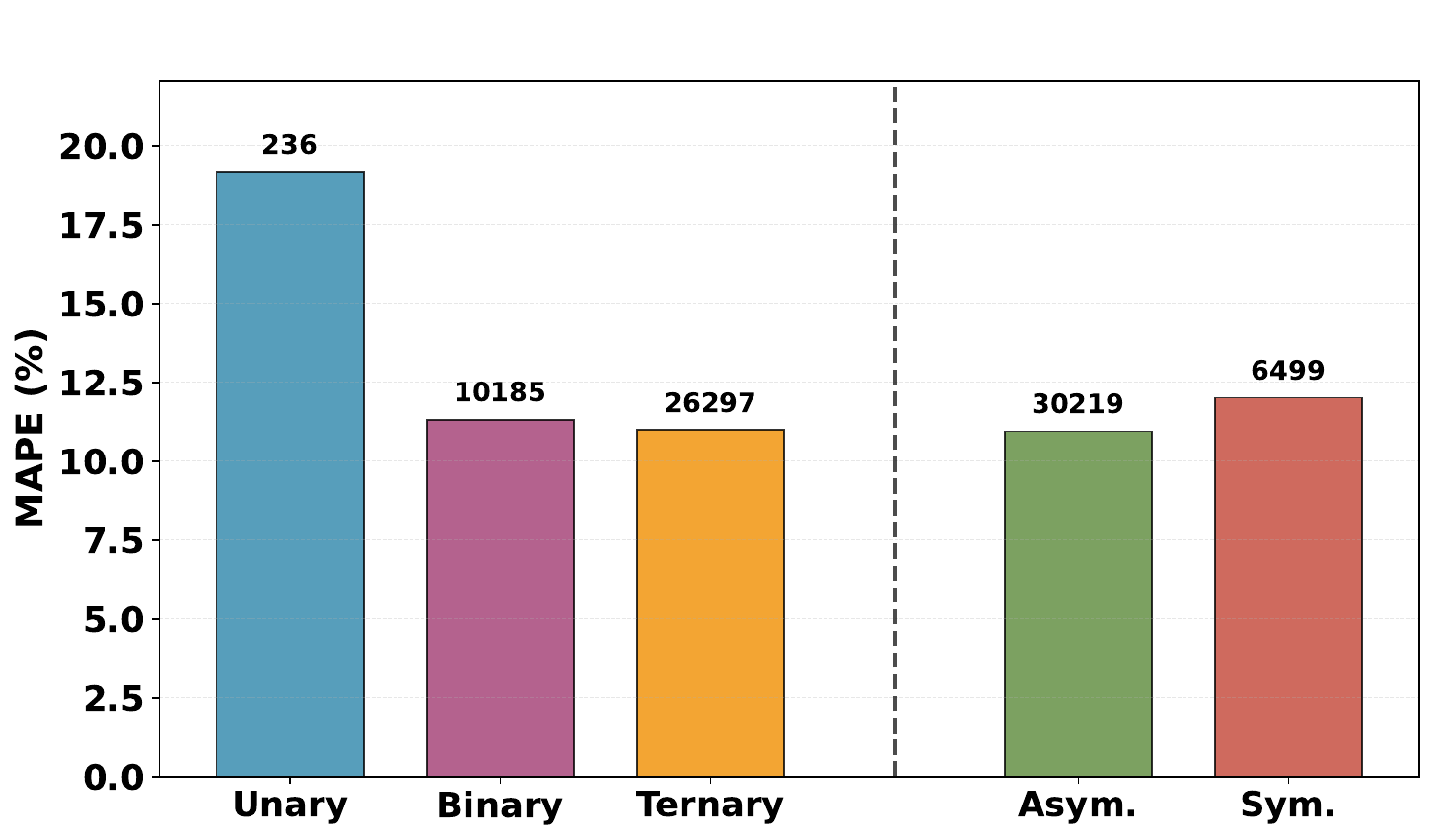}
\caption{\textbf{MAPE analysis by compositional complexity and surface symmetry for AlphaNet--OMA.} 
Bar chart showing prediction errors grouped by compositional complexity (unary, binary, ternary) and surface symmetry (asymmetric, symmetric), with sample counts indicated above each bar. The dashed line separates composition-based analysis (left) from symmetry-based analysis (right).}
\label{fig:alphanet_oma_composition_symmetry}
\end{figure}

\begin{figure}[ht]
\centering
\includegraphics[width=0.8\columnwidth]{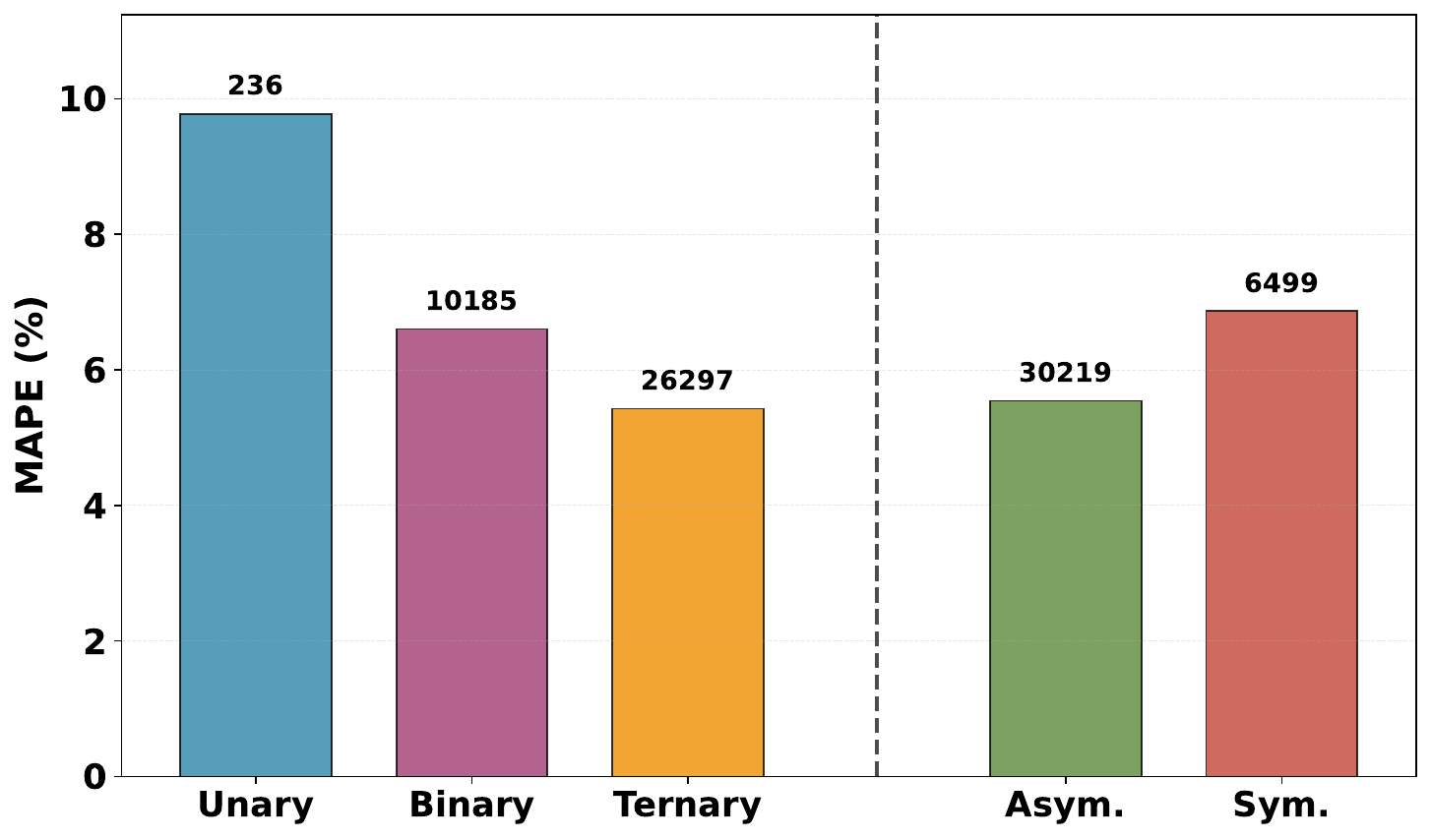}
\caption{\textbf{MAPE analysis by compositional complexity and surface symmetry for UMA--m--1p1--OMAT24.} 
Bar chart showing prediction errors grouped by compositional complexity (unary, binary, ternary) and surface symmetry (asymmetric, symmetric), with sample counts indicated above each bar. The dashed line separates composition-based analysis (left) from symmetry-based analysis (right).}
\label{fig:uma_m_1p1_composition_symmetry}
\end{figure}

\clearpage

\bibliography{references} 
\bibliographystyle{rsc} 